\documentclass[12pt,letterpaper,showpacs,english,aps,preprint,nofootinbib]{revtex4}
\usepackage{subfigure}
\usepackage{color}
\usepackage{graphicx}
\usepackage{amssymb}
\makeatletter

\providecommand{\tabularnewline}{\\}


\usepackage{babel}
\makeatother
\begin{document}
\begin{flushright}
MSUHEP-040903 \\
hep-ph/0409040
\end{flushright}

\newcommand{\oalphas}{O(\alpha_{s})}
\newcommand{\met}{\not\!\! E_{T}}

\title{Next-to-Leading Order Corrections to Single Top Quark Production
and Decay at the Tevatron: 1. s-channel Process}
\author{Qing-Hong Cao}
\email{cao@pa.msu.edu}
\author{Reinhard Schwienhorst}
\email{schwier@pa.msu.edu}
\author{C.-P. Yuan}
\email{yuan@pa.msu.edu}
\affiliation{
{Department of Physics $\&$ Astronomy, \\
Michigan State University,\\
East Lansing, MI, 48824, USA.\\ }}

\vspace{0.15in}

\begin{abstract}
We present a study of s-channel single top quark production at the
upgraded Tevatron $p\bar{p}$ collider, including the next-to-leading
order (NLO) QCD corrections to the production and decay of the
top quark. The ``modified'' narrow width approximation was adopted
to preserve the spin of the top quark in its production and decay.
We discuss the effect of the different $\oalphas$ contributions on
the inclusive cross section as well as various kinematical distributions
after imposing the relevant cuts to select s-channel single top signal
events. In particular the $\oalphas$ decay contribution, while small
in size, has a significant impact on several distributions. With the
help of the best-jet algorithm to reconstruct the top quark we demonstrate
that it is possible to study kinematical and spin correlations in
s-channel single top events. We furthermore compare top quark spin
measurements in two different basis and show how NLO corrections have
to be taken into consideration in searches for the Higgs boson through
$W^{\pm}H$ associated production at the Tevatron. 
\end{abstract}


\maketitle

\section{Introduction}

At the Fermilab Tevatron, the dominant single top quark production
mechanisms are the s-channel process $q\bar{q}'\rightarrow W^{*}\rightarrow t\bar{b}$~\cite{Cortese:1991fw,Stelzer:1995mi,Smith:1996ij,Mrenna:1997wp,Tait:1997fe}
and the t-channel process $bq\rightarrow tq'$ (including $b\bar{q}'\rightarrow t\bar{q}$,
also referred to as the $W$-gluon fusion process)~ \cite{Willenbrock:1986cr,Yuan:1989tc,Ellis:1992yw,Carlson:1993dt,Bordes:1994ki,Heinson:1996zm,Stelzer:1997ns}.
It has been shown that it is possible to disentangle the single top
signals from various backgrounds\cite{Amidei:tev2000}. While top
quark pair production is primarily a QCD process, single top quark
production occurs dominantly in electroweak processes. It thus provides
a probe of electroweak properties of the top quark. For example, one
can study the Cabibbo-Kobayashi-Maskawa 
(CKM) matrix element $V_{tb}$, new $W$-$t$-$b$ couplings
beyond the Standard Model (SM)~\cite{Kane:1991bg,Carlson:1994bg,Rizzo:1995uv,Datta:1996gg,Li:1996ir,Tait:1996dv,Li:1997qf,Whisnant:1997qu,Hikasa:1998wx,Boos:1999dd,Tait:2000sh,Espriu:2001vj},
and \textcolor{black}{C}P violation~\cite{Yuan:1994fn,Atwood:1996pd,Bar-Shalom:1997si}
in single top quark production and decay. The angular correlations
among the decay products of polarized top quarks also provide a useful
handle on these couplings~\cite{Mahlon:1996pn,Mahlon:1998uv,Mahlon:1999gz,vanderHeide:2000fx,Fischer:2001gp,Espriu:2002wx,Boos:2002xw,delAguila:2002nf}.
The t-channel production process can also provide us with information
about the $b$ quark distribution function inside the proton. The
s-channel process is also a significant background to Higgs searches
at the Tevatron in the production process $q\bar{q}'\rightarrow WH$
with decay $H\rightarrow b\bar{b}$~\cite{Stange:1993ya,Stange:1994bb,Belyaev:1995gb}
and other new physics search~\cite{Cao:2003tr}. The expected single top
production rate, associated with a $W$-boson, via the process $bg\rightarrow tW^{-}$~\cite{Moretti:1997ng,Ladinsky:1990ut,Tait:1999cf,Belyaev:2000me,Zhu:2001hw}
is too small to be observed at the Tevatron and therefore not considered
in this paper. 

Searches for single top production were performed by both D\O~\cite{Abbott:2000pa}
and CDF~\cite{Acosta:2004er} in Run~I. At the 95\% confidence level,
the D\O~ limit on the s-channel production is 17 picobarn (pb) and
the CDF limit is 18 pb. At the same confidence level, the D\O~limit
on the t-channel production cross section is 22 pb and the CDF limit
is 13 pb. The s-channel and t-channel single top quark processes can
be probed separately at the Tevatron by taking advantage of $b$-tagging
using displaced vertices and differences in the kinematical distributions
of the $b$-tagged jet. Usually, only one $b$-tagged jet can be expected
in the t-channel case while two $b$-tagged jets can be expected in
the s-channel case. This is because the $\bar{b}$ quark produced
with the top quark tends to be collinear with the initial gluon in
the t-channel, therefore having a transverse momentum ($p_{T}$) that
is too low to be observed. We shall focus on the s-channel single
top quark process in this paper, the t-channel results will be discussed
elsewhere~\cite{Cao:2004Pri1}.

The extraction of a signal is more challenging for single top production
than for top pair production since there are fewer objects in the
final state and the overall event properties are less distinct from
the large $W$+jets background. Therefore, an accurate calculation
including higher order QCD corrections is needed. The next-to-leading
order (NLO) $\oalphas$ corrections to single top quark production
have already been carried out in Refs.~\cite{Smith:1996ij,Bordes:1994ki}.
Similarly, studies of the $\oalphas$ corrections to the top quark
decay have been done~\cite{Jezabek:1994zv}. However, the effect
of the decay of the top quark and the top quark width have not been
included in the previous studies of kinematics which focused on the
differential cross section in on-shell production~\cite{Harris:2002md,Sullivan:2004ie}.
Since top quark production and decay do not occur in isolation from
each other, a fully differential calculation should include both kinds
of corrections together and maintain all spin and angular correlations.
In Ref.~\cite{Cao:2004ky}, a formalism is presented which calculates
the complete next-to-leading order QCD corrections to production and
decay of single top quarks. In that calculation, the ``modified''
narrow width approximation is used to separate the production of the
top quarks from their decay. Top quark width effects are also included.
In order to study the spin properties of the top quark, the helicity
amplitude method is used in the matrix element calculation, thus keeping
full spin information for all final state objects. Ref.~\cite{Cao:2004ky}
adopts the phase space slicing method (with one cutoff scale) to calculate
the NLO differential cross section. A preliminary phenomenology study
using this method was presented in Ref.~\cite{Cao:2004Pheno}. A
similar study using the subtraction method has also been published
recently~\cite{Campbell:2004ch}. 

In this study, we construct a next-to-leading order Monte Carlo calculation
which treats $\oalphas$ corrections in a consistent way to both production
and decay of the top quarks at the upgraded Fermilab Tevatron.%
\footnote{A similar study of top quark pair production and decay in $e^{+}e^{-}$
annihilation was given in Ref.~\cite{Schmidt:1995mr}.%
} We examine various kinematical event properties in s-channel single
top quark production. Our approach will allow experimentalists to
compare their results with the theoretical predictions directly. A
detailed phenomenological study, including background processes will
be discussed elsewhere~\cite{Cao:2004Pri2}. The paper is organized
as follows: In Sec.~\ref{sec:InclXS}, we present the inclusive cross
section of the s-channel single top quark processes, including an
examination of the theoretical cutoff dependence. The acceptance for
s-channel single top events under various scenarios of kinematical cuts
is presented in Sec.~\ref{sec:Acceptance}. We study final state
kinematical distributions in Sec.~\ref{sec:EventDistr}. In Sec.~\ref{sec:Higgs},
we consider the s-channel single top quark processes as a background
to SM Higgs boson searches in the channel $q\bar{q}'\rightarrow WH$ with
decay $H\rightarrow b\bar{b}$. Finally, we give our conclusions in
Sec.~\ref{sec:Conclusions}.

\section{Cross Section (Inclusive Rate)\label{sec:InclXS}}

In order to be able to compare theoretical predictions with experimental
results it is important to not only determine the total production
rate of single top events but also the rate of events passing kinematical
cuts due to detector acceptances or the need for background suppression.
Furthermore, reconstructing the top quark is critical to many of the
physics goals of the Tevatron and the LHC. At the Tevatron Run II,
experiments may hope for an accuracy of 2 GeV~\cite{Amidei:tev2000}
in the top mass measurement. But the ability to achieve this accuracy
depends on how well systematic effects - especially those associated
with gluon radiation - are understood and controlled. It is therefore
crucial to properly simulate final state particle distributions in
single top events. In this section, we discuss inclusive production
rates while the next section focuses on the acceptance of s-channel
single top events under various kinematical cuts. The following section
is devoted to a discussion of distributions of final state particles
with an emphasis on the effects of gluon radiation.

We present numerical results for the production of single top events
considering the leptonic decay of the $W$-boson from the top quark
decay at the upgraded Tevatron (a 1.96~TeV $p\bar{p}$ collider).
Unless otherwise specified, we use the NLO parton distribution function
set CTEQ6M~\cite{Pumplin:2002vw}, defined in the $\overline{MS}$
scheme, and the NLO (2-loop) running coupling $\alpha_{s}$ with $\Lambda_{\overline{MS}}$
provided by the PDFs. For the CTEQ6M PDFs, $\Lambda_{\overline{MS}}^{(4)}=0.326~{\textrm{GeV}}$
for four active quark flavors. The values of the relevant electroweak
parameters are: $\alpha=1/137.0359895$, $G_{\mu}=1.16637\times10^{-5}\,{\rm GeV}^{-2}$,
$m_{t}=178\,{\textrm{GeV}}$~\cite{Azzi:2004rc,Abazov:2004cs}, $m_{W}=80.33\,{\textrm{GeV}}$,
$m_{Z}=91.1867\,{\rm GeV}$, and $\sin^{2}\theta_{W}=0.231$. Thus,
the square of the weak gauge coupling is $g^{2}=4\sqrt{2}m_{W}^{2}G_{\mu}$.
Here, we focus our attention on the positively charged electron lepton
(i.e., positron) only, though our analysis procedure also applies
to the $\mu$ lepton. Including the $\oalphas$ corrections to $W\to\bar{q}q^{\prime}$,
the decay branching ratio of the $W$ boson into leptons is $Br(W\to\ell^{+}\nu)=0.108$~\cite{Cao:2004yy}.
Unless otherwise specified, we will choose the top quark mass to be
178 GeV and the renormalization scale ($\mu_{R}$) as well as the
factorization scale ($\mu_{F}$) to be equal to the top quark mass. 
The top quark mass and scale dependences of single top events
are investigated in the second and third part of this section.

\subsection{Theoretical Cutoff Dependence}

The total cross section up to NLO can be obtained by adding up the
contributions from the Born level contribution and the virtual correction
and the real correction with either soft or hard gluons. We adopt
the phase space slicing method with one cutoff scale ($s_{min}$)
and the universal crossing functions~\cite{Giele:1991vf,Giele:1993dj,Keller:1998tf}
to regularize both soft and collinear singularities.
The cutoff $s_{min}$ serves to separate the real correction phase
space into two regions: (1) the resolved region in which the amplitude
has no divergences and can be integrated numerically by Monte Carlo
(MC) methods and (2) the unresolved region in which the amplitude
contains all the soft and collinear divergences and can be integrated
out analytically. The resulting divergences will be canceled by the
virtual corrections or absorbed into the parton structure functions
in the factorization procedure. Since this cutoff is introduced in
the calculation only for this technical reason and is unrelated
to any physical quantity, the result must not depend on it. In other
words, the sum of all contributions, virtual, resolved, and unresolved
corrections must be independent of $s_{min}$. This is the case as
long as $s_{min}$ is small enough so that the soft and collinear
approximations are valid. However, the numerical cancellation in the
MC integration becomes unstable if $s_{min}$ is too small. Furthermore,
the jet-finding algorithm and other infrared-safe experimental observables
also should be defined in a way that is consistent with the choice
of $s_{min}$. In practice, one wants to choose the largest $s_{min}$
possible within these constraints in order to minimize the running
time of the MC integration program. For our study, we found a value
of $s_{min}=5\,{\textrm{GeV}}^{2}$ to be appropriate. A detailed
discussion of the phase space slicing method can be found in the Ref.~\cite{Cao:2004ky}.

\begin{figure}
\includegraphics[%
  scale=0.5]{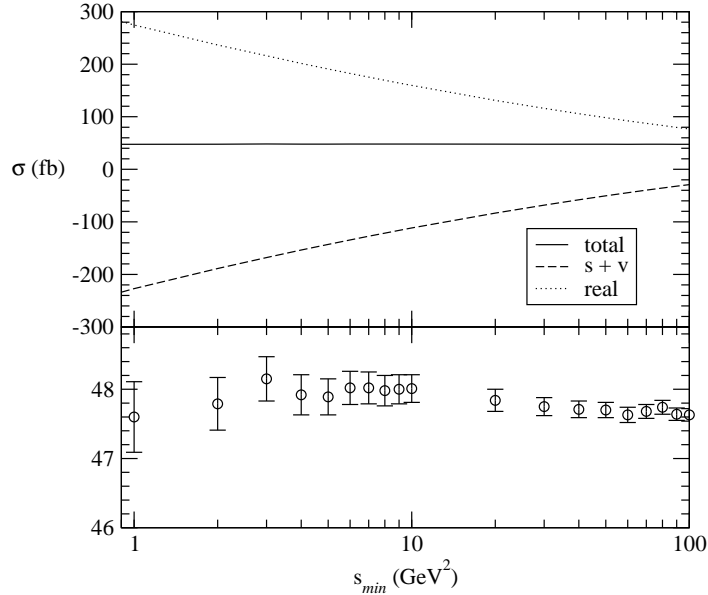}

\caption{The theoretical cutoff $s_{min}$ dependence of the inclusive s-channel
single top quark cross section at the Tevatron with $\mu_{R}=\mu_{F}=m_{t}$
for $m_{t}=178\,{\rm GeV}$. The decay branching ratio $t\rightarrow bW^{+}(\rightarrow e^{+}\nu)$
has been included.\label{fig:deps-smin}}
\end{figure}

That the total rate is indeed insensitive to the value of $s_{min}$
for a large range is illustrated in Fig.~\ref{fig:deps-smin}, which
shows the sum of the virtual and unresolved real corrections ($s+v$)
as well as the resolved contribution ($real$) to the s-channel single
top quark process as a function of $s_{min}$. Although the contributions
from the individual pieces vary, their sum ($total$) remains essentially
constant for a large range of $s_{min}$\emph{.} With the choices
of $\mu_{R}=\mu_{F}=m_{t}$ at the Tevatron, we obtain an inclusive
cross section of the s-channel single top ($t$ only) processes (with
$W$-boson decay branching ratio) as 47.9 fb, which agrees with Ref.~\cite{Campbell:2004ch}
with the understanding that different electroweak parameters are used. \emph{}

\subsection{Top Quark Mass and Renormalization/Factorization Scale Dependence}

To test the standard model and measure the CKM matrix element $V_{tb}$,
one needs an accurate prediction of the single top quark production
and decay to reduce the theoretical uncertainty. Examining the top
quark mass dependence can provide information about how accurately
the top quark mass must be measured in $t\bar{t}$ events in order to 
reduce the uncertainty
on the extraction of the CKM element $V_{tb}$ from the measured 
single top cross section. 
Besides the top quark mass, the choices of renormalization and factorization
scales also contribute to the uncertainty of the theoretical prediction.
The renormalization scale $\mu_{R}$ is introduced when redefining
the bare parameters in terms of the renormalized parameters, while
the factorization scale $\mu_{F}$ is introduced when absorbing the
collinear divergence into the parton distribution functions. Therefore,
both $\mu_{R}$ and $\mu_{F}$ are unphysical and the final predictions
should not depend on them. However, since we work at a fixed order
in perturbation theory, we indeed see dependences of the predicted
cross section on $\mu_{R}$ and $\mu_{F}$. The change due to varying
the scale is formally of higher order. Since the single top rate is
small at the Tevatron, it is very important to reduce the scale
uncertainty in order to compare the theory prediction with experimental
data. Here, we examine the top quark mass and the scale dependence
of s-channel single top quark events.

\begin{figure}
\includegraphics[scale=0.6]{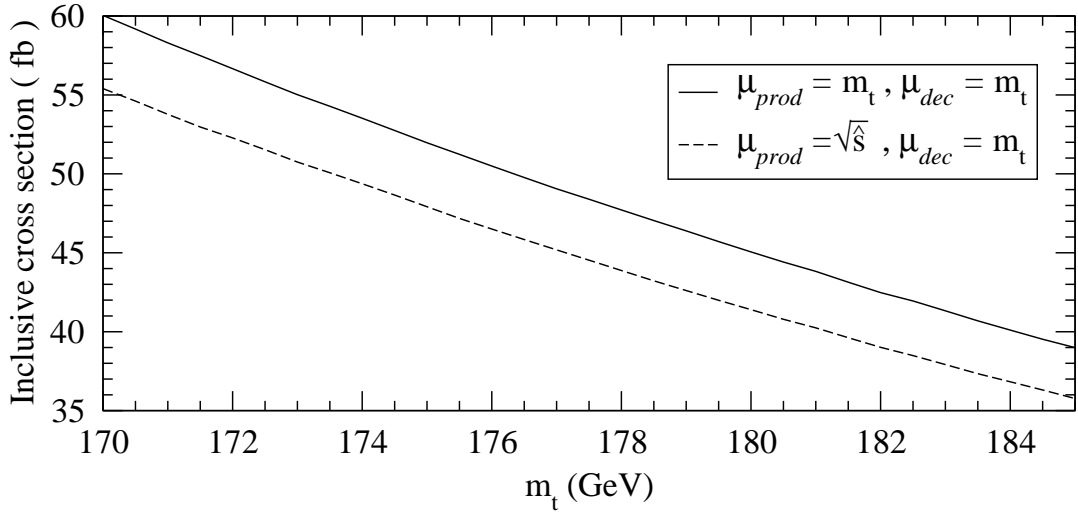}
\caption{Top quark mass as well as renormalization and factorization scale
dependence of the inclusive s-channel single top quark cross section
at Tevatron. The decay branching ratio of $t\to bW^{+}(\to e^{+}\nu)$
has been included. \label{fig:deps-mt}}
\end{figure}

As shown in Fig.~\ref{fig:deps-mt}, the cross section changes by
about $\pm10\%$ when the top quark mass $m_{t}$ is varied by its
current uncertainty of about $\mp5\,{\rm GeV}$. Measuring the top
quark mass to an uncertainty of $1-2\,{\rm GeV}$ will thus reduce
the theoretical uncertainty on the single top cross section. 
The reduction in uncertainty of the s-channel single top production 
rate will improve the measurement of the CKM matrix element $V_{tb}$.

\begin{figure}
\includegraphics[scale=0.7]{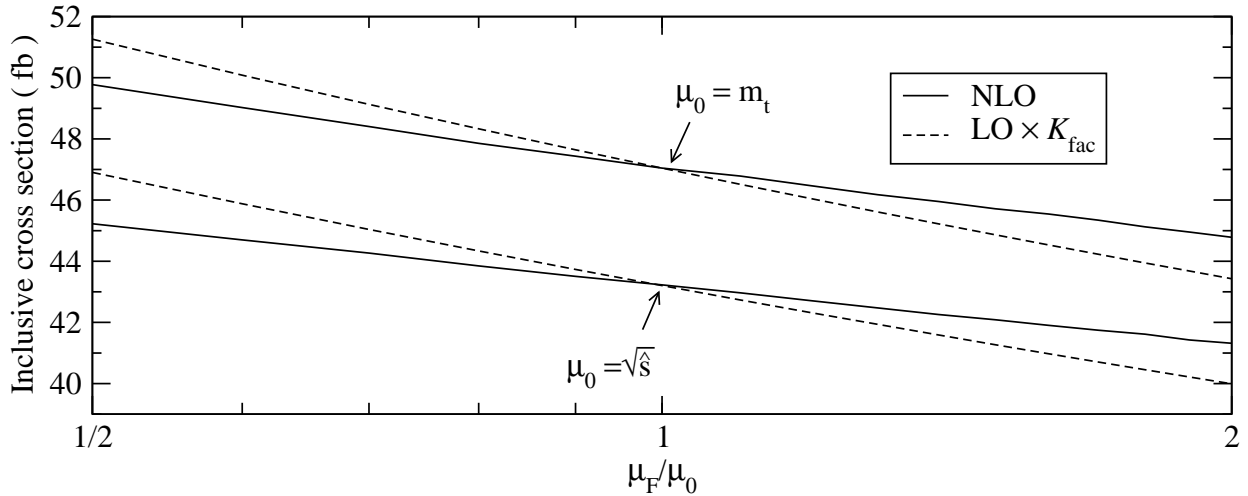}
\caption{Inclusive s-channel single top quark cross section
at Tevatron for $m_t=178$ GeV, versus the ratio of the factorization scale
$\mu_F$ to its typical value $\mu_0$, where $\mu_0=m_t$ and $\mu_0=\sqrt{\hat{s}}$, respectively. 
The decay branching ratio of $t\to bW^{+}(\to e^{+}\nu)$
has been included. \label{fig:varyscale}}
\end{figure}

To examine the scale dependence of the s-channel single 
top quark production
rate, we show in Fig.~\ref{fig:deps-mt} the results of
two typical scales; one is
 the top quark mass ($\mu_{F}=\mu_{R}^{prod}=m_{t}$),
shown as the solid-line, and another is the total invariant mass of the event
($\mu_{F}=\mu_{R}^{prod}=\sqrt{\hat{s}}$), shown as the dashed-line.
For the decay of top quark, we take 
$\mu_{R}^{dec}=m_{t}$ which gives similar results as the choice of 
$\mu_{R}^{dec}=M_W$. 
The band constrained by these two $\mu_{F}$ scales represents 
a range of uncertainty due to the NLO predictions.  
The usual practice for estimating the yet-to-be calculated higher order 
QCD correction to a perturbative cross
section is to vary around the typical scale by a factor of 2, though in
principle the ``best'' scale to be used for estimation cannot be determined 
without completing the higher order calculation.
In Fig.~\ref{fig:varyscale} we show the total cross section of the
s-channel single top production for a range of the scale $\mu_{F}$ 
which, for simplicity, is set to be equal to $\mu_{R}$. 
We have multiplied a constant factor (the ``K-factor'') of 1.52 
to the LO cross sections (shown as the dashed-line) 
in order to compare to the NLO ones (shown as the solid-line). 
It is clear that the NLO calculation reduces the scale
dependence. For example, around $\mu_F=m_t$, the LO rate varies by 
$+8.9\%$ to $-5.7\%$ when scale is changed from $\mu_F/2$ to $2 \mu_F$,
while the NLO rate varies by $+7.6\%$ to $-4.7\%$. Similar results also
hold for varying the scale around $\mu_F=\sqrt{\hat{s}}$.

\subsection{Inclusive Cross Section}

Due to the electroweak nature of single top quark production, the
higher-order QCD corrections can be divided into three separate gauge
invariant sets: the corrections from the initial state of top quark
production (INIT), the corrections from the final state of top quark
and $\bar{b}$ quark (FINAL), and the corrections from the top quark
decay (SDEC). The explicit diagrams and definitions for these three
sets can be found in Ref.~\cite{Cao:2004ky}. The inclusive cross
section as well as the individual $\oalphas$ contributions are listed
in Table~\ref{tab:inclusive}. The effect of the finite widths of
the top quark and $W$-boson has been included. We use a ``modified''
narrow width approximation (MNWA) in the calculation instead of the
usual narrow width approximation (NWA). In the usual NWA, the effect
of the Breit-Wigner resonance propagator reduces to a delta-function
in the limit of vanishing top quark width, i.e. 
\begin{equation}
\int dp^{2}\frac{1}{\left(p^{2}-m_{t}^{2}\right)^{2}+m_{t}^{2}\Gamma_{t}^{2}}
=\frac{\pi}{m_{t}\Gamma_{t}}\delta\left(p^{2}-m_{t}^{2}\right).
\label{eq:nwa_phase}
\end{equation}
Therefore, the invariant mass of the top quark decay particles
will be exactly equal to $m_t$ (a fixed value).
To model the reconstructed top quark invariant mass from its decay particles
with a Breit-Wigner resonance shape to reflect the non-vanishing decay 
width of the top quark (for being an unstable resonance), we introduced
the MNWA method in our numerical calculation in
which we generate a Breit-Wigner distribution for the top quark invariant
mass in the phase space generator and then calculate the squared matrix 
element with $m_t$ being the invariant mass 
generated by the phase space generator on the event-by-event basis.
We find that the total event rate 
and the distributions of various kinematics variables 
(except the distribution of the reconstructed top quark invariant mass) 
calculated using the ``modified NWA'' method agree well with 
that using the NWA method. 
In the NWA method, the reconstructed top quark invariant mass
distribution is a delta-function, while in the 
``modified NWA'' method, it is almost a Breit-Wigner distribution. 
The reason that the ``modified NWA'' method does not generate a perfect 
Breit-Wigner shape in the distribution of the top quark invariant mass 
is because the initial state 
parton luminosities (predominantly due to valence quarks) 
for the s-channel single-top process drop rapidly 
at the relevant Bjorken-$x$ range, where 
$\left\langle x\right\rangle \simeq \frac{m_t}{\sqrt{s}} \sim 0.1$. 
As shown in Table~\ref{tab:inclusive}, the total
$\oalphas$ QCD corrections increase the Born level cross section
of the single top quark event by $54\%$. The INIT correction dominates over the FINAL  and SDEC 
corrections, due to the enhancement from collinear physics. The contribution
from the SDEC corrections is furthermore suppressed by almost a factor
of 20 compared to the INIT corrections. This can be understood from
Eqs. (146)-(149) in Ref.~\cite{Cao:2004ky} and is explained below.

When calculating the NLO QCD corrections to the top quark decay process,
we must use the NLO top quark decay width in order to restore the
correct top quark decay branching ratio. With the narrow width approximation,
the NLO differential cross section of the top quark decay process
is given by\begin{equation}
d\sigma_{{\rm NLO}}^{{\rm decay}}=d\sigma_{{\rm LO}}^{{\rm production}}\times\frac{\pi}{\Gamma_{t}^{{\rm NLO}}}\times d\Gamma_{{\rm NLO}}^{{\rm decay}},\label{eq:NLO_SDEC}\end{equation}
where \begin{eqnarray*}
d\Gamma_{{\rm NLO}}^{{\rm decay}} & = & d\Gamma_{0}^{{\rm decay}}+d\Gamma_{1}^{{\rm decay}},\\
\Gamma_{t}^{{\rm NLO}} & = & \Gamma_{t}^{0}+\Gamma_{t}^{1}.\end{eqnarray*}
Here $\Gamma_{t}^{0}$ denotes the Born level top quark decay width,
and $\Gamma_{t}^{1}$ denotes the $O(\alpha_s)$ correction to the top quark
decay width. In our calculation, $\Gamma_{t}^{0}=1.695\,{\rm GeV}$
and $\Gamma_{t}^{{\rm NLO}}=1.558\,{\rm GeV}$~\cite{Li:1990qf}.
The $\oalphas$ SDEC correction includes only the $O(\alpha_{s})$ QCD
correction in Eq.~(\ref{eq:NLO_SDEC}),\begin{equation}
d\sigma_{O(\alpha_{s})}^{{\rm SDEC}}=d\sigma_{{\rm LO}}^{{\rm production}}\times\frac{\pi}{\Gamma_{t}^{{\rm NLO}}}\times d\Gamma_{1}^{{\rm decay}}.\label{eq:DCS_SDEC}\end{equation}
 We obtain\begin{equation}
{\displaystyle \frac{\pi}{\Gamma_{t}^{{\rm NLO}}}=\frac{\pi}{\Gamma_{t}^{0}+\Gamma_{t}^{1}}=\frac{\pi}{\Gamma_{t}^{0}\left(1+{\displaystyle \frac{\Gamma_{t}^{1}}{\Gamma_{t}^{0}}}\right)}\approx\frac{\pi}{\Gamma_{t}^{0}}\left(1-{\displaystyle \frac{\Gamma_{t}^{1}}{\Gamma_{t}^{0}}}\right)+O(\alpha_{s}^{2}).}\label{eq:top_prop}\end{equation}
Expanding Eq.~(\ref{eq:NLO_SDEC}) up to $O(\alpha_{s})$, we get\begin{eqnarray}
d\sigma_{{\rm NLO}}^{{\rm decay}} & = & d\sigma_{{\rm LO}}^{{\rm production}}\times\frac{\pi}{\Gamma_{t}^{{\rm 0}}}\left(1-{\displaystyle \frac{\Gamma_{t}^{1}}{\Gamma_{t}^{0}}}\right)\times\left(d\Gamma_{0}^{{\rm decay}}+d\Gamma_{1}^{{\rm decay}}\right)\nonumber \\
 & = & d\sigma_{{\rm LO}}^{{\rm production}}\times\frac{\pi}{\Gamma_{t}^{{\rm 0}}}\times d\Gamma_{0}^{{\rm decay}}\nonumber \\
 &  & -d\sigma_{{\rm LO}}^{{\rm production}}\times\frac{\pi}{\Gamma_{t}^{{\rm 0}}}\times\frac{\Gamma_{t}^{1}}{\Gamma_{t}^{0}}\times d\Gamma_{0}^{{\rm decay}}+d\sigma_{{\rm LO}}^{{\rm production}}\times\frac{\pi}{\Gamma_{t}^{{\rm 0}}}\times d\Gamma_{1}^{{\rm decay}}+O(\alpha_{s}^{2}).\label{eq:NLO_expand}\end{eqnarray}
 The first term in Eq.~(\ref{eq:NLO_expand}) belongs to the Born
level contribution, and both the second and third terms belong to
the $O(\alpha_{S})$ decay contribution. The net contribution of the
second and third terms should be exactly zero in order to respect
the top quark decay branching ratio (to be one), after we integrate
out the phase space of the top quark decay products. As shown in Table~\ref{tab:inclusive},
there is a small remaining SDEC contribution, which is due to higher
order contributions:

\begin{itemize}
\item The $O(\alpha_{s}^{2})$ term, which is dropped in the expansion in
Eq.~(\ref{eq:NLO_expand}), will give about a $1\%$ contribution.
\item A mismatch exists between $\int d\Gamma_{1}^{{\rm decay}}$ and $\Gamma^{1}$,
since we use a Breit-Wigner distribution of the top quark mass modulated
by its decay width in the MNWA method rather than a fixed mass
(which indicates an on-shell top quark). 
\end{itemize}
This SDEC contribution will become sizable once we use a jet finding
algorithm to define infrared-safe observables and impose the kinematical
cuts. This is especially the case for three-jet events which can only
come from the real emission corrections. However, the decay contribution
cannot be very large because the large top quark mass has already
set the scale for the decay of the top quark which will be shown later.

\begin{table}
\begin{center}\begin{tabular}{c|c|c}
\hline 
&Cross Section (fb) &
Fraction of NLO ($\%$)
\tabularnewline
\hline
Born Level&
 31.2&
 65.0
\tabularnewline
\hline
INIT&
10.7&
22.3
\tabularnewline
FINAL&
5.5&
11.5
\tabularnewline
SDEC&
0.57&
1.19
\tabularnewline
\hline
$O(\alpha_{s})$ sum&
16.8&
35.0
\tabularnewline
\hline 
NLO&
47.9&
100\tabularnewline
\hline
\end{tabular}\end{center}

\caption{Inclusive single top production cross section for different subprocesses,
including the top quark decay branching ratio $t\rightarrow bW^{+}(\rightarrow e^{+}\nu)$.
 \label{tab:inclusive}}
\end{table}

\section{Single Top Acceptance Studies\label{sec:Acceptance}}

In this section we explore the final state objects of s-channel single
top events. Although the $W$-boson from the top quark decay can decay
in both leptonic and hadronic modes, we only focus on the leptonic
decay mode in this study because the all-jet production mode of single
top events is difficult to observe experimentally. Therefore, the
signature of s-channel single top events that is accessible experimentally
consists of one charged lepton, missing transverse energy, and
two or three jets. Since we are studying the effect of NLO QCD
radiative corrections on the production rate and the kinematical distributions
of single top events at the parton level in this paper, we are not
including any detector effects, such as jet energy resolution or b-tagging
efficiency. Only the kinematical acceptance of the detector is considered.
Since the Tevatron is a $p\bar{p}$ collider, and $p\bar{p}$ is a CP-even
state, the production cross sections for $b\bar{t}(j)$ at the Tevatron
are the same as those for $t\bar{b}(j)$ (when ignoring the small
CP violation effect induced by the CKM mixing matrix in quarks). For
simplicity we therefore only show distributions of s-channel single top
events in this work in which a $t$ quark (not including $\bar{t}$)
decays into $W^{+}(\rightarrow\ell^{+}\nu)$ and $b$ quark. 

To meaningfully discuss the effect of gluon radiation in single top
events, we have to define a jet as an infrared-safe observable. The
numerical stability of several jet algorithms commonly used in experiments
has been investigated in Ref.~\cite{Kilgore:1996sq}. In this study,
we adopt the cone-jet algorithm~\cite{Alitti:1990aa} which is organized
as follows:

\begin{enumerate}
\item Build a sorted list of all clusters (in this case the final state
$b$ and $\bar{b}$ quarks plus a light quark or gluon from the $O(\alpha_{s})$
correction), decreasing in transverse energy $E_{T}$.
\item Select the highest $E_{T}$ cluster from the cluster list and draw
a cone of radius $R$ around its axis. Call this a jet and calculate
the transverse jet energy and a new jet axis by summing over the 4-momenta
of all clusters inside the cone.
\item Remove all clusters in the cone from the cluster list and move the
newly constructed jet to the jet list.
\item Apply the appropriate minimum transverse energy and rapidity cuts
to the entries in the jet list to create the final state list of jets.
\end{enumerate}
More specifically, we adopt the $E$-scheme cone-jet (4-momenta of
particles in one cone are simply added to form a jet) with radius
$R=\sqrt{\Delta\eta^{2}+\Delta\phi^{2}}$ to define $b$, $\bar{b}$
and possibly extra $g$ or $q$ (or $\bar{q}$) jets, where $\Delta\eta$
and $\Delta\phi$ are the separation of particles in the pseudo-rapidity
$\eta$ and the azimuthal angle $\phi$. For reference, we shall consider
both $R=0.5$ or $R=1.0$. The same $R$-separation will also be applied
to the separation between the lepton and each jet.

The kinematical cuts imposed on the final state objects are:\begin{eqnarray}
P_{T}^{\ell}\ge15\,{\rm GeV} & , & \left|\eta_{\ell}\right|\le\eta_{\ell}^{max},\nonumber \\
\met\ge15\,{\rm GeV} & ,\nonumber \\
E_{T}^{j}\ge15\,{\rm GeV} & , & \left|\eta_{j}\right|\le\eta_{j}^{max},\nonumber \\
\Delta R_{\ell j}\ge R_{cut} & , & \Delta R_{jj}\ge R_{cut}\label{eq:cuts}\end{eqnarray}
where the jet cuts are applied to both the $b$- and $\bar{b}$-jet
as well as any gluon or light anti-quark jet in the final state. Two
lepton pseudo-rapidity cuts are considered here: a loose version with
$\eta_{\ell}^{max}=2.5$; and a tight version with $\eta_{\ell}^{max}=1.0$.
Similarly, loose and tight cuts are also considered for the jet pseudo-rapidity,
$\eta_{j}^{max}=3.0$ and $\eta_{j}^{max}=2.0$, respectively. The
minimum transverse energy cuts on the lepton as well as the jets is
15~GeV. Each event is furthermore required to have at least one lepton
and two jets passing all selection criteria. The cut on the separation
in $R$ between the lepton and the jets as well as between different
jets is given by $R_{cut}$ which is is chosen to be $0.5$ (or $1.0$).

\begin{table}
\begin{center}\begin{tabular}{c|c|c|c|c|c|c|c|c}
\hline 
\multicolumn{2}{c|}{}&
\multicolumn{5}{c|}{Cross section (fb)}&
\multicolumn{2}{c}{Acceptance (\%)}\tabularnewline
\cline{3-7} \cline{8-9} 
\multicolumn{2}{c|}{}&
LO&
 NLO &
INIT&
FINAL &
SDEC&
LO&
NLO\tabularnewline
\hline
(a)&
 $t\bar{b}+t\bar{b}j$ &
 22.7 &
 32.3 &
6.6&
3.0&
0.16&
72&
67.4\tabularnewline
&
$t\bar{b}+t\bar{b}j$, $\, E_{Tj}>30\,{\textrm{GeV}}$&
17.8&
24.8&
5.5&
1.7&
-0.16&
57&
52\tabularnewline
 &
 $t\bar{b}j$&
 &
 10.6 &
 6.2 &
 2.6 &
 1.8&
&
22.1\tabularnewline
&
$t\bar{b}j,\, E_{Tj}>30\,{\textrm{GeV}}$&
&
3.5&
2.2&
0.92&
0.33&
&
7.3\tabularnewline
\hline
(b)&
 $t\bar{b}+t\bar{b}j$&
19.0&
21.7&
2.0&
1.4&
-0.54&
60.9&
45.3\tabularnewline
&
$t\bar{b}+t\bar{b}j$, $\, E_{Tj}>30\,{\textrm{GeV}}$&
14.8&
16.7&
1.8&
0.70&
-0.48&
47.4&
34.9\tabularnewline
&
 $t\bar{b}j$&
&
5.6&
3.9&
1.1&
0.56&
&
11.7\tabularnewline
&
 $t\bar{b}j,\, E_{Tj}>30\,{\textrm{GeV}}$&
&
1.9&
1.4&
0.40&
0.08&
&
4.0\tabularnewline
\hline
(c)&
$t\bar{b}+t\bar{b}j$ &
14.7&
21.4&
4.4&
2.2&
0.13&
47&
44.7\tabularnewline
&
$t\bar{b}+t\bar{b}j$, $\, E_{Tj}>30\,{\textrm{GeV}}$&
12.2&
16.9&
3.6&
1.3&
-0.17&
39&
35.3\tabularnewline
&
$t\bar{b}j$&
&
6.4&
3.6&
1.7&
1.1&
&
13.4\tabularnewline
&
$t\bar{b}j,\, E_{Tj}>30\,{\textrm{GeV}}$&
&
2.3&
1.5&
0.65&
0.23&
&
4.8\tabularnewline
\hline
\end{tabular}\end{center}

\caption{The s-channel single top production cross section (in fb) and acceptance
at the Tevatron for different subprocesses under various
scenario: (a) is for the loose cuts with $\eta_{l}^{max}=2.5$,
$\eta_{j}^{max}=3.0$, and $R_{cut}=0.5$, (b) is also for the loose cuts, but
with a larger jet clustering cone size (and jet-lepton separation
cut) of $R_{cut}=1.0$, and (c) is for the tight cuts with
$\eta_{l}^{max}=1.0$, $\eta_{j}^{max}=2.0$, and $R_{cut}=0.5$.
The first five columns show the cross section at Born level and NLO as well
as the different $O(\alpha_{s})$ contributions. The last two columns show the 
acceptance at Born level and NLO.
The decay branching ratio $t\rightarrow bW^{+}(\rightarrow e^{+}\nu)$
is included. \label{tab:total}}
\end{table}

In Table~\ref{tab:total}, we show the single top production cross
sections in femtobarns (fb) as well as acceptances for different subprocesses for
several sets of cuts. We apply the $E_T$ cuts listed
in Eq.~(\ref{eq:cuts}) and study three separate sets of values:
\begin{itemize}
\item[(a)] loose cuts with small $R_{cut}$: $\eta_{l}^{max}=2.5$,
$\eta_{j}^{max}=3.0$, and $R_{cut}=0.5$,
\item[(b)] loose cuts with large $R_{cut}$: $\eta_{l}^{max}=2.5$,
$\eta_{j}^{max}=3.0$, and $R_{cut}=1.0$,
\item[(c)] tight cuts with small $R_{cut}$: $\eta_{l}^{max}=1.0$, 
$\eta_{j}^{max}=2.0$, and $R_{cut}=0.5$.
\end{itemize}
A larger value
for $R_{cut}$ reduces the acceptance significantly mainly because
more events fail the lepton-jet separation cut. While this is only
a 16\% reduction at Born level, the difference grows to 33\% at NLO.
Hence, a smaller cone size is preferred to keep the acceptance at
a high level. For events with at least three jets, imposing a harder
cut on the transverse momentum of the third jet ($E_{Tj}>30\,{\rm GeV}$)
only decreases the contribution from INIT and FINAL corrections by
a factor of 3. By contrast, this cut reduced the contribution from
the SDEC correction by a factor of more than five because the top quark
mass sets the scale for the top quark decay contribution rather than
the invariant mass of system. A tighter cut on jet $E_{T}$ also reduces
the relative SDEC contribution in the 3-jet bin.
The $E_{T}$ spectrum of the gluon from the SDEC contribution is softer
because the available phase space is limited by the top quark mass,
which can also be seen below in Fig.~\ref{fig:ptetaJet3}.

We note that the acceptances at Born level and NLO  
for inclusive 2-jet events differ by 7\% for the loose set of cuts 
(case (a) in Table~\ref{tab:total}) and by 5\% for the tight set of cuts 
(case (c) in Table~\ref{tab:total}). In both cases, the acceptance is 
reduced at NLO compared to Born level. This reduction grows to
over 30\% when a larger value for $R_{cut}$ of $1.0$ is used. Since 
the acceptances are quite different between Born level and NLO, 
we cannot use a constant K-factor to scale the Born level distributions
to NLO. The kinematical differences between Born level and NLO that result in different
acceptances will need to be taken into account properly in the measurement of $V_{tb}$. 

\begin{figure}
\subfigure[]{\includegraphics[%
  scale=0.3]{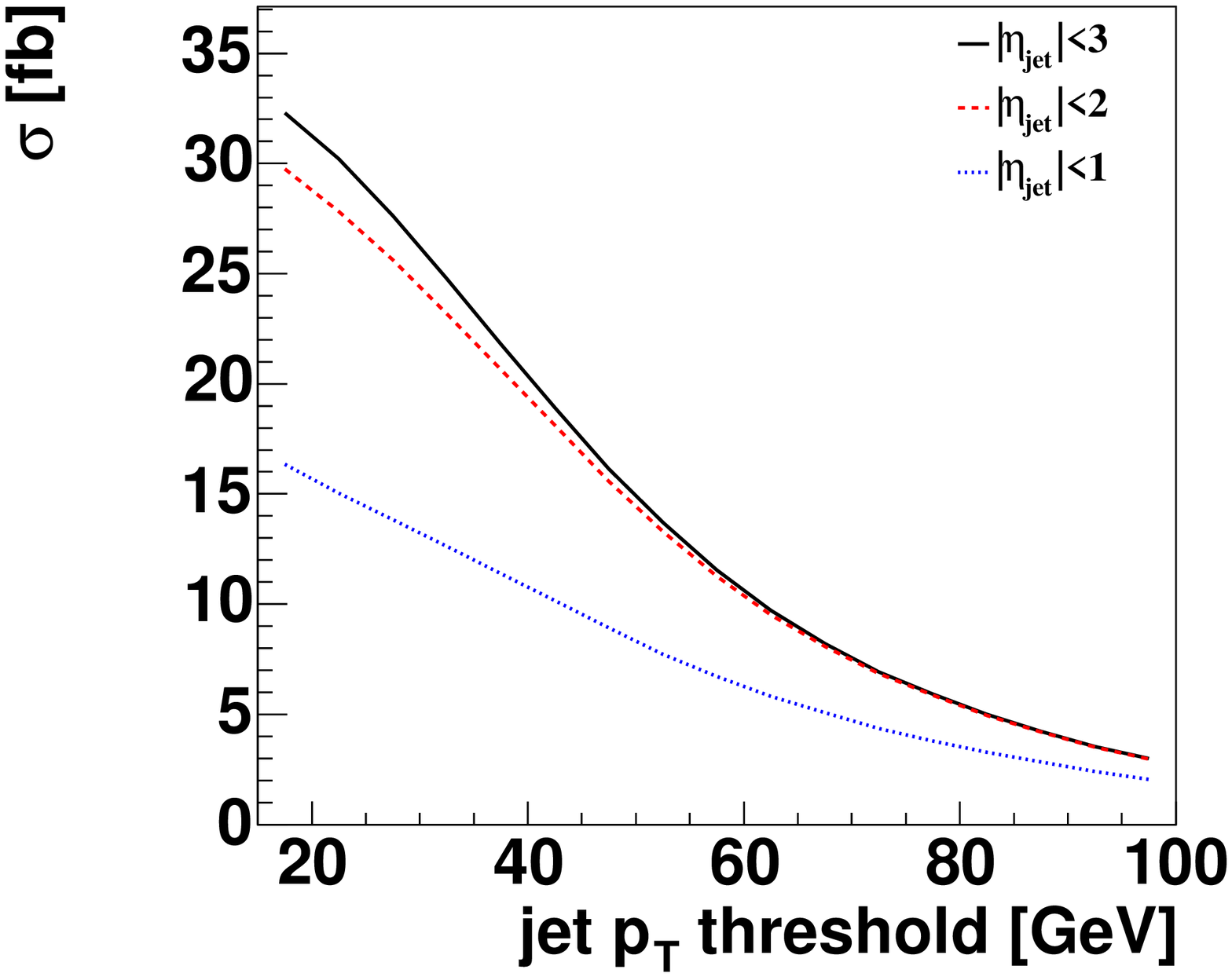}}\subfigure[]{\includegraphics[%
  scale=0.3]{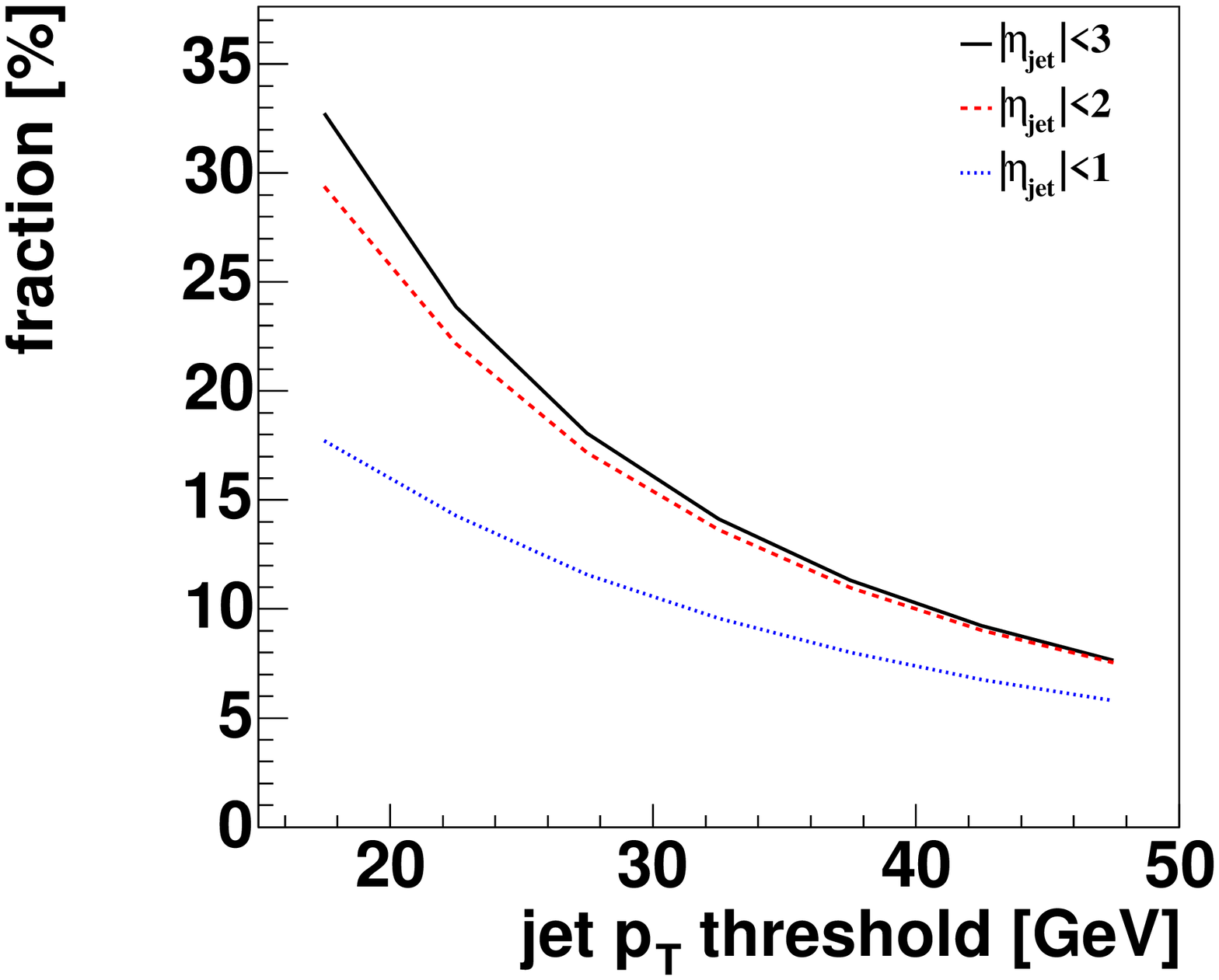}}

\caption{Cross section and fraction of 3-jet events at NLO for varying jet
$E_{T}$ cuts, requiring only that $n_{jets}\geq2$, and not making
any cuts on the electron or neutrino. The left-hand figure shows the
total cross section for events with 2 or 3 jets as a function of the
jet $E_{T}$ cut for three different jet pseudo-rapidity cuts. The
right-hand figure shows the fraction of 3-jet events as a function
of the jet $E_{T}$ for different jet pseudo-rapidity cuts. The lowest
threshold considered is 15 GeV. \label{fig:njets_jet_pt}}
\end{figure}

Fig.~\ref{fig:njets_jet_pt} shows how the observed cross section
changes as a function of the jet $E_{T}$ cut when applying the loose
set of cuts, including a requirement of at least 2 jets. The cross
section (and thus the acceptance) doesn't change very much until the
jet $E_{T}$ threshold reaches about 25 GeV because the $b$- and
$\bar{b}$-jets typically have high $E_{T}$. The figure also shows
that a jet pseudo-rapidity cut of $|\eta|<2$ or $|\eta|<3$ does
not impact the overall acceptance; the cross section only decreases
significantly when restricting jets to the very central region ($|\eta|<1$).
This implies not only that $\eta$ cuts on jets should be large when
separating single top events from the large backgrounds, but also
that $b$-tagging needs to be efficient over a large $\eta$ range
(at least $|\eta|\lesssim2$) to be able to detect s-channel single
top signal events. 

The dependence of the fraction of 2-jet events and 3-jet events on
the jet $E_{T}$ cut is also shown in Fig.~\ref{fig:njets_jet_pt}.
At Born level, there are only 2-jet events, whereas $O(\alpha_{s})$
corrections can produce an additional soft jet. The fraction of events
with these additional jets is low only for very high jet $E_{T}$
thresholds. For typical jet $E_{T}$ thresholds considered by experiments
of 15~GeV to 25~GeV these jets add a significant contribution. As
expected, the effect is not quite as large when only jets within a
very small $\eta$ range are considered because the extra jet typically
has higher $\eta$. In order to study $\oalphas$ effects it is thus
important to set the jet $\eta$ cut as high as possible and the jet
$E_{T}$ cut as low as possible. 

As mentioned before, the event rate for single top events is small
and it is important for experiments to maximize their acceptance.
We will thus use the loose set of cut values for the following discussion:
$\eta_{l}^{max}=2.5$, $\eta_{j}^{max}=3.0$, and $R_{cut}=0.5$,
$E_{Tj}^{min}=15$~GeV, cf. Eq.~(\ref{eq:cuts}).

\section{Single Top Event Distributions\label{sec:EventDistr}}

In this section, we study the kinematical properties of single top quark
events. For the s-channel process at the Born level, two $b$-jets
(the $b$ from the top quark decay and the $\bar{b}$ produced with
the top) appear in the final state and cannot be distinguished from
each other experimentally. A prescription is needed to identify which
of the two jets corresponds to the $b$ quark from the top decay.
At NLO, an addition jet is radiated, which further complicates the
reconstruction of the top quark. This is because the additional jet
can come from either the production or the decay of the top quark.
Production-stage emission occurs before the top quark goes on shell
and decay-stage emission occurs only after the top quark goes on shell.
In production emission events, the $W$ boson and $b$ quark momenta
will combine to give the top quark momentum, while in decay emission
events the gluon momentum must also be include to reconstruct the top
quark momentum. To find the best prescription of classifying the gluon
jet correctly, we first examine various kinematical distributions of
the final state particles. We then investigate two top quark reconstruction
prescriptions: the leading jet algorithm and the best-jet algorithm.
Having chosen a prescription, the effects of NLO corrections on distributions
concerning the reconstructed top quark are studied, in particular
spin correlations between the final state particles. Finally, we explore
the impact of the radiated jet in exclusive 3-jet events.

\subsection{Final State Object Distributions}

In this section we examine various kinematical distributions of final
state objects after event reconstruction and after applying the loose
set of cuts, cf. Table~\ref{tab:inclusive}(a) and Eq.~(\ref{eq:cuts}).
We study inclusive 2-jet events in this section because they give
more reliable infrared-safe predictions. 

\begin{figure}
\includegraphics[%
  scale=0.3]{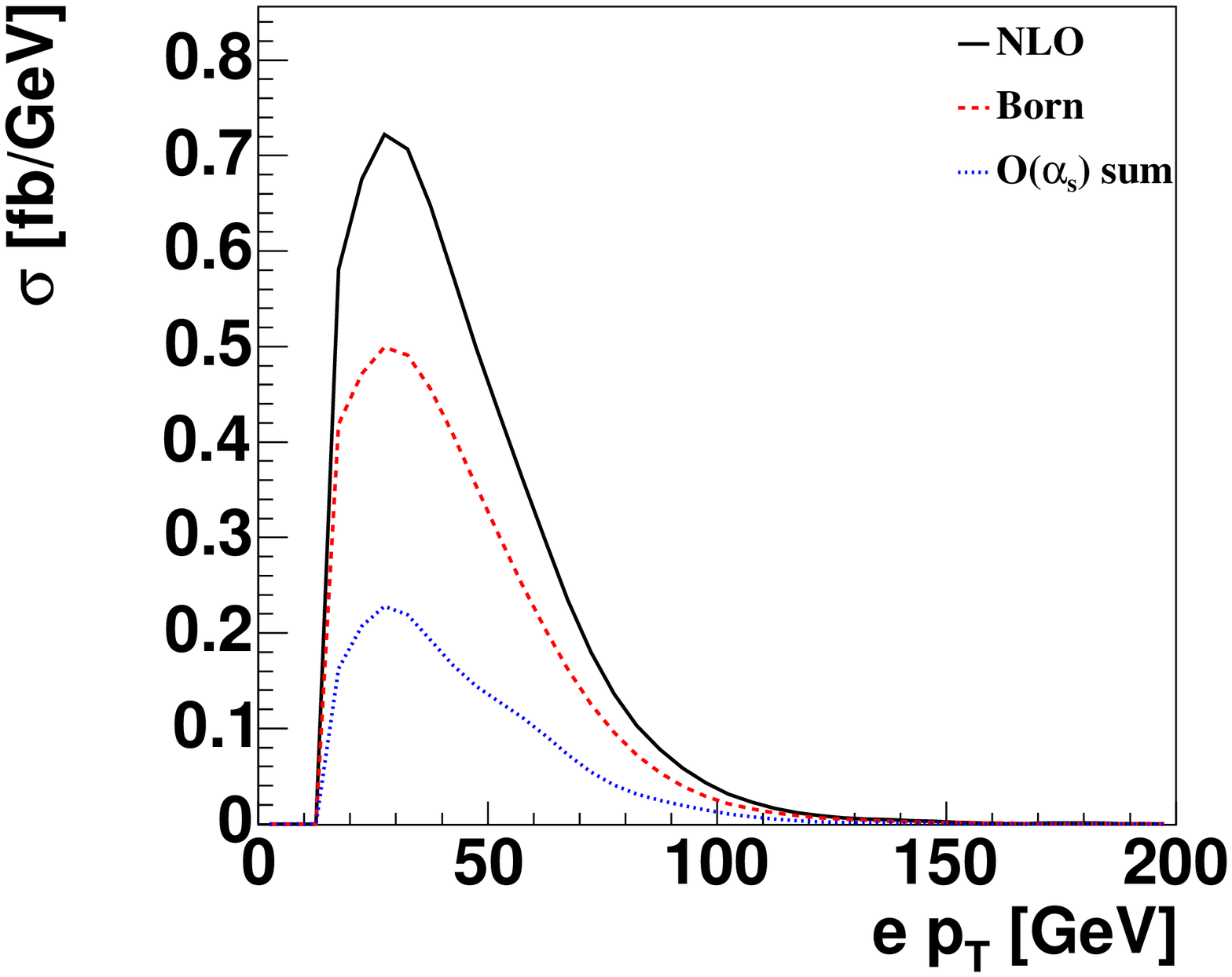}\includegraphics[%
  scale=0.3]{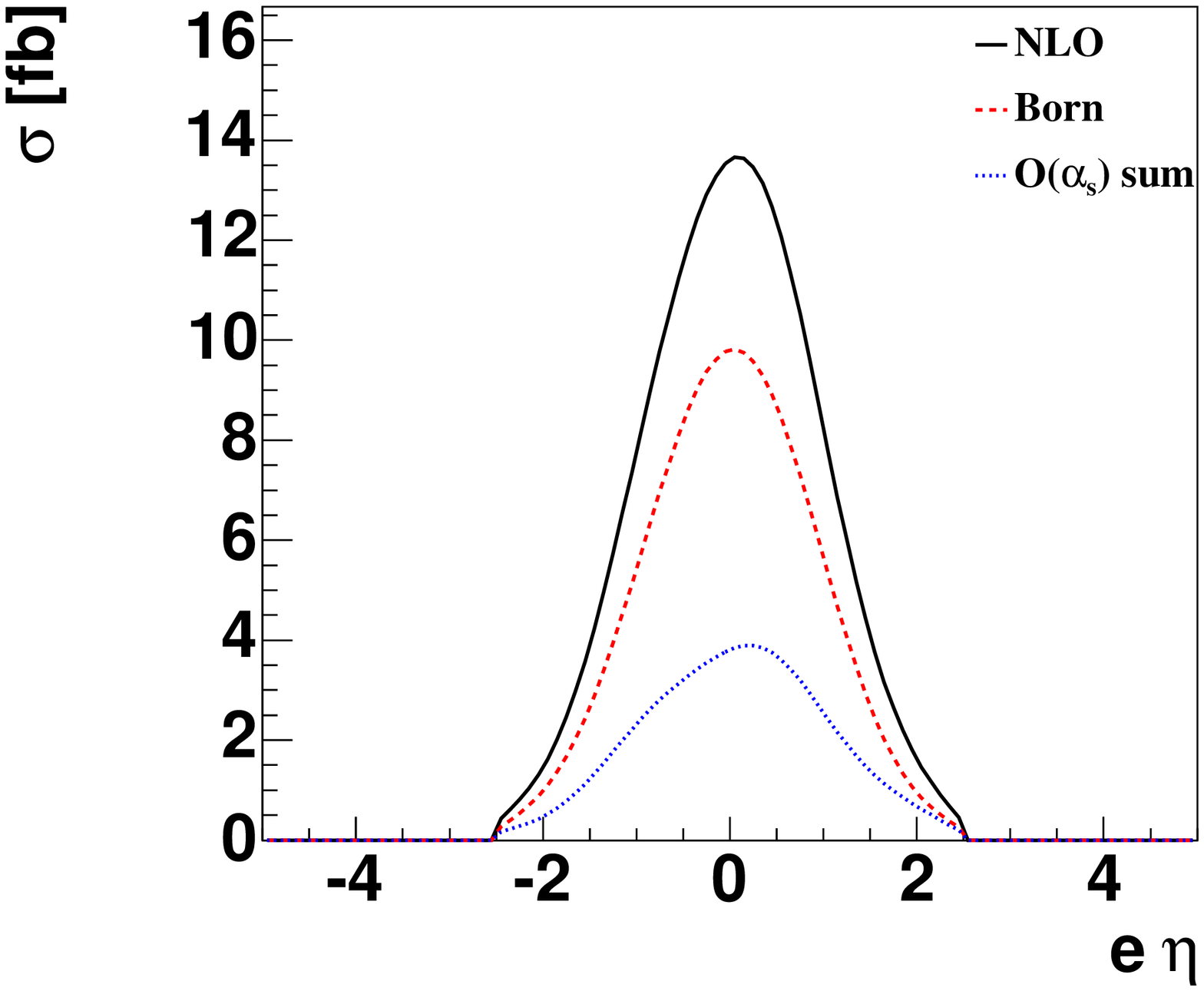}\includegraphics[%
  scale=0.3]{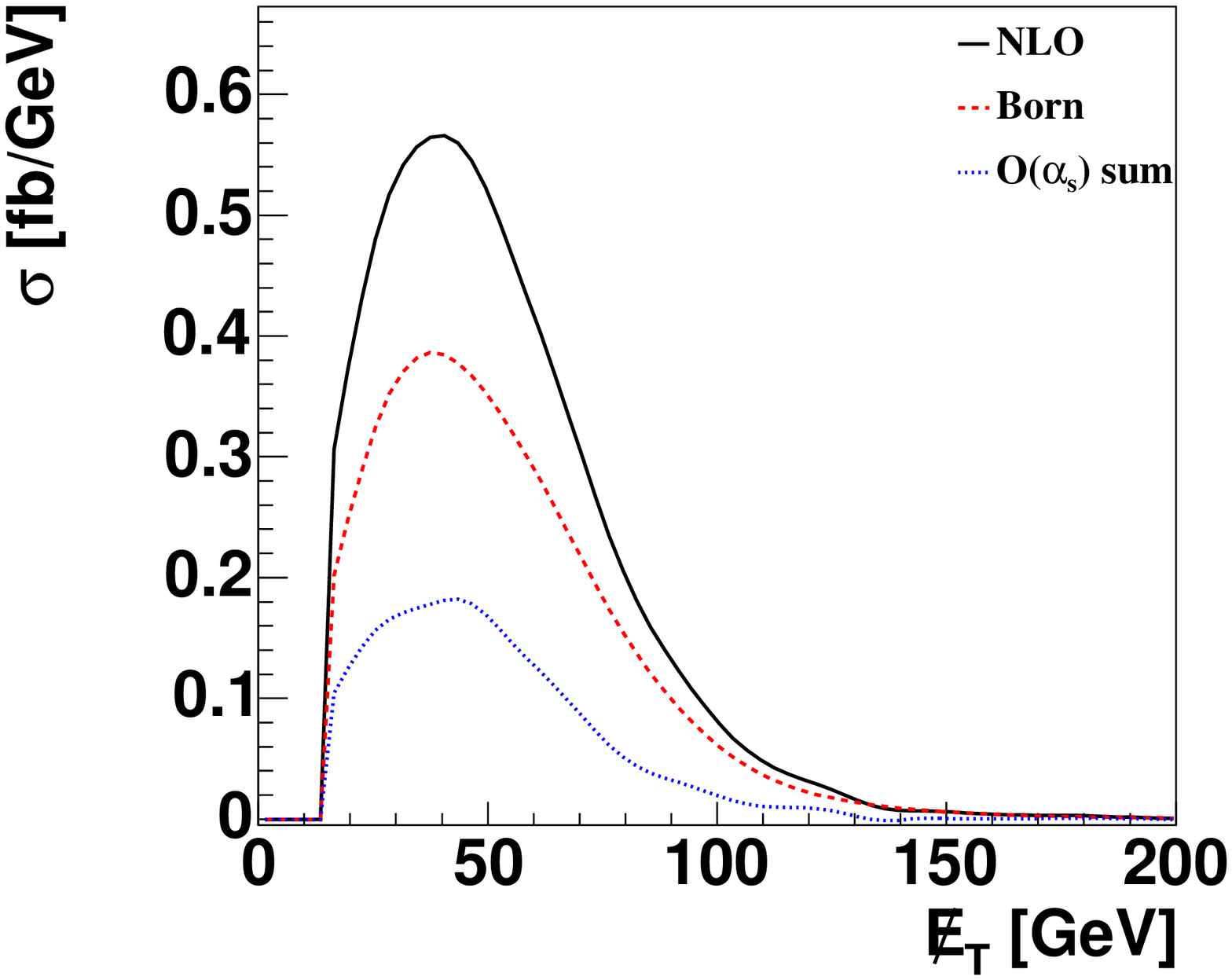}

\caption{The transverse momentum $p_{T}$ (left) and pseudo-rapidity $\eta$
(middle) distributions of the electron and the missing transverse
energy $\met$ (right) after selection cuts, comparing Born level
to $\oalphas$ and full NLO.\label{fig:pte-etae}}
\end{figure}

Fig.~\ref{fig:pte-etae} shows the transverse momentum of the electron
and the missing transverse momentum $\met$. As expected because they
are leptons and not quarks, the change in shape when going from Born level
to $\oalphas$ is not very large for the electron or the neutrino
where the $\oalphas$ corrections have the same shape and peak position
as the Born level contribution. The pseudo-rapidity distribution of
the electron is also given in Fig.~\ref{fig:pte-etae}. This distribution
widens at $\oalphas$ and shifts slightly to more positive values
of $\eta$, but again the effect is not very large. We note that the
peak position of the $\met$ distribution is at a higher value than
that of the electron $p_{T}$ because the neutrino from the $W$-boson
decay moves preferentially along the direction of the top quark. This
is due to the left-handed nature of the charged current interaction
and can easily be seen when examining the spin correlation between
the charged lepton and the top quark in the top quark rest frame.
We will comment more on this subject in Sec.~\ref{sub:Object-Correlations}. 

\begin{figure}
\subfigure[]{\includegraphics[scale=0.3]{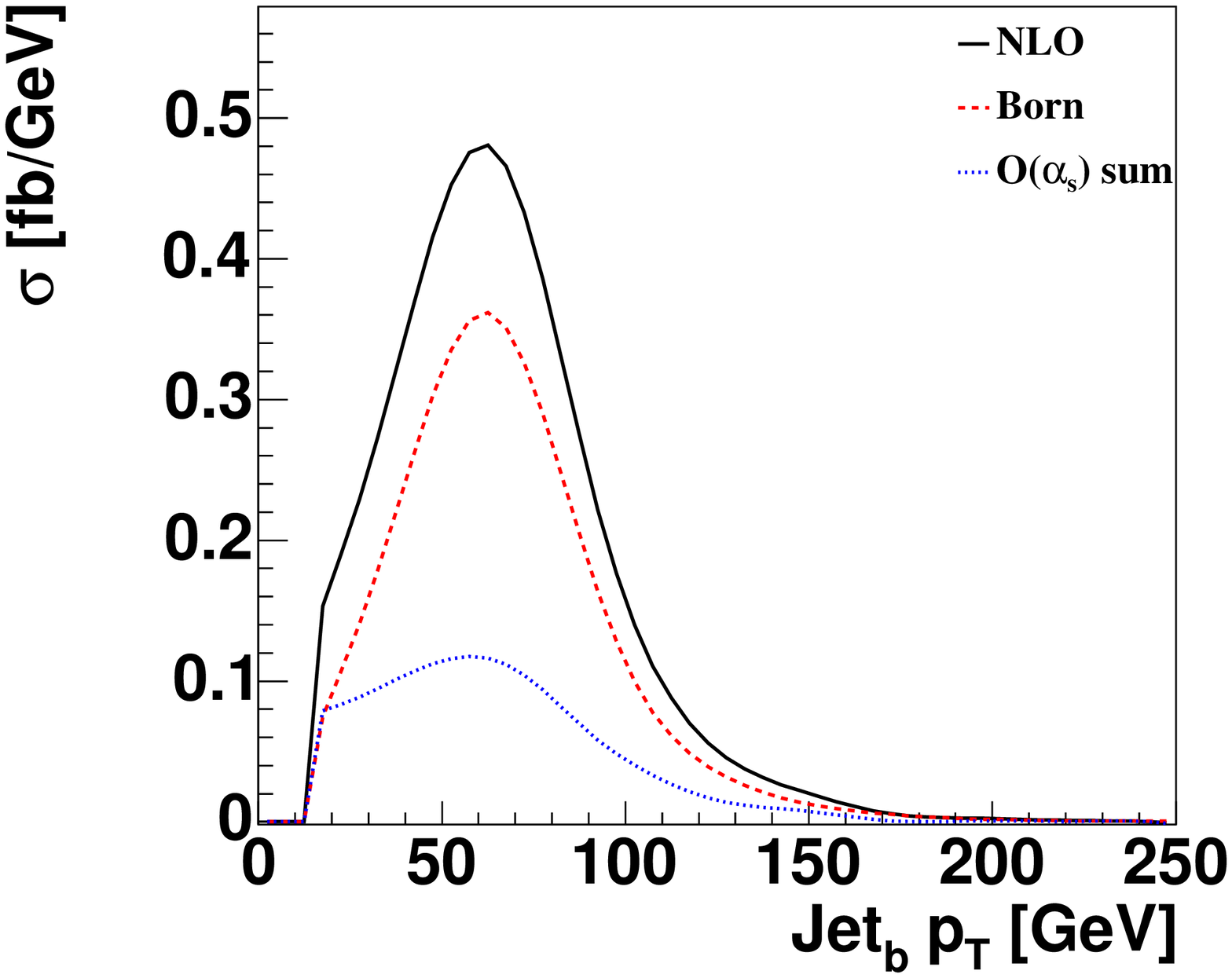}}
\subfigure[]{\includegraphics[scale=0.3]{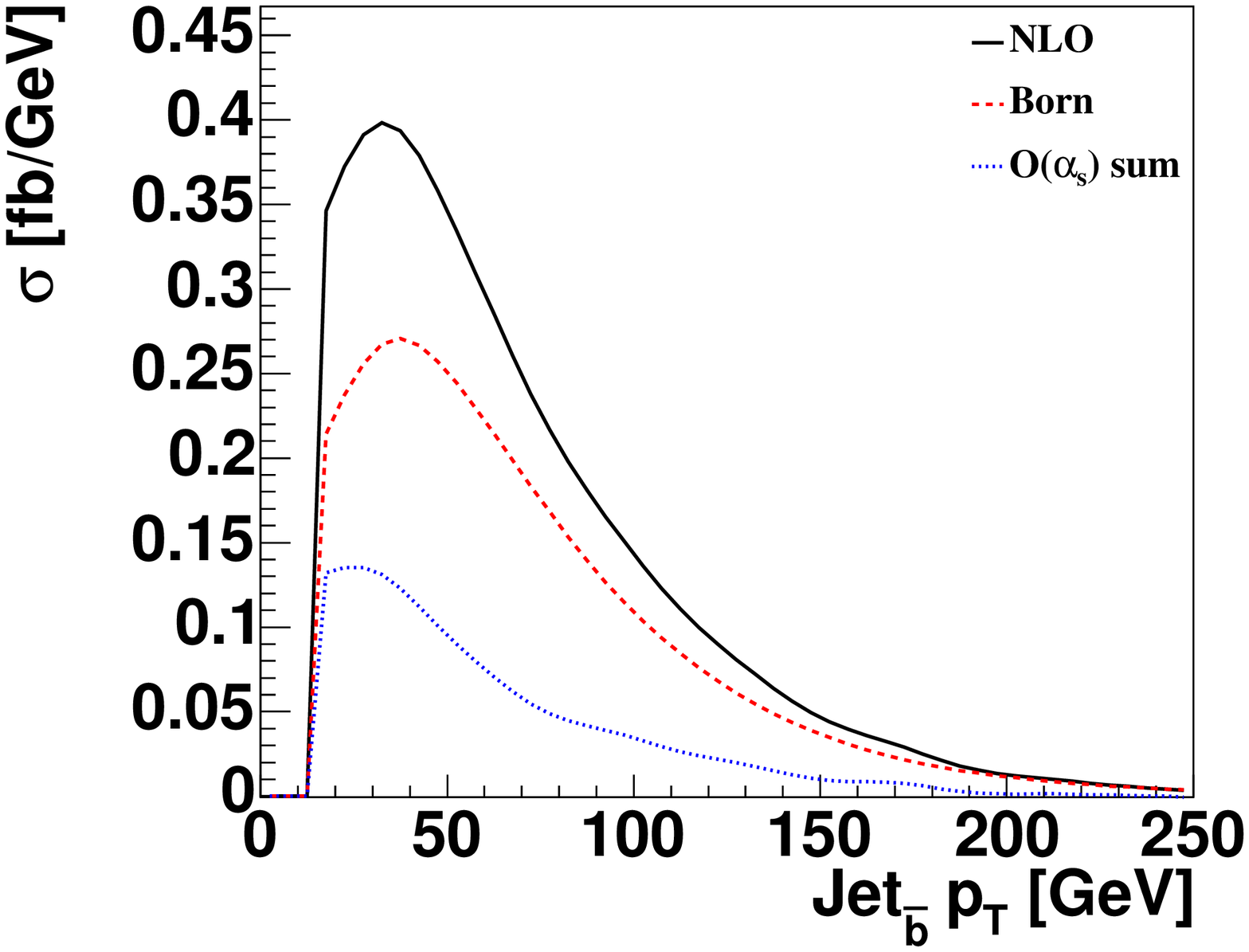}}

\subfigure[]{\includegraphics[scale=0.3]{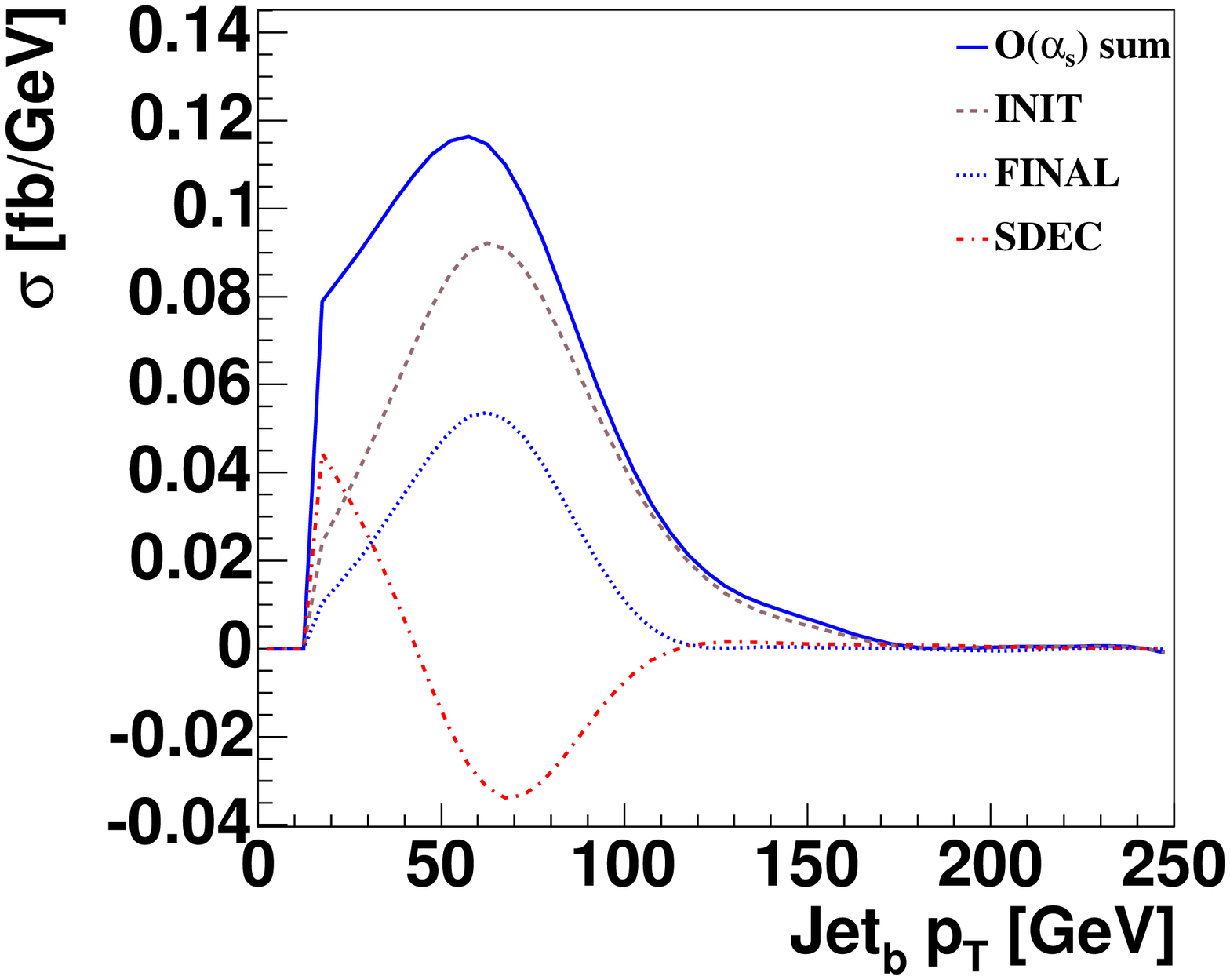}}
\subfigure[]{\includegraphics[scale=0.3]{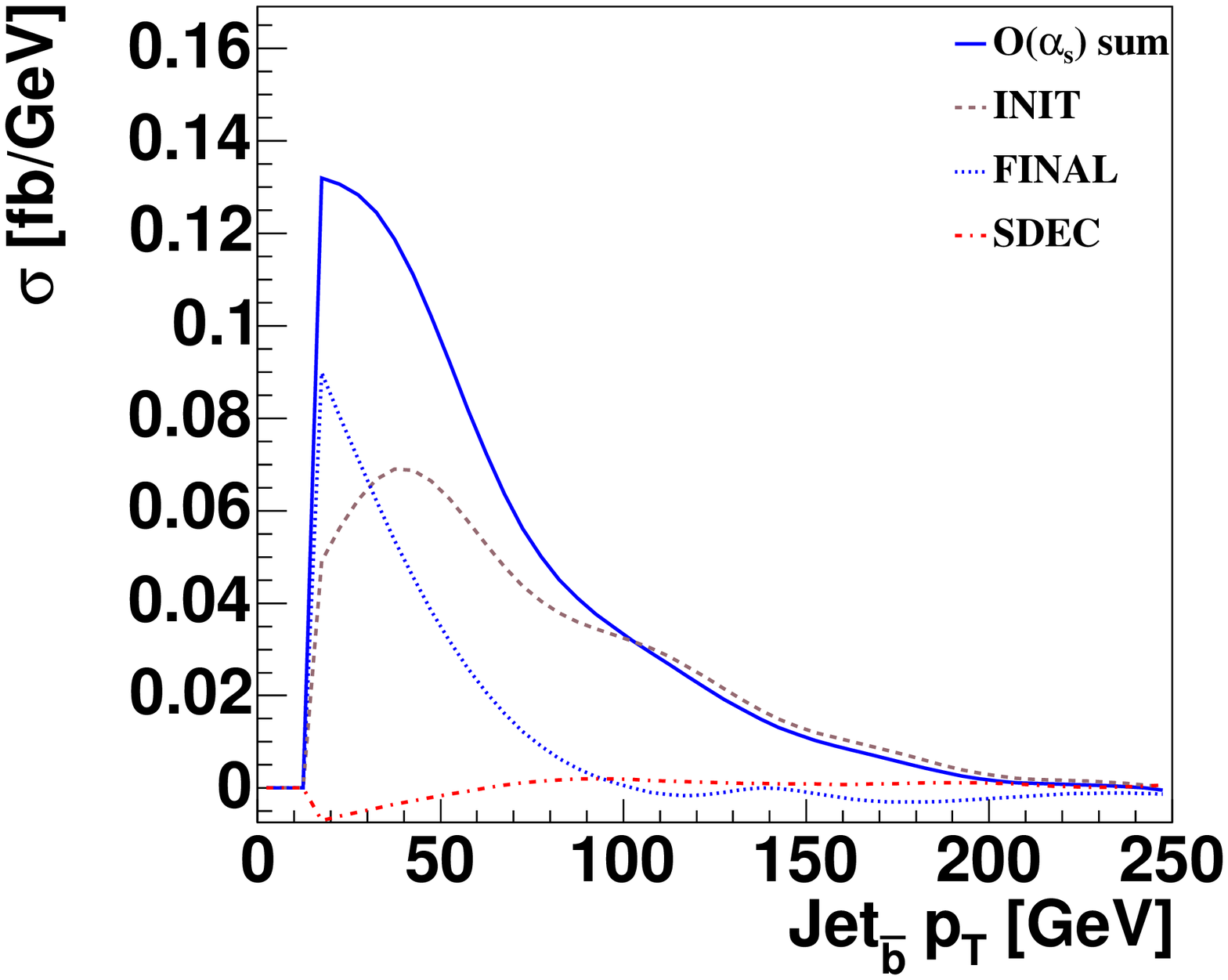}}

\caption{Transverse momentum $p_{T}$ of the $b$ jet (left) and the $\bar{b}$
jet (right) after selection cuts, comparing Born level to $\oalphas$
corrections. The top row shows the NLO, Born, and sum of $\oalphas$ contributions,
the bottom row shows the individual $\oalphas$ contributions.\label{fig:ptb}}
\end{figure}

Compared to the lepton and neutrino, the effect of the $\oalphas$
corrections on the reconstructed $b$-jet and $\bar{b}$-jet is more
pronounced. Fig.~\ref{fig:ptb} shows a comparison of the $E_{T}$
distributions of these two jets between the Born level and the $\oalphas$
contributions. Though it is not possible to distinguish between the
$b$- and $\bar{b}$-jet experimentally, it is nevertheless instructive
to consider their $E_{T}$ distributions individually. At the Born level,
the transverse momentum of the $b$-jet peaks higher and drops off
faster than that of the $\bar{b}$-jet, resulting in similar mean
$E_{T}$ for the two (68~GeV for the $b$-jet and 71~GeV for the
$\bar{b}$-jet). Since the $\bar{b}$-jet is produced in association
with the heavy top quark, it has a long tail into the high $p_{T}$
region to balance the top quark. The $p_{T}$ distribution of the
$b$-jet (from top quark decay) on the other hand is predominantly
controlled by the top quark mass and therefore peaks at $\sim m_{t}/3$.
The NLO QCD corrections broaden the LO transverse momentum distributions
and shift the peak positions to lower values. The result for the mean
of the distribution depends on the cuts that are applied because the
different $\oalphas$ corrections have different effects; in the case
of the loose cuts both distributions have a mean of 66~GeV at $\oalphas$.
The INIT correction shifts the mean of the $b$-jet $p_{T}$ distribution
up while the FINAL and SDEC corrections tend to shift it down. The
$\bar{b}$-jet $p_{T}$ distribution receives a large contribution
from the FINAL correction and a negligible contribution from the
SDEC correction, as expected. When a gluon is radiated from ($t,\,\bar{b}$)
quark line, it prefers to move along the $\bar{b}$-jet direction
due the collinear enhancement and therefore shifts the $\bar{b}$-jet
$p_{T}$ distribution to the small $p_{T}$ region, as shown in Fig.~\ref{fig:ptb}(d).
For the same reason, the SDEC correction to the $b$-jet $p_{T}$ distribution
also peaks around 60 GeV (the Born level peak position), but it provides
a negative contribution, cf. Fig.~\ref{fig:ptb}(c). Furthermore,
because the FINAL correction is larger in size than the SDEC correction,
the resulting shift to the low $p_{T}$ region is stronger for the
$\bar{b}$-jet than the $b$-jet.

\begin{figure}
\includegraphics[scale=0.3]{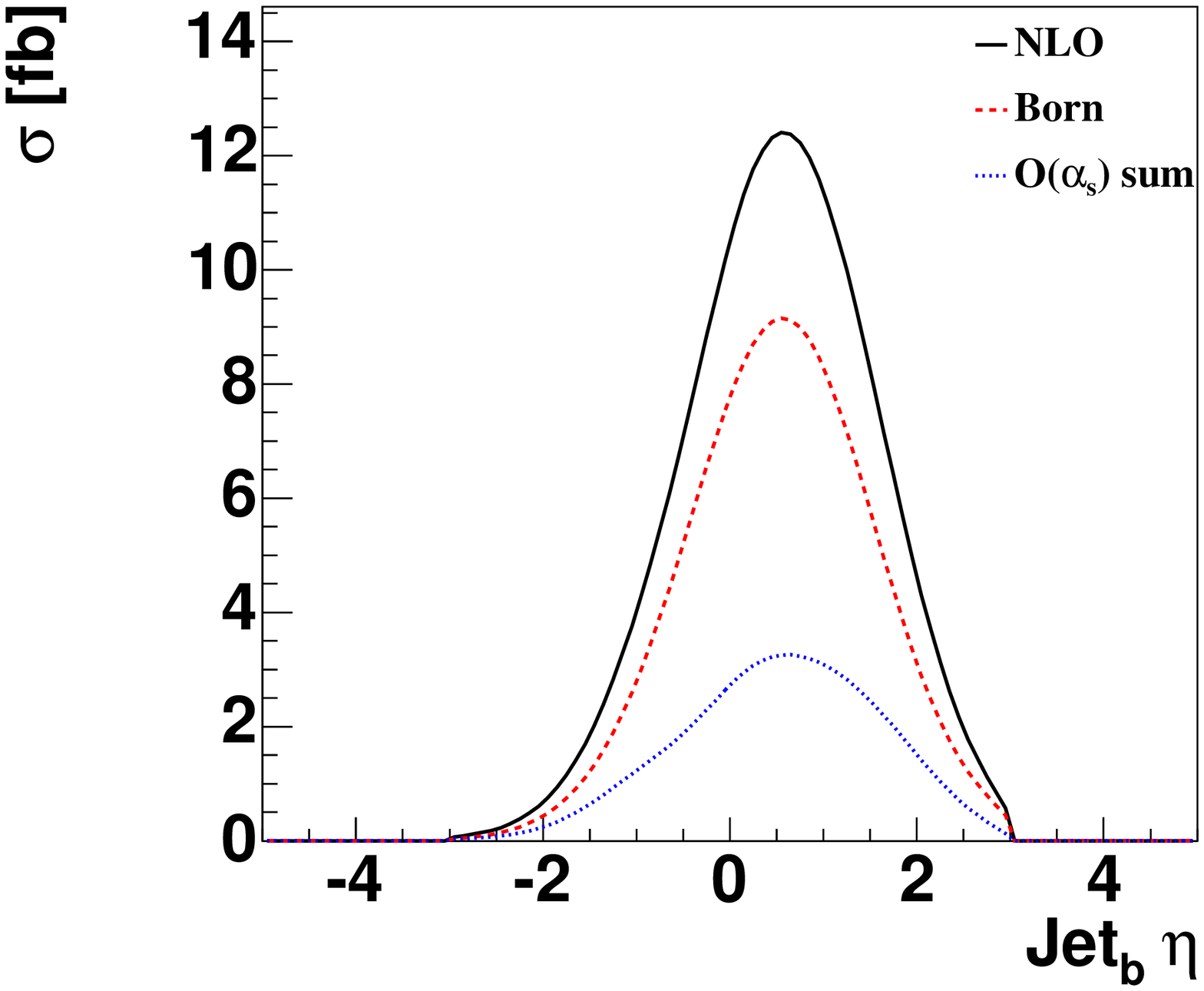}
\includegraphics[scale=0.3]{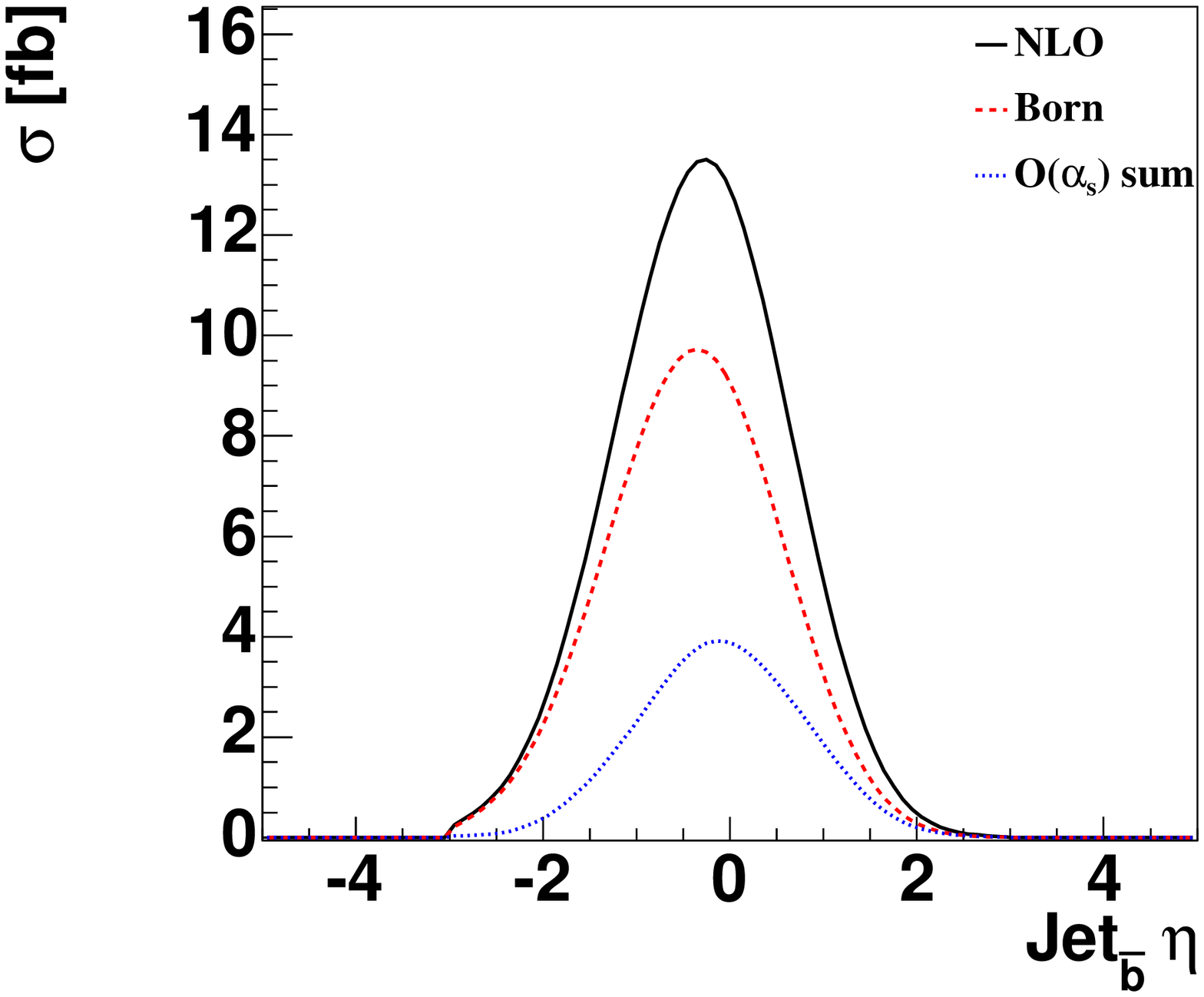}

\caption{Pseudo-rapidity $\eta$ of the $b$-jet (left) and the $\bar{b}$-jet
(right) after selection cuts, comparing Born level to $\oalphas$
corrections. \label{fig:etab}}
\end{figure}

This shift in the $\eta$ distribution of the $\bar{b}$-jet, when
comparing the Born level to $\oalphas$, is also evident in Fig.~\ref{fig:etab},
where in particular the FINAL correction shifts the distribution to
more central pseudo-rapidities. The top quark is so heavy that it
is mostly produced in the central rapidity region and thus its partner
$\bar{b}$-jet also peaks around a pseudo-rapidity of zero. We note
that the pseudo-rapidity distribution of the $\bar{b}$-jet is slightly
asymmetric at the Born level, denoted by the red dotted line in Fig.~\ref{fig:etab},
favoring negative pseudo-rapidity. This is similar to the lepton pseudo-rapidity
asymmetry observed in the $p\bar{p}\rightarrow W^{+}\rightarrow\ell^{+}\nu$
events due to the ratio of down-quark and up-quark parton distributions
inside the proton and anti-proton. The $b$-jet coming from the top
quark decay follows the top quark moving direction and therefore favors
a slightly positive pseudo-rapidity, denoted by the solid line in
the Fig.~\ref{fig:etab}. The $\oalphas$ corrections, in particular
the FINAL correction, shift the $\bar{b}$-jet to more central regions.
The shape of the $b$-jet pseudo-rapidity distribution remains almost
unchanged compared to Born level because it comes from the top quark
decay. Therefore we expect to see an asymmetry between $\eta_{b}$
and $\eta_{\bar{b}}$, shown below after full event reconstruction.

The impact that the different $\oalphas$ corrections have on the
$p_{T}$ of the jets is also reflected in event-wide energy variables
such as the total transverse energy ($H_{T}$) in the event, defined
as\begin{equation}
H_{T}=p_{T}^{lepton}+\met+\sum_{jets}p_{T}^{jet}.\label{eq:HT}\end{equation}
\begin{figure}
\includegraphics[%
  scale=0.3]{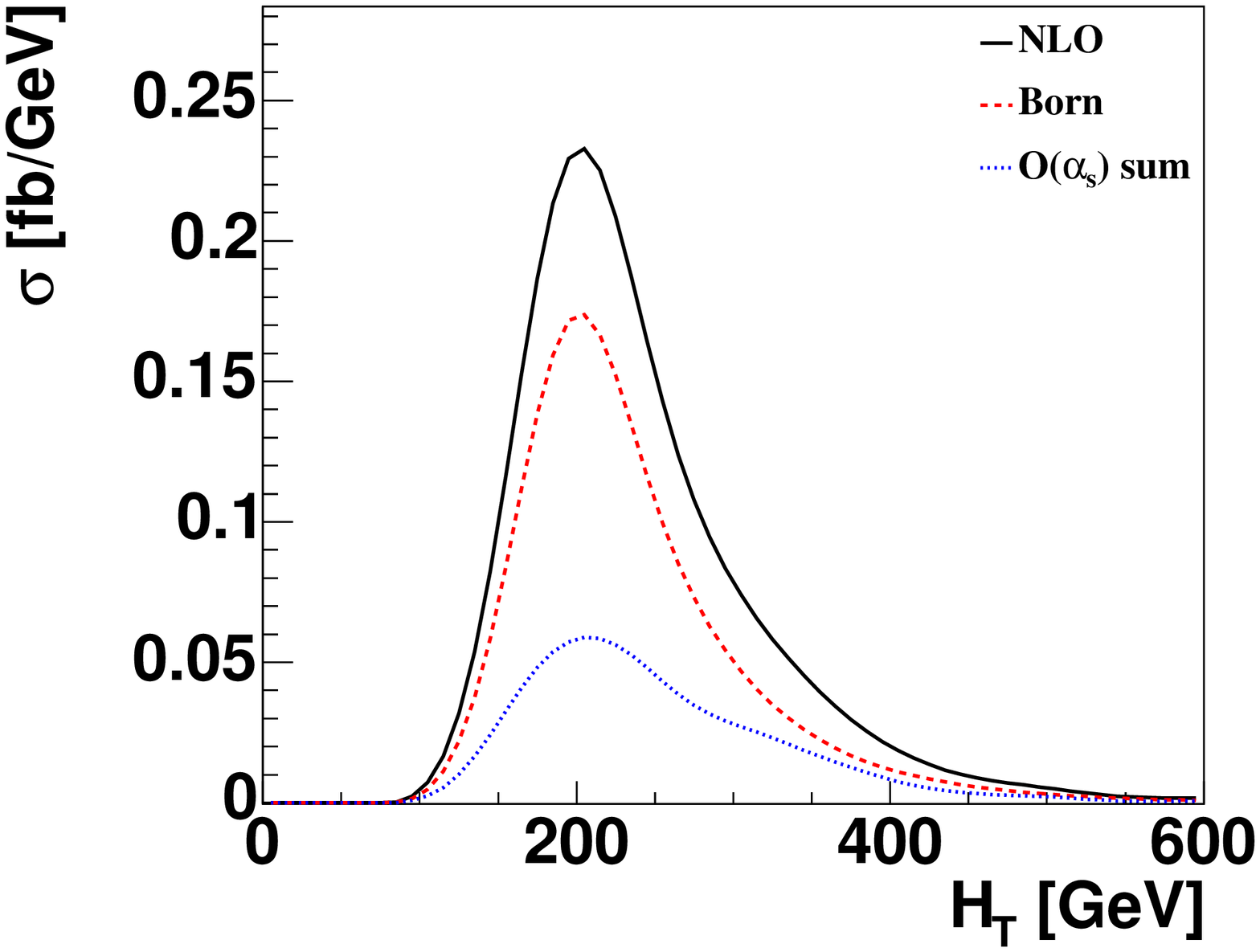}\includegraphics[%
  scale=0.3]{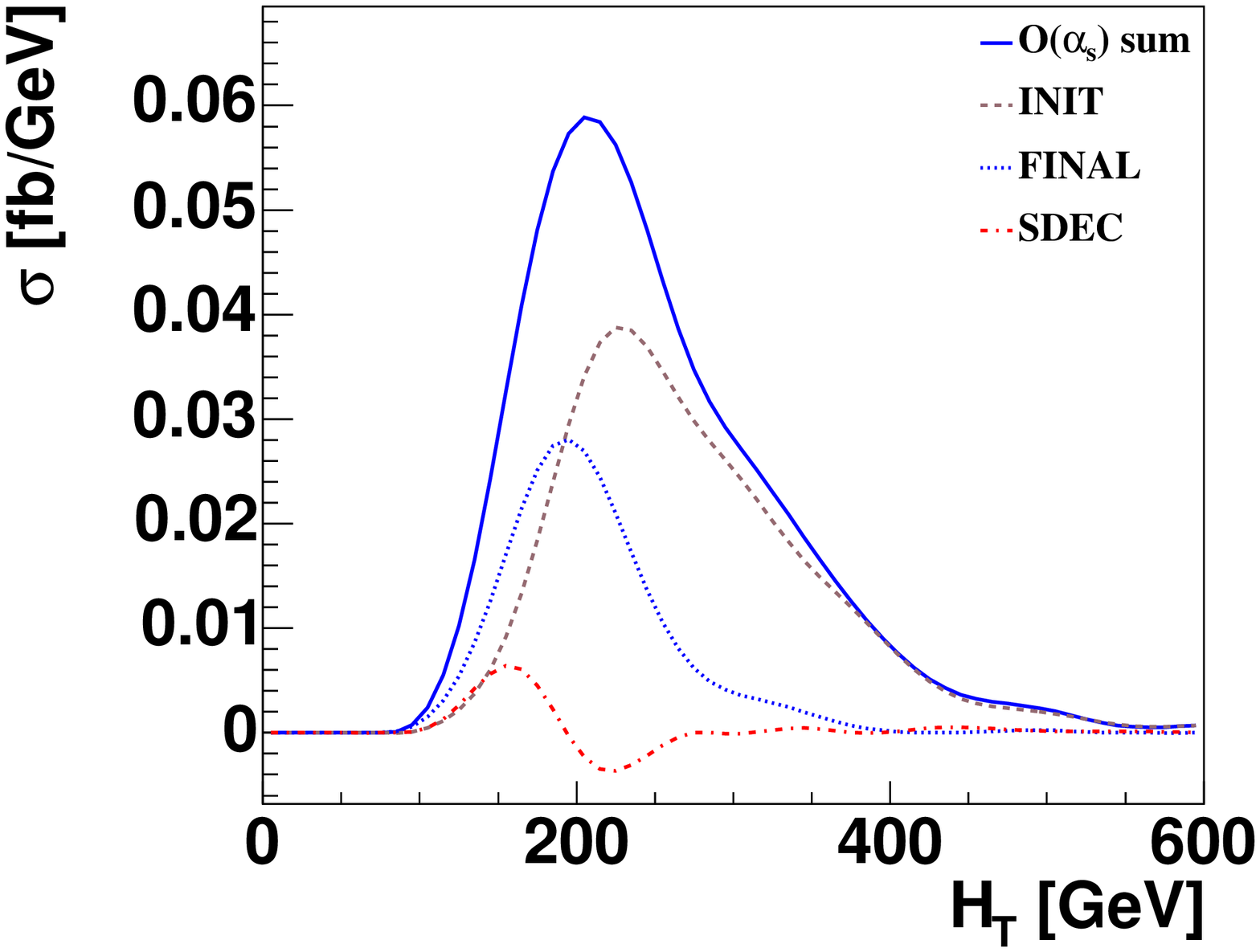}

\caption{Total event transverse energy $H_{T}$ after selection cuts, comparing
Born level to $\oalphas$ corrections.\label{fig:HT}}
\end{figure}
The distribution of $H_{T}$ for s-channel single top events is
shown in Fig.~\ref{fig:HT}. Similar to the $b$- and $\bar{b}$-jet
$p_{T}$ distributions, the FINAL and SDEC contributions shift the
$H_{T}$ distributions down, while the INIT contribution shifts it
up. This is expected because the INIT correction contributes additional
energy to the event in the form of a third jet.

\subsection{Event Reconstruction}

When selecting single top events we would like to take advantage not
only of simple single-object kinematics but also of correlations between
objects. For that we need to reconstruct the event completely, not
just the final state jets but also intermediate particles, in particular
the $W$-boson and the top quark. 

The $W$-boson can be reconstructed from the final state electron
and the observed missing transverse energy $\met$. The lack of information
about the beam-direction component of the neutrino momentum ($p_{z}^{\nu}$)
that would prevent this reconstruction is overcome by requiring the
invariant mass of the electron-neutrino system to be equal to the
mass of the $W$-boson. This additional constraint yields two possible
solutions for $p_{z}^{\nu}$, and typically, both of them are physical
solutions for a signal event. In our analysis, we follow the prescription
given in Ref.~\cite{Kane:1989vv} to choose the solution which has
the smaller $\left|p_{Z}^{\nu}\right|$. This picks the correct $p_{z}^{\nu}$
in about $70\%$ of the events passing the loose set of cuts. 

To reconstruct the top quark, the reconstructed $W$-boson then needs
to be combined with the $b$-jet from the top quark decay. The challenge
that has to be overcome in this case is to identify the correct jet
as the $b$-jet. Several different methods have been used in the past
to select this jet. The simplest method is to choose the highest-$E_{T}$
jet in the event (leading jet). Another possibility that can be used
when $b$-tagging is available experimentally is to choose the $b$-tagged
jet in the event. However, neither of these methods is very successful
in s-channel single top production because both the $b$ quark from
the top decay and the $\bar{b}$ quark produced with the top quark
have high $E_{T}$ and central pseudo-rapidities. While the $E_{T}$
distributions are different for $b$- and $\bar{b}$-jets, Fig.~\ref{fig:ptb}
shows that they differ both in peak position and in shape. For jet
$E_{T}$ values above 100~GeV, the leading jet is actually more likely
to be from the $\bar{b}$ quark. The leading jet corresponds to the
$b$ from the top about 55\% of the time in the sample after loose
cuts. 

A more effective method to identify which jet corresponds to the $b$
quark from the top decay is to make use of the known top quark mass
that is measured in $t\bar{t}$ events. In this ``best-jet'' algorithm,
the $Wj$ combination that gives an invariant mass closest to the
true top mass is chosen as the reconstructed top quark. The jet thus
identified is called ``best-jet''~\cite{Abbott:2000pa}.

If an additional jet is produced in the decay $\oalphas$ contribution,
then the best-jet method will not be able to reconstruct the top quark
correctly. We therefore need to extend the best-jet algorithm to also
include 2-jet pairs when forming a top quark. In other words, candidates
include not only $Wj$ but also $Wjj$. The list of systems for which
the invariant mass is evaluated and from which the best-jet is chosen
thus consists of: $(W,jet_{1})$, $(W,jet_{2})$, $(W,jet_{1},jet_{3})$,
$(W,jet_{2},jet_{3})$.%
\footnote{The combination $(W,jet_{1},jet_{2})$ is not considered because it
is very unlikely for the gluon jet to be one of the leading two jets
due to its generally low $p_{T}$ and because the gluon jet will not
be b-tagged. %
} The jet (or 2-jet system) that is thus chosen will be called the
``best-jet''. Once we have identified the $b$-jet from the top
quark decay, we can assign one of the other jets to the $\bar{b}$
quark produced with the top. We choose this ``non-best-jet'' also
from the leading two jets and assign the label to whichever of the
two jets was not identified as the ``best-jet''.

\begin{figure}
\includegraphics[%
  scale=0.3]{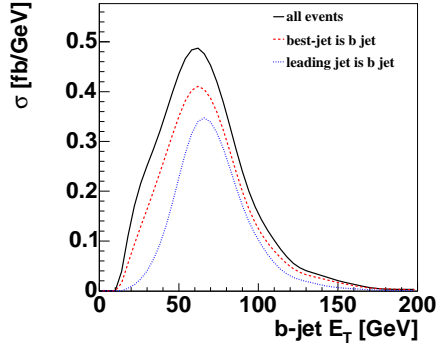}

\caption{Transverse momentum distribution of the $b$ jet from the top quark
decay, for all events that pass the loose set of cuts (solid line),
only for those events in which the $b$ jet is also the leading jet
(dashed line), and only for events in which the $b$ jet is the best-jet
(dotted line). \label{fig:bJetIDEff}}
\end{figure}

The best-jet algorithm identifies the $b$-jet properly about 80\% of the time.
The effectiveness of this method is mainly limited by the
efficiency of the $W$-boson identification method; if the $W$-boson
is not reconstructed properly, then identifying the $b$ quark from
the top decay becomes a random pick. Fig.~\ref{fig:bJetIDEff} shows
the dependence of this efficiency on the transverse momentum of the
$b$-jet. As expected, the leading jet corresponds to the $b$ quark
from the top quark decay mostly when the transverse momentum of the
$b$-jet is very large. The best-jet algorithm in comparison shows
high efficiency for all transverse momenta.

\begin{figure}
\includegraphics[%
  scale=0.3]{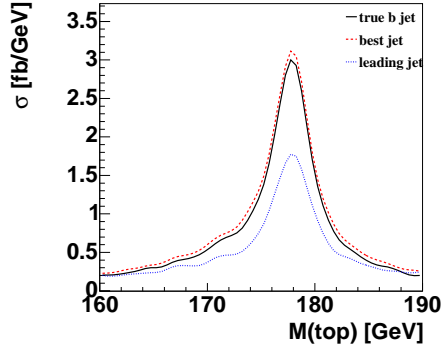}

\caption{Invariant mass of the reconstructed $W$ and a jet object. This jet
object is: the jet containing the $b$ quark from the top decay, using
parton information (solid line); the best-jet identified using the
method outlined in the main text (dashed line); the leading jet (dotted
line).\label{fig:TopM}}
\end{figure}

Fig.~\ref{fig:TopM} shows the invariant mass of the reconstructed
top quark, comparing the different methods to identify the $b$ quark
from the top decay. For the $b$-jet curve in Fig.~\ref{fig:TopM},
parton-level information is used to identify which of the final state
jets contains the $b$ quark from the top decay. If this $b$ quark
produced a gluon in decay-stage emission, this resulting extra jet
is also included when reconstructing the top quark. For the leading
jet curve, only the highest $E_{T}$ jet in each event is used. 

The top quark width is larger than it would be at parton level even
though no smearing was applied in this study. This is the result of
using the reconstructed kinematics of the $W$- boson (in particular
the neutrino $z$-momentum) rather than parton-level information.
Since the same reconstructed $W$-boson was used in each
case, differences between the individual curves are solely due to
which jet is chosen to reconstruct the top quark. Furthermore, because
parton-level information is used for the $b$-jet curve, it functions
as a reference and upper limit for the other methods. 

As expected, using the leading jet gives a peak at the invariant mass
of the top quark with a height of about half the $b$-jet peak. The
best-jet algorithm by contrast shows approximately the same width
and height as the $b$-jet curve. This is the result of two competing
effects, to be discussed in order. On the one hand the best-jet algorithm
identifies the correct jet only in some fraction of the events because
the $W$-boson itself is mis-reconstructed part of the time, thus
reducing the height of the curve. On the other hand the algorithm
chooses an invariant mass close to 178 GeV and thus tends to bring
mis-reconstructed events closer to 178 GeV. The latter effect is slightly
larger than the former, hence the best-jet curve is slightly higher
than the $b$-jet curve in Fig.~\ref{fig:TopM}.

\begin{figure}
\subfigure[]{\includegraphics[scale=0.3]{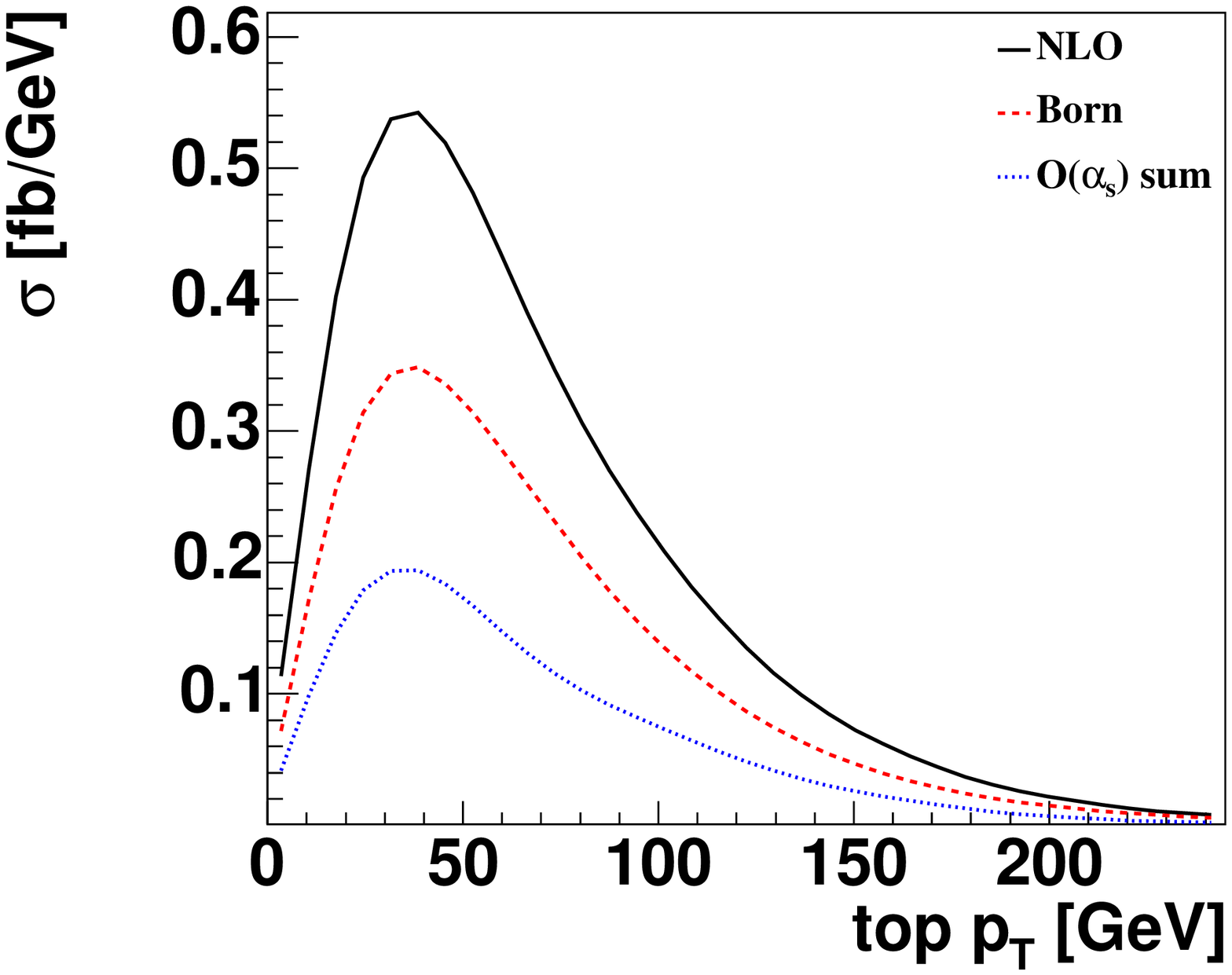}}
\subfigure[]{\includegraphics[scale=0.3]{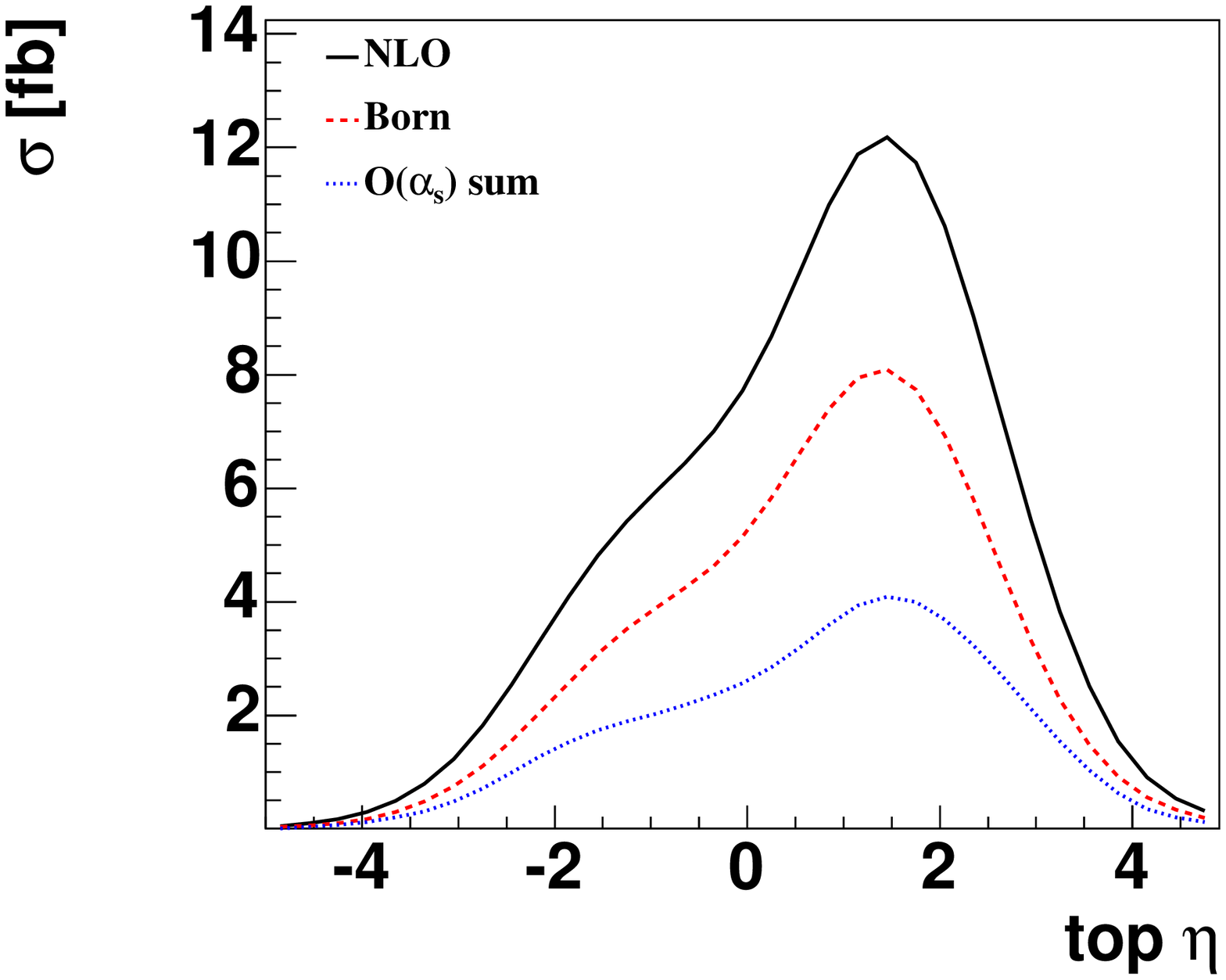}}
\subfigure[]{\includegraphics[scale=0.3]{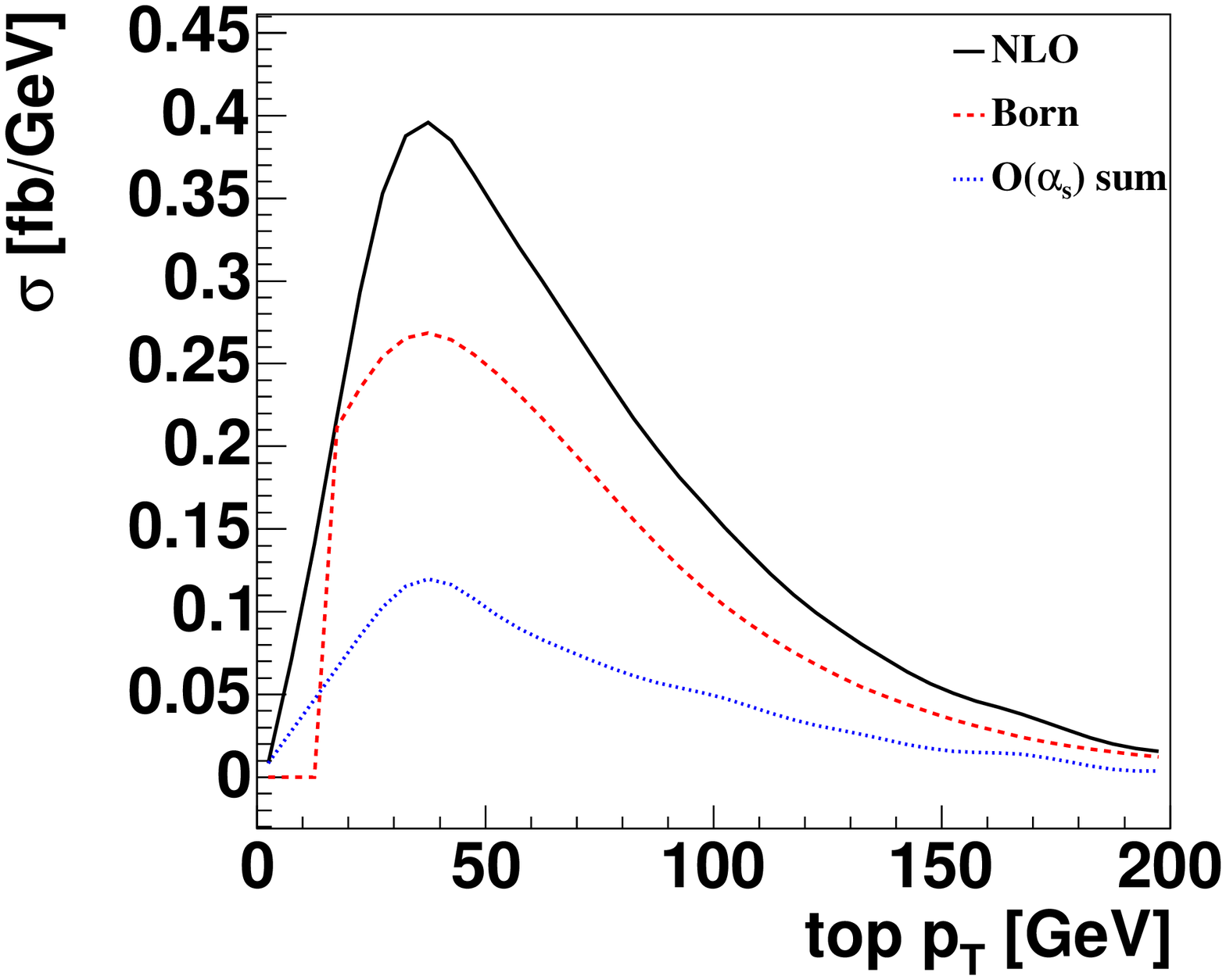}}
\subfigure[]{\includegraphics[scale=0.3]{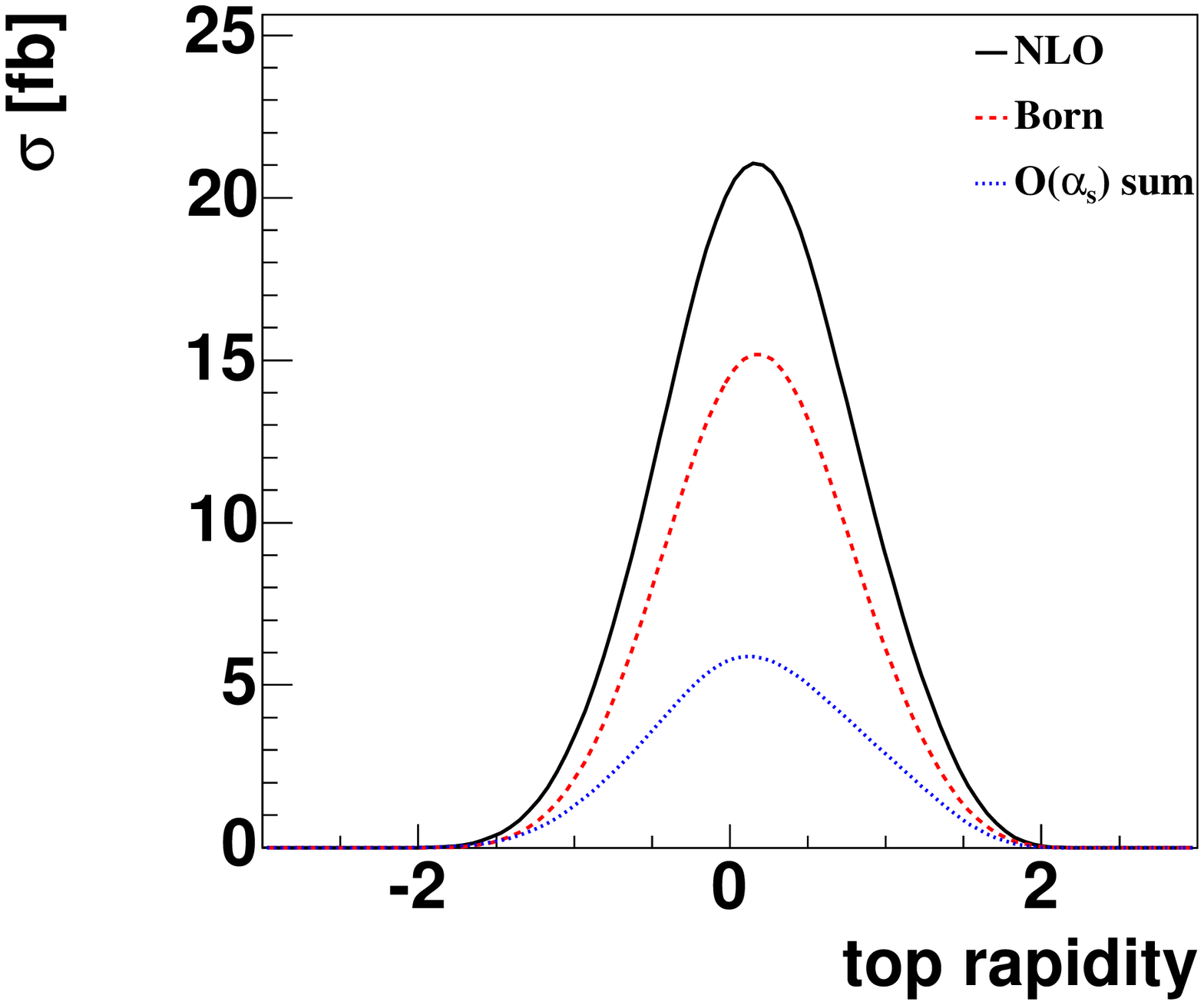}}
\caption{Transverse momentum and rapidity distributions 
of top quark at the 
parton level without any kinematical cuts $(a,b)$,
 and after reconstructing the top quark
from the final state particles with kinematic cuts $(c,d)$,  
comparing Born level to $\oalphas$ corrections.\label{fig:TopEtaPt}}
\end{figure}

Fig.~\ref{fig:TopEtaPt} shows the transverse 
momentum and rapidity distributions 
of the top quark at the parton level without any kinematic cuts
$(a, b)$ and after reconstructing the top quark
from the the final state particles ($W$ boson and the best-jet) 
with kinematic cuts $(c,d)$.
Our parton level result, without imposing any 
kinematical cut, is in good agreement with that in Ref.~\cite{Sullivan:2004ie}. 
After event reconstruction, the $\oalphas$ QCD corrections change the 
parton level top quark transverse momentum distribution, 
especially in the low $p_T$ region. The corrections also shift the top 
quark rapidity distribution to the more central region. We note that using
the best-jet algorithm to reconstruct top quark
 results in distributions that are very similar
to those obtained using the true $b$-jet from top quark decay.

\subsection{Kinematical and Spin Correlations\label{sub:Object-Correlations}}

\begin{figure}
\includegraphics[scale=0.3]{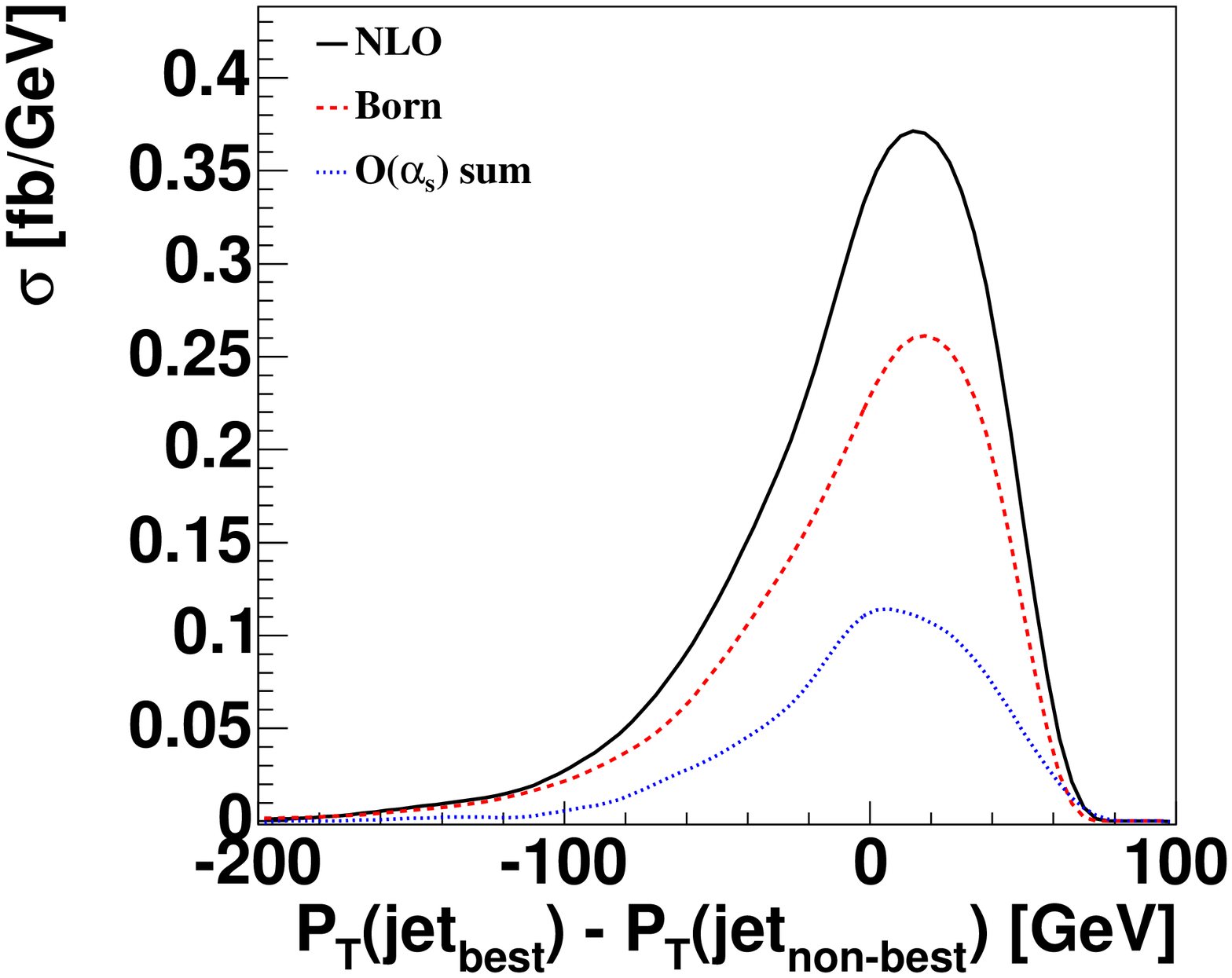}
\includegraphics[scale=0.3]{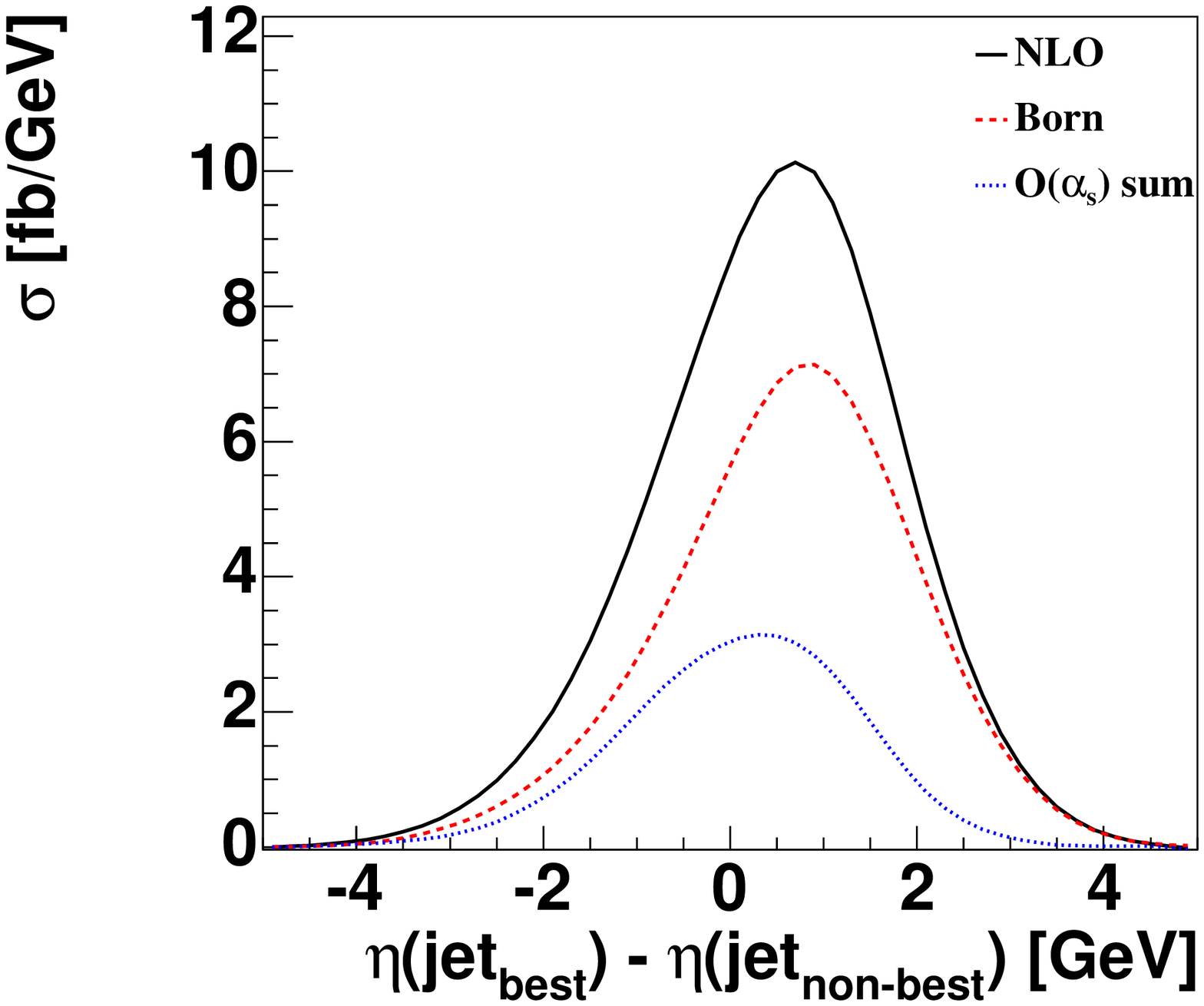}

\caption{Transverse momentum difference between the best-jet and the non-best-jet
(left) and pseudo-rapidity difference between the two (right) after
selection cuts, comparing Born level to $\oalphas$ corrections.\label{fig:bJetMbbarJet}}
\end{figure}

Having identified the $b$-jet from the top quark decay through the
best-jet algorithm, we can now study correlations expected from event
kinematics. In s-channel single top events, there is a strong correlation
between the kinematics of the $b$ and $\bar{b}$ quarks at the Born
level. This correlation is modified by gluon radiation in the production
and decay of the top quark. Fig.~\ref{fig:bJetMbbarJet} shows how
$\oalphas$ effects change the momentum difference and the pseudo-rapidity
difference between the best-jet and the non-best-jet.

The transverse momentum difference is not affected very much; while
the $\oalphas$ corrections affect both the $b$-jet and the $\bar{b}$-jet
and tend to make the distribution broader, this is a small change
compared to the overall width of the distribution. The $\oalphas$
corrections have a larger affect on the pseudo-rapidity difference,
shifting the distribution to more central values and making it broader.
Both effects are mainly due to the FINAL and SDEC corrections, which
tend to weaken the correlation between the $b$-jet and the $\bar{b}$-jet.

\subsubsection{Top Polarization}

\begin{figure}
\includegraphics[scale=0.3]{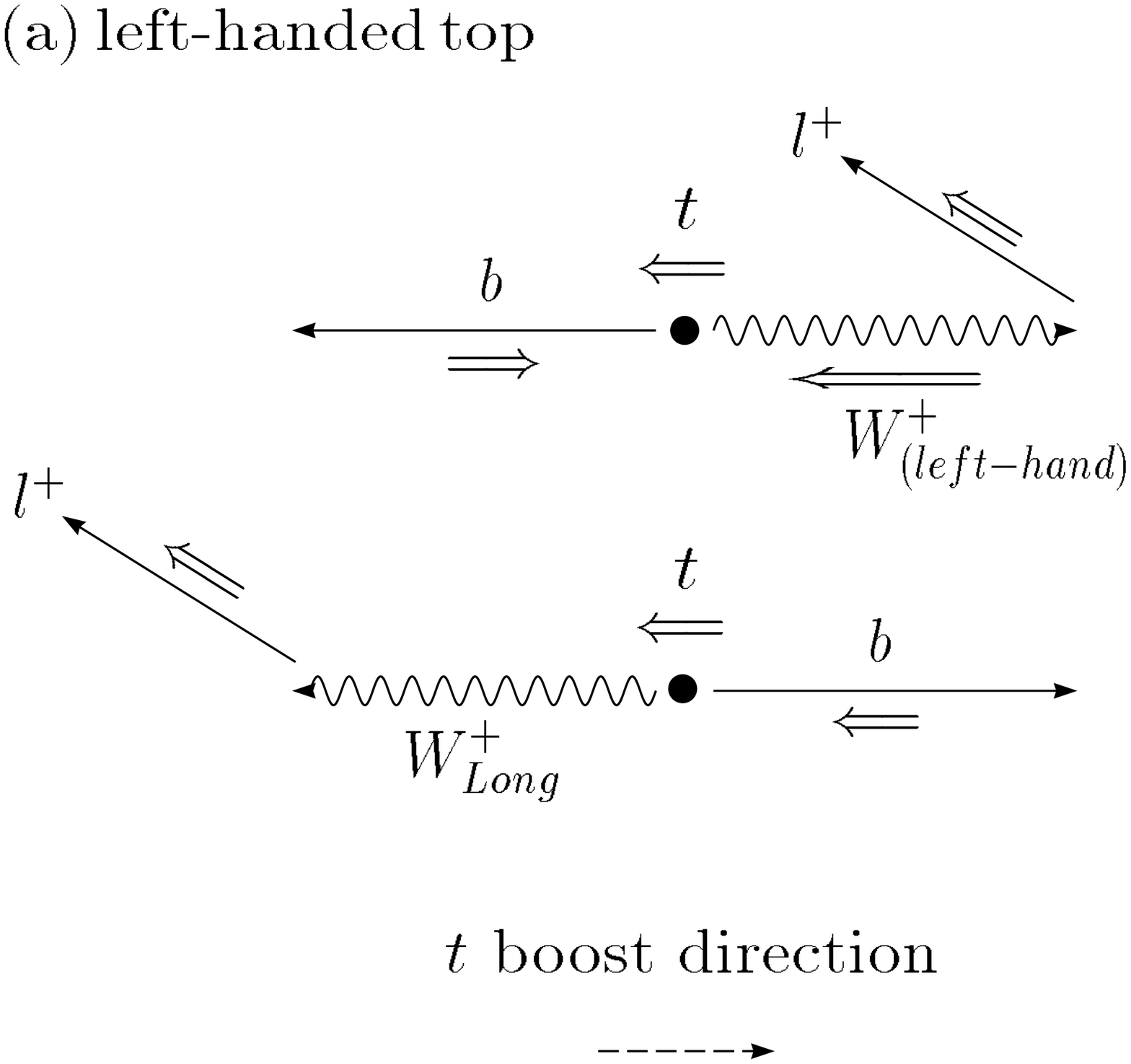}~~~~~~~~~~~~~~
\includegraphics[scale=0.3]{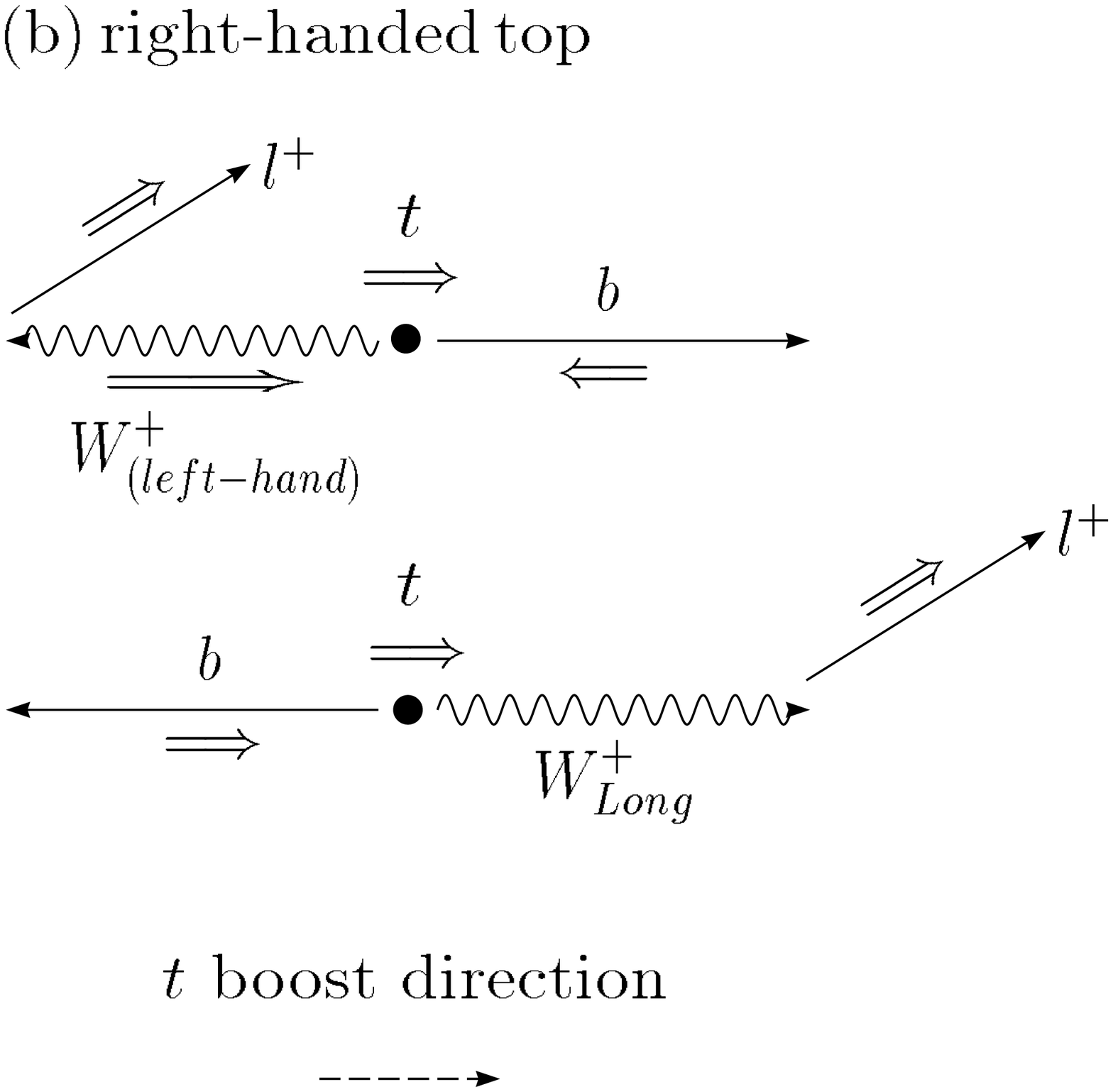}

\caption{Illustration of the correlation between the charged lepton from the
top decay and the top quark spin, in the top quark rest frame. \label{fig: spin-correlation}}
\end{figure}

In the SM, the top quark produced in single top events is highly polarized,
and this polarization can in principle be measured. Two definitions
for the polarization have been studied in the literature, differing
by the reference frame used to define the polarization: One calculation
uses the helicity basis, another the so-called ``optimal'' basis~\cite{Mahlon:1995zn,Parke:1996pr}.
In the more common helicity basis the top quark spin is measured along
the top quark direction of motion in the center of mass (c.m.) frame
which is chosen as the frame of the (reconstructed top quark, non-best-jet)
system after event reconstruction. In the optimal basis (beamline
basis) we can maximize the spin correlations by taking advantage of
the fact that the top quark produced through the s-channel single
top quark processes is almost $100\%$ polarized along the direction
of the $d$-type quark, which comes predominantly from the anti-proton.
In the discussion below, we will examine the
polarization of single top quark events in both cases. The particular
strengths and weaknesses of both methods were discussed in Ref.~\cite{Tait:2000sh}.

It has been pointed out that among the decay products of top quark
the charged lepton is maximally correlated with the top quark spin~\cite{Mahlon:1995zn,Parke:1996pr}.
We can thus obtain the most distinctive distribution by plotting the
angle between the spin axis and the charged lepton in the top quark
rest frame. Before examining this spin correlation effect for a particular
process or in a particular basis, it is worthwhile to give a schematic
picture. Due to helicity conservation and the left-handedness of charged
current interactions, the $W$-boson from the top quark decay can
be either longitudinally or left-handed polarized. This is true for
a massless $b$ quark and diagrammatically shown in Fig.~\ref{fig: spin-correlation}.
The figure shows the preferred moving direction of the lepton from
a polarized $W$-boson in the rest frame of a polarized top quark.
When the $W^{+}$ is left-handed polarized, the fact that the $b$
quark must be left-handed forces it to move along the direction of
the top quark polarization, cf. the upper part of Fig.~\ref{fig: spin-correlation}(a).
The $W^{+}$ thus moves against this direction. When the $W^{+}$
decays, the charged anti-lepton ($e^{+}$) must be right-handed, hence
it prefers to move against the $W^{+}$ direction, in the same direction
as the top quark polarization. When the $W^{+}$ is longitudinally
polarized, it prefers to move in the same direction as the top spin,
cf. the lower part of Fig.~\ref{fig: spin-correlation}(a). Its
decay products prefer to align along the $W^{+}$ polarization, and
since the $W^{+}$ is boosted in the direction of the top quark polarization,
the charged anti-lepton again prefers to move along the top quark
spin axis. Since in both cases the charged anti-lepton moves against
the top quark moving direction, the neutrino from the $W^{+}$ decay
will be harder than the charged anti-lepton, as shown in Fig.~\ref{fig:pte-etae}.
Similarly, the preferred moving direction of the charged anti-lepton
from a right-handed polarized top quark is illustrated in Fig.~\ref{fig: spin-correlation}(b).
For simplicity, we will continue to use the phrase ``charged lepton''
when referring to the charged anti-lepton from the $W$ decay.

\begin{figure}
\includegraphics[scale=0.3]{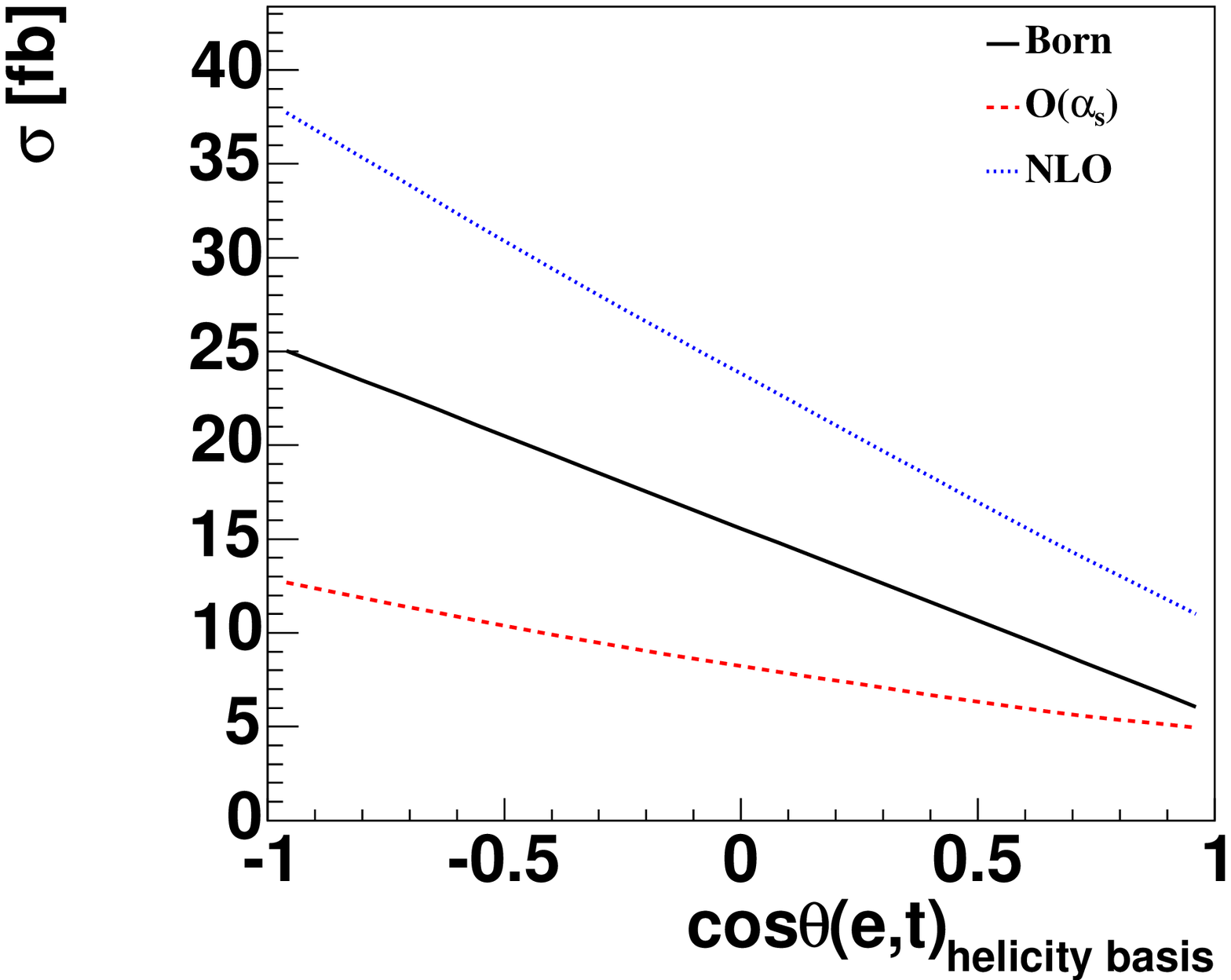}
\includegraphics[scale=0.3]{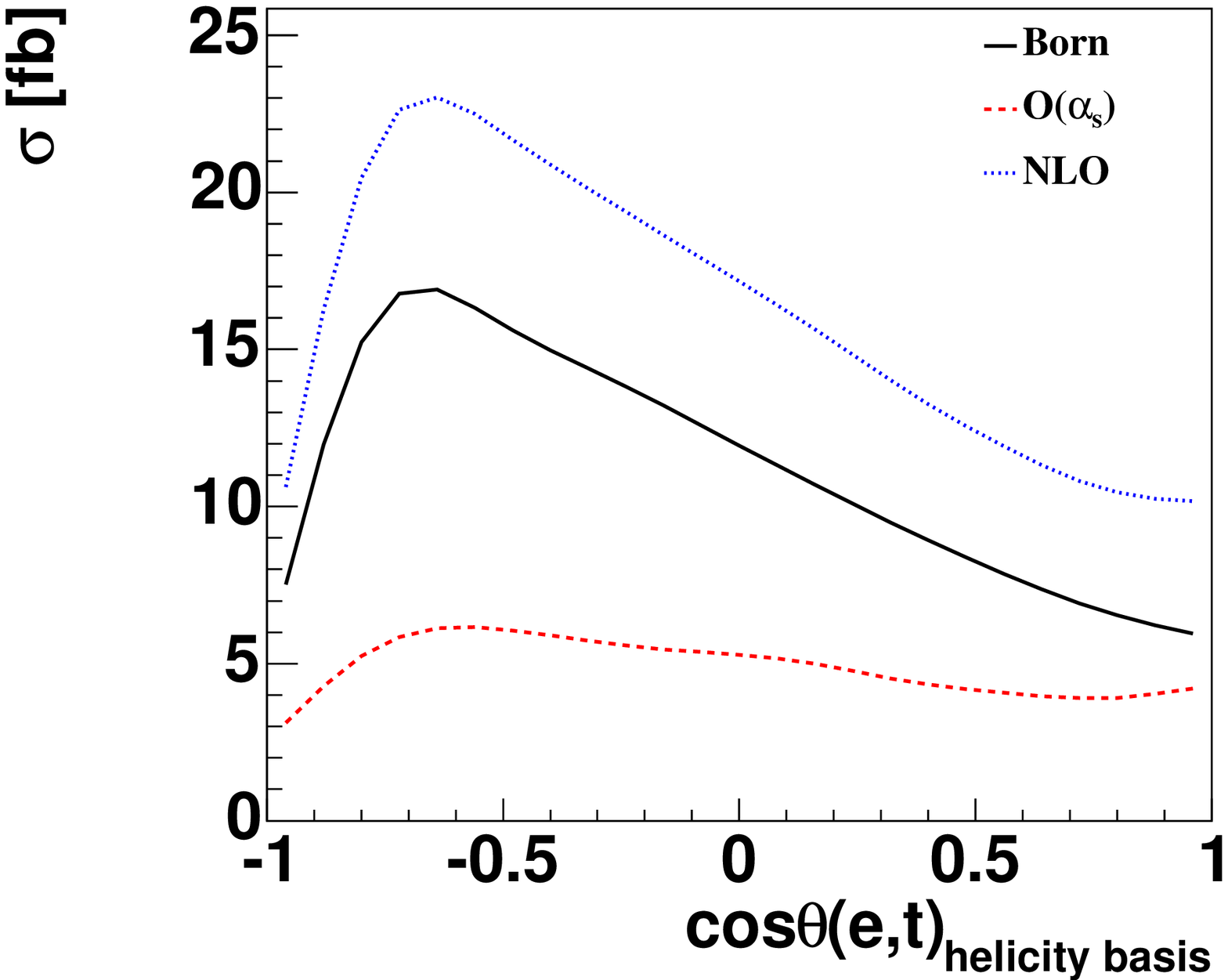}

\caption{Top quark polarization in the helicity basis using the full parton
information (left) and after event reconstruction (right) with selection
cuts, comparing Born level to $\oalphas$ corrections.\label{fig:TopPolHel}}
\end{figure}

In the helicity basis, the polarization of the top quark is examined
as the angular distribution ($\cos\theta_{hel}$) of the lepton in
the c.m. frame of the incoming partons relative to the moving direction
of the top quark in the same frame. The angular correlation in this
frame is given by \begin{eqnarray}
\cos\theta_{hel}=\frac{\vec{p}_{t}\cdot\vec{p}_{\ell}^{*}}{|\vec{p}_{t}||\vec{p}_{\ell}^{*}|},\end{eqnarray}
 where $\vec{p}_{t}$ is the top quark three-momentum defined in the
c.m. frame of the two incoming partons, and $\vec{p}_{\ell}^{*}$
is the charged lepton three-momentum defined in the rest frame of
the top quark. For a left-handed top quark, the angular correlation
of the lepton $\ell^{+}$ is given by $(1-\cos\theta_{hel})/2$, and
for a right-handed top quark, it is $(1+\cos\theta_{hel})/2$. Fig.~\ref{fig:TopPolHel}
shows that this linear relationship for $\cos\theta_{hel}$ is indeed
a valid description for s-channel single top events at the parton
level. The figure also shows that the top quark is not completely
polarized in the helicity basis, and that this polarization is weakened
further when including $\oalphas$ corrections.

The c.m. frame after event reconstruction is chosen as the frame of
the (reconstructed top quark, non-best-jet) system. After reconstruction
the effect of the lepton-jet separation cut can be seen as a drop-off
of the $\cos\theta_{hel}$ distribution close to a value of $-1$.

\begin{figure}
\includegraphics[scale=0.3]{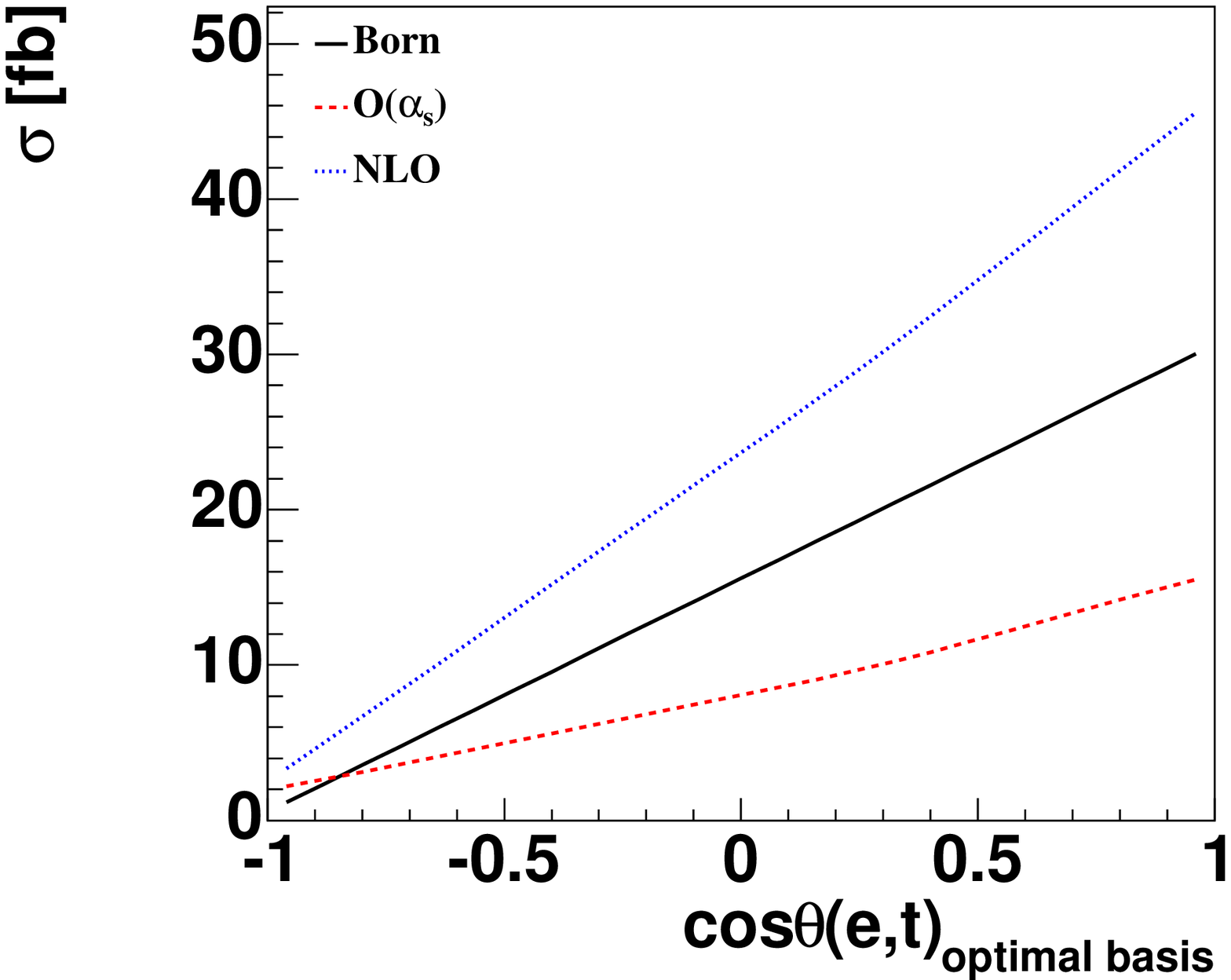}
\includegraphics[scale=0.3]{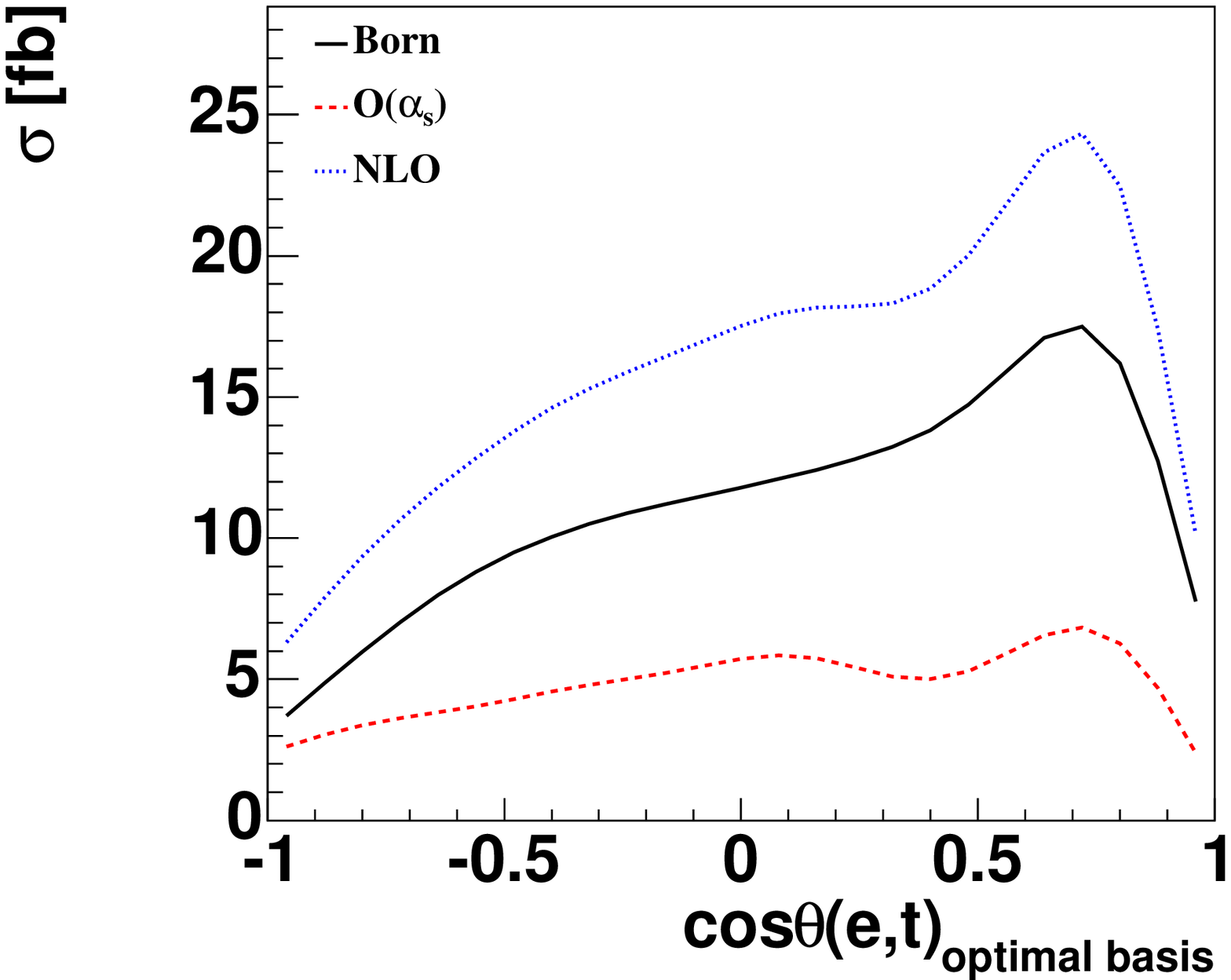}

\caption{Top quark polarization in the ``optimal basis'' using the full parton
information (left) and after event reconstruction(right) with selection
cuts, comparing Born level to $\oalphas$ corrections.\label{fig:TopPolOpt}}
\end{figure}

In the ``optimal'' basis, the relevant angular correlation for
the s-channel process is $\cos\theta_{opt}$, defined as \begin{eqnarray}
\cos\theta_{opt}=\frac{\vec{p}_{\bar{p}}^{*}\cdot\vec{p}_{\ell}^{*}}{|\vec{p}_{\bar{p}}^{*}||\vec{p}_{\ell}^{*}|}\,,\end{eqnarray}
 where $\vec{p}_{\bar{p}}^{*}$ is the anti-proton three-momentum
in the top quark rest frame and $\vec{p}_{\ell}^{*}$ is the lepton
three-momentum in the top quark rest frame. In this analysis, we orient
the coordinate system such that the protons travel in the positive
$z$ direction; the antiprotons travel in the negative $z$ direction.
For a top quark polarized along the anti-proton moving direction,
the angular distribution of the lepton $\ell^{+}$ is $(1+\cos\theta_{opt})/2$,
while for a top quark polarized along the proton moving direction
it is $(1-\cos\theta_{opt})/2$. Fig.~\ref{fig:TopPolOpt} shows
that this linear relationship for $\cos\theta_{opt}$ is a valid description
for s-channel single top events at the parton level. Moreover, in
contrast to the helicity basis, the top quark is almost completely
polarized in the ``optimal basis'' at parton level, and the $\oalphas$
corrections don't change this picture very much.

However, in this case the event reconstruction itself has a significant
effect on the distribution. The reconstructed $\cos\theta_{opt}$
distribution also shows a drop-off, in this case at high $\cos\theta_{opt}$,
which is due to the $\eta$ cut on the lepton.

To better quantify the change in polarization, it is useful to define
the degree of polarization $\mathcal{D}$ of the top quark. This is
given as the ratio\[
\mathcal{D}=\frac{N_{-}-N_{+}}{N_{-}+N_{+}},\]
 where $N_{-}$ ($N_{+}$) is , the number of left-hand (right-hand)
polarized top quarks in the helicity basis. Similarly, in the optimal
basis, $N_{-}$ ($N_{+}$) is the number of top quarks with polarization
against (along) the direction of the anti-proton three momentum in
the top quark rest frame $\vec{p}_{\bar{p}}^{*}$. The angular distribution
is then given by~\cite{Mahlon:1998uv}\textbf{\begin{eqnarray*}
\frac{1}{\sigma}\frac{d\sigma}{d(\cos\theta)} & = & \frac{N_{-}}{N_{-}+N_{+}}\frac{1+\cos\theta}{2}+\frac{N_{+}}{N_{-}+N_{+}}\frac{1-\cos\theta}{2}\\
 & = & \frac{1}{2}\left(1+D\cos\theta_{i}\right).\end{eqnarray*}
}A simple algebra leads to the following identity: \begin{eqnarray}
\mathcal{D} & = & -3\int_{-1}^{1}x\,{\frac{d\sigma}{\sigma dx}}\, dx\,,\label{dpola}\end{eqnarray}
 where ${\displaystyle \frac{d\sigma}{\sigma dx}}$ is the normalized
differential cross section as a function of the polar angle $x$.
Here, $x$ denotes $\cos\theta_{hel}$ in the helicity basis and $\cos\theta_{opt}$
in the ``optimal basis''. Based on the degree of polarization
$\mathcal{D}$, we can easily get the spin fractions $\mathcal{F}_{\pm}$
as:\begin{eqnarray*}
\mathcal{F}_{-} & = & \frac{N_{-}}{N_{-}+N_{+}}=\frac{1+\mathcal{D}}{2},\\
\mathcal{F}_{+} & = & \frac{N_{+}}{N_{-}+N_{+}}=\frac{1-\mathcal{D}}{2}.\end{eqnarray*}
Note that $\mathcal{F}_{-}$($\mathcal{F}_{+}$) is the fraction of
left-handed (right-handed) polarized top quarks in the helicity basis.
Similar, in the optimal basis, $\mathcal{F}_{-}$($\mathcal{F}_{+}$)
is the fraction of top quarks with polarization against (along) the
direction of the anti-proton three momentum in the top quark rest
frame.

We can also define the asymmetry ${\mathcal{A}}$ of the distribution
as \begin{eqnarray}
{\mathcal{A}} & = & \frac{\int_{-1}^{0}d\sigma(\cos\theta)-\int_{0}^{1}d\sigma(\cos\theta)}{{\int_{-1}^{0}d\sigma(\cos\theta)+\int_{0}^{1}d\sigma(\cos\theta)}}.\end{eqnarray}
It is easy to check that without imposing any kinematical cuts, $D=2{\mathcal{A}}$.
Furthermore, the ratio of top quarks with spin along the basis direction
will be $r_{\uparrow}=0.5-{\mathcal{A}}$ when no cuts are applied.
However, when cuts are imposed, the two relations break down.

\begin{table}
\begin{center}\begin{tabular}{cc|c|c|c|c|c|c|c|c|c}
\hline 
&
&
\multicolumn{3}{c|}{$\mathcal{D}$ }&
\multicolumn{3}{c|}{\textbf{$\mathcal{F}$}}&
\multicolumn{3}{c}{$\mathcal{A}$}\tabularnewline
\cline{3-5} \cline{6-8} \cline{9-11} 
&
&
LO&
$O(\alpha_{s})$&
NLO&
LO&
$O(\alpha_{s})$&
NLO&
LO&
$O(\alpha_{s})$&
NLO\tabularnewline
\hline
Helicity Basis: &
Parton level&
0.63&
0.48&
0.58&
0.82&
0.74&
0.79&
0.32&
0.24&
0.29\tabularnewline
&
Reconstructed events&
0.46&
0.17&
0.37&
0.73&
0.58&
0.68&
0.26&
0.11&
0.21\tabularnewline
\hline
Optimal basis: &
Parton level&
-0.96&
-0.83&
-0.92&
0.98&
0.92&
0.96&
-0.48&
-0.42&
-0.46\tabularnewline
&
Reconstructed events&
-0.48&
-0.28&
-0.42&
0.74&
0.64&
0.71&
-0.24&
-0.14&
-0.21\tabularnewline
\hline
\end{tabular}\end{center}

\caption{Degree of polarization $\mathcal{D}$, polarization fraction $\mathcal{F}$,
and asymmetry $\mathcal{A}$ in s-channel single top events. Here,
$\mathcal{F}$ corresponds to $\mathcal{F}_{-}$ in the helicity basis
for left-handed top quarks and to $\mathcal{F}_{+}$ in the optimal
basis for top quarks with polarization along the direction of the
anti-proton three momentum. \label{tab:toppol}}
\end{table}

Table~\ref{tab:toppol} shows that the relationship $\mathcal{D}=2{\mathcal{A}}$
indeed holds at parton level (within rounding errors) and is still
approximately true at $O(\alpha_{s})$. We present our result at the
parton level, both before and after event reconstruction. We found
that at the parton level stage the ``optimal basis'' presents
a significant improvement over the the helicity basis at both the
Born level and NLO. The top quark is almost completely polarized in
the optimal basis. However, after event reconstruction both bases
give almost the same spin fractions. The gain from using the optimal basis is lost
again and the helicity basis is equivalent to the optimal basis in
terms of polarization. This is a result of the jet reconstruction
and lepton-jet separation cut and of the constraint
on the $W$ mass when reconstructing the neutrino. 
Furthermore, the $O(\alpha_{s})$ corrections
reduce the polarization already at the parton level in both the helicity
and optimal basis.

Before concluding this section, we note that the relative amounts
of production- and decay-stage emission depend sensitively on the
kinematical cuts applied. Also, we do not show the transverse momentum
distribution of the $t$ $\bar{b}$ pair or their azimuthal angle separation
because those distributions are sensitive to multiple gluon radiations,
and they can only be reliably predicted using a resummed calculation~\cite{Mrenna:1997wp}.

\subsection{Distributions for three-jet event}

As shown in Fig.~\ref{fig:njets_jet_pt}, a large fraction of the
events passing the loose selection cuts contains three jets. In this
section we focus on the properties of the additional jet.

\subsubsection{Kinematical Distribution of the Extra Jet}

Initial- and final-state emission of additional gluons occurs before
the top quark goes on shell and can thus be considered as ``production-stage
emission'', while decay-stage emission occurs only after the top
quark goes on shell. In principle, an event with an extra jet can
thus be classified as production-stage or decay-stage by looking at
the invariant mass of the decay products. In production-stage emission
events, the $W$-boson and $b$-quark momenta will combine to give
the top momentum. In decay-stage emission events, the gluon momentum
must also be included to reconstruct the top momentum. This interpretation
is exact at the parton level in the narrow width approximation. Finite
top width effects can blur it due to interference between production- and
decay-stage emission. This classification is nevertheless still useful
in our case because the top width of 1.5~GeV is small compared to
the hard gluon $E_{T}$ cut imposed in the MC calculations. It should
be kept in mind that in an experiment, the production-decay distinction
is further blurred by the experimental jet energy resolution and ambiguities
associated with properly assigning partons to jets and the like. 

\begin{figure}
\includegraphics[scale=0.3]{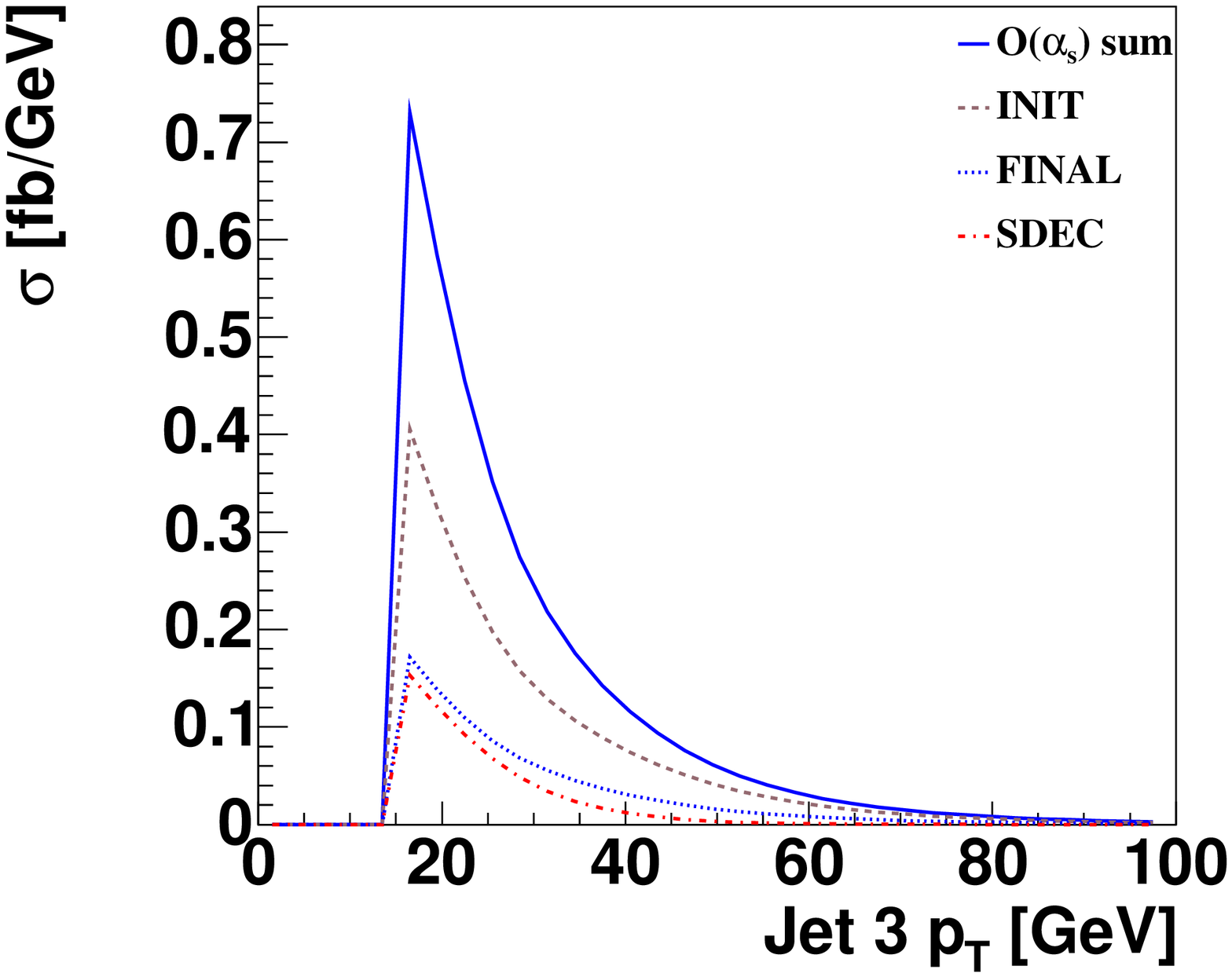}
\includegraphics[scale=0.3]{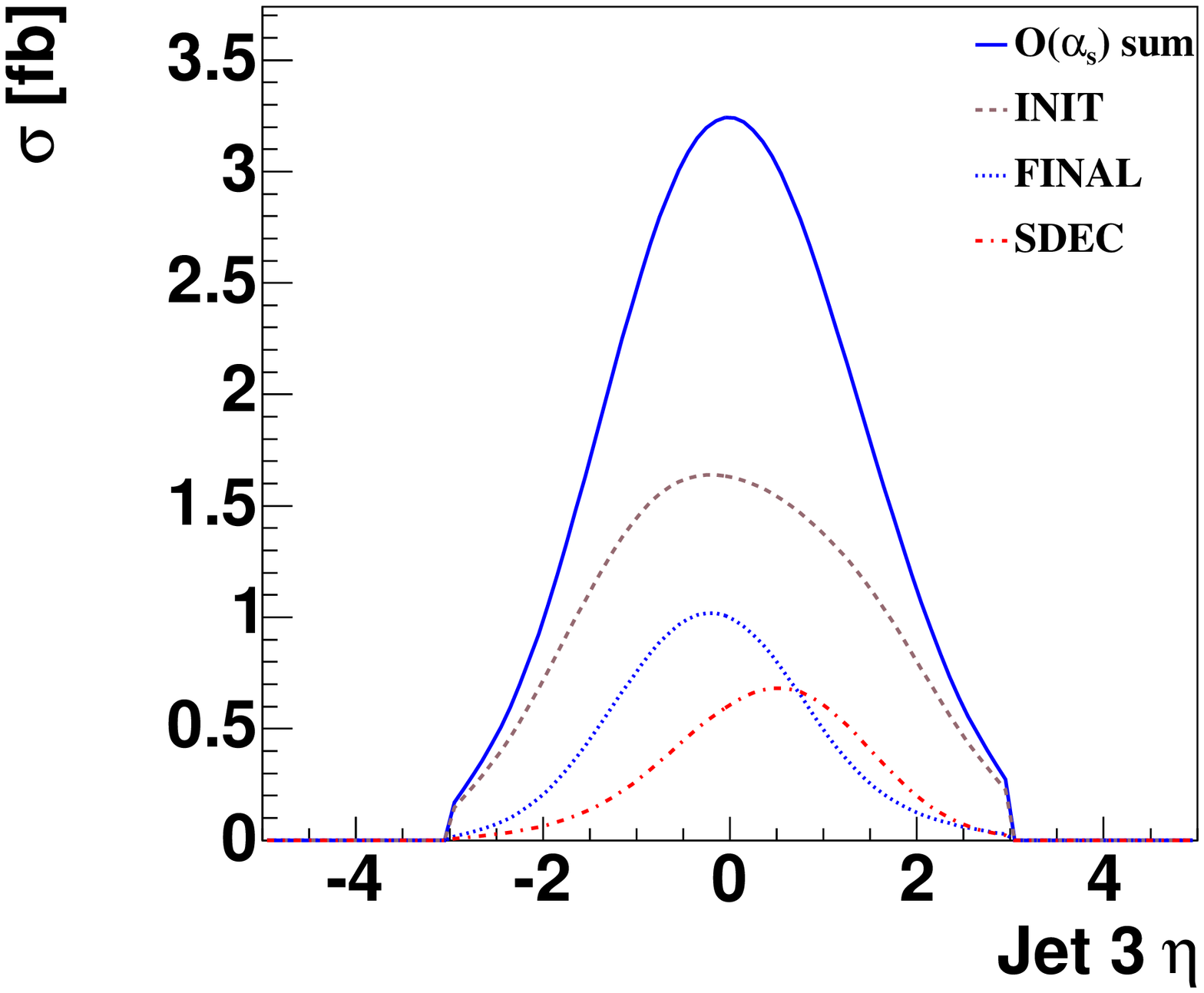}

\caption{Transverse momentum (left) and pseudo-rapidity of the third jet after
selection cuts for the various $\oalphas$ contributions.\label{fig:ptetaJet3}}
\end{figure}

Fig.~\ref{fig:ptetaJet3} shows the transverse momentum distribution
as well as the pseudo-rapidity distribution for the third jet in 3-jet
events. This jet corresponds to the gluon in about 70\% of the events
after the loose set of cuts.

Note that production-stage emission is dominant over decay-stage emission
because the decay contribution is determined not by the collider energy,
but by the phase space of the top quark decay. As expected,
the $E_{T}$ distribution is steeply falling for all contributions,
but it extends to much higher $E_{T}$ values for production-stage
emission. The smaller values of $E_{T}$ to which decay emission is
constrained are again a consequence of the top decay kinematics. Recall
that the top quarks are produced with relatively modest transverse
momentum (cf. Fig.~\ref{fig:TopEtaPt}), so that gluons from the
decay do not receive much of a boost. Note also that an increase in
the $E_{T}$ cut on the jet would result in a further reduction in
relative size of the decay contribution compared to production. Fig.~\ref{fig:ptetaJet3}
also shows the distribution in pseudo-rapidity of the extra jet. Initial-state
emission is relatively flat in pseudo-rapidity, as compared to the
more central decay emission. This is consistent with our intuition
that decay-stage radiation, being associated with the final-state
particles - which tend to appear in the central pseudo-rapidity region
- is also likely to be produced centrally. But this decay contribution
is small and production-stage radiation dominates even in the central
region. Also, final-state emission and decay emission have a tendency
to follow the direction of the $\bar{b}$-jet and $b$-jet respectively,
hence the decay $\oalphas$ contribution peaks at a positive $\eta$
value while the final state $\oalphas$ contribution peaks at a negative
$\eta$ value, cf. Fig.~\ref{fig:etab}. 

\begin{figure}
\includegraphics[scale=0.3]{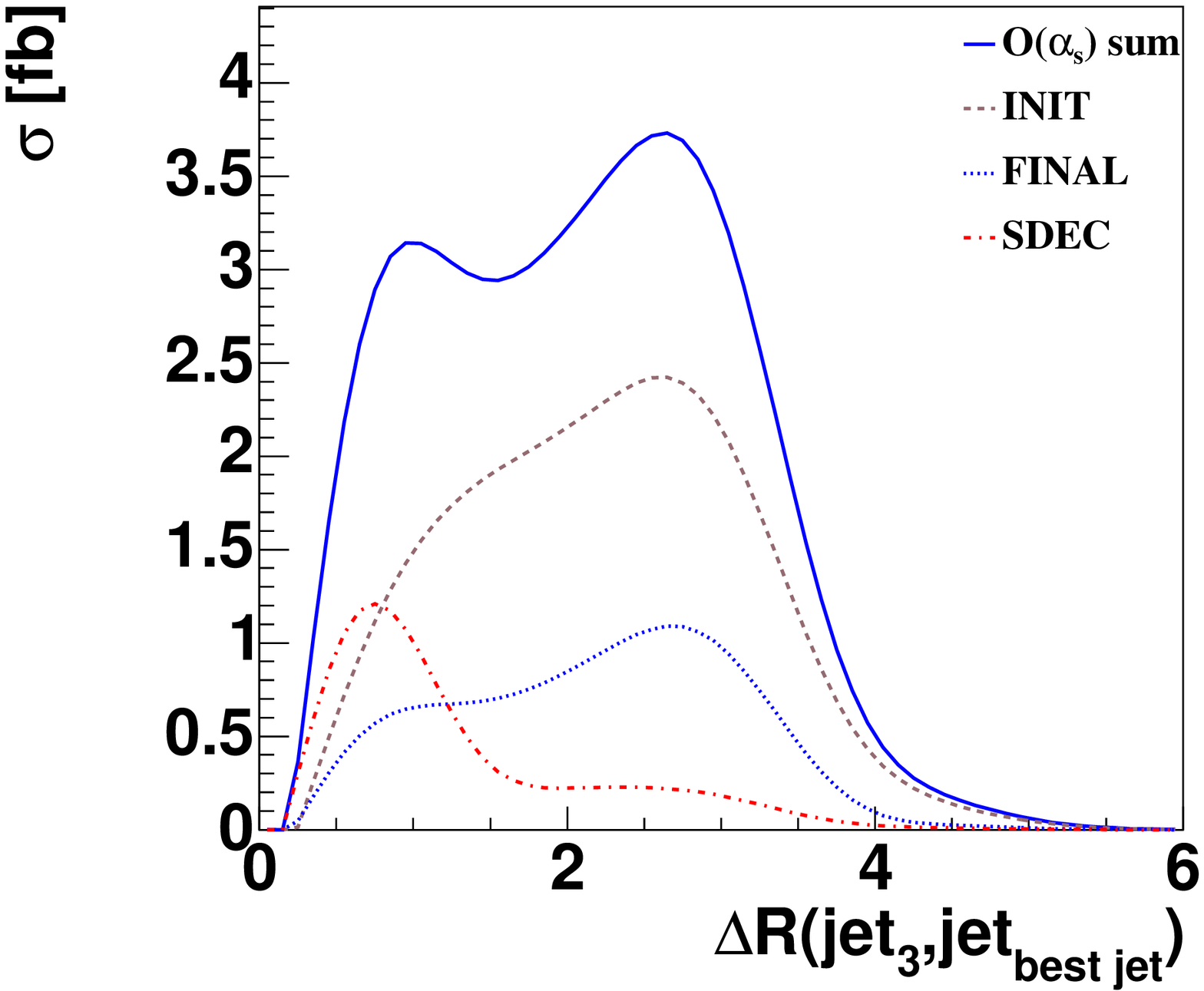}
\includegraphics[scale=0.3]{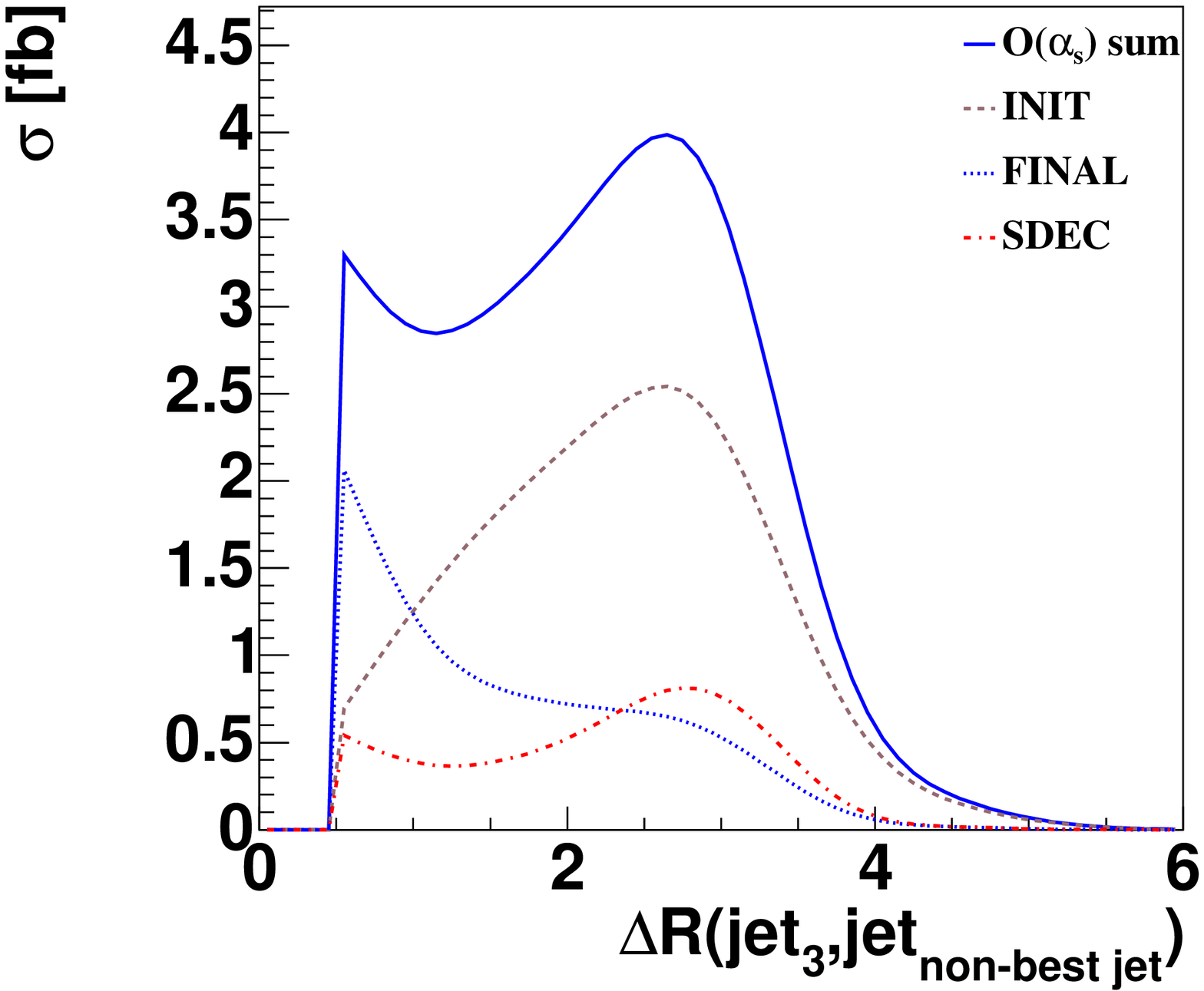}

\caption{Distance between the third jet and the best-jet (left) and distance
between the third jet and the non-best-jet (right) after selection
cuts for the various $\oalphas$ corrections. \label{fig:dRJet3bJet}}
\end{figure}
This tendency of decay-stage radiation to be associated with the final-state
$b$ quarks might lead one to expect that if the extra jet is ``near''
the $b$-jet it should be included in the mass reconstruction, and
if it is not then it should be excluded. Fig.~\ref{fig:dRJet3bJet},
which shows the distribution in $\Delta R$ between the extra-jet
and the best-jet, confirms that the decay-stage radiation peaks close
to the best-jet, and production-stage radiation peaks farther away.
Unfortunately, the production contribution is so large that it dominates
even at the low $\Delta R$ cutoff. A higher $E_{T}$ cut on the jet
would make this situation even worse. The best choice of what is ``near''
the $b$ quark will therefore balance the competing effects of decay
emission falling outside the cone and production emission falling
inside the cone. In Fig.~\ref{fig:dRJet3bJet}, the equivalent distribution
in $\Delta R$ between the extra jet and the non-best-jet is also
shown, where the production-stage radiation peaks close to the non-best-jet
due to the effect of gluon radiation in the final state. 

Tuning a prescription for dealing with the extra jet in s-channel
single top events is further complicated because effects of multiple
emission, hadronization, and detector resolutions will affect the
result.

Finally, we point out that the above conclusion does not strongly
depend on the choice of jet algorithm. We have checked that at the
parton level using the Durham $k_{T}$ algorithm~\cite{Catani:1992zp,Ellis:1993tq}
leads to a similar conclusion on the relative importance of the production-
and decay-stage gluon emission.

\subsubsection{Angular Correlation Between the Extra Jet and the Best-Jet}

While we cannot use the angular distance between the gluon jet and
the $b$-jet to distinguish production-stage emission from decay-stage
emission, it is nevertheless possible to separate the two using the
best-jet algorithm. 

\begin{figure}
\includegraphics[scale=0.3]{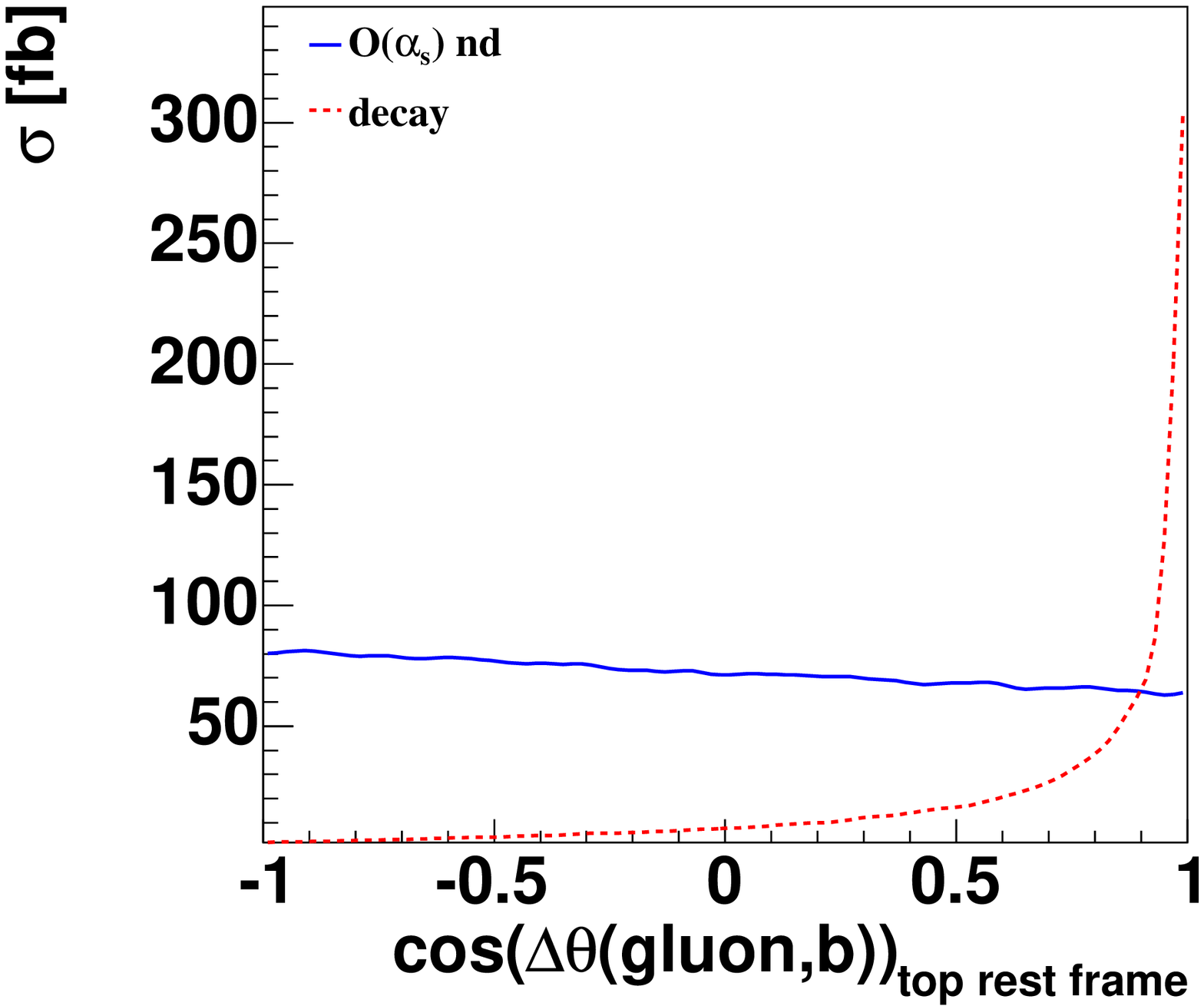}
\includegraphics[scale=0.3]{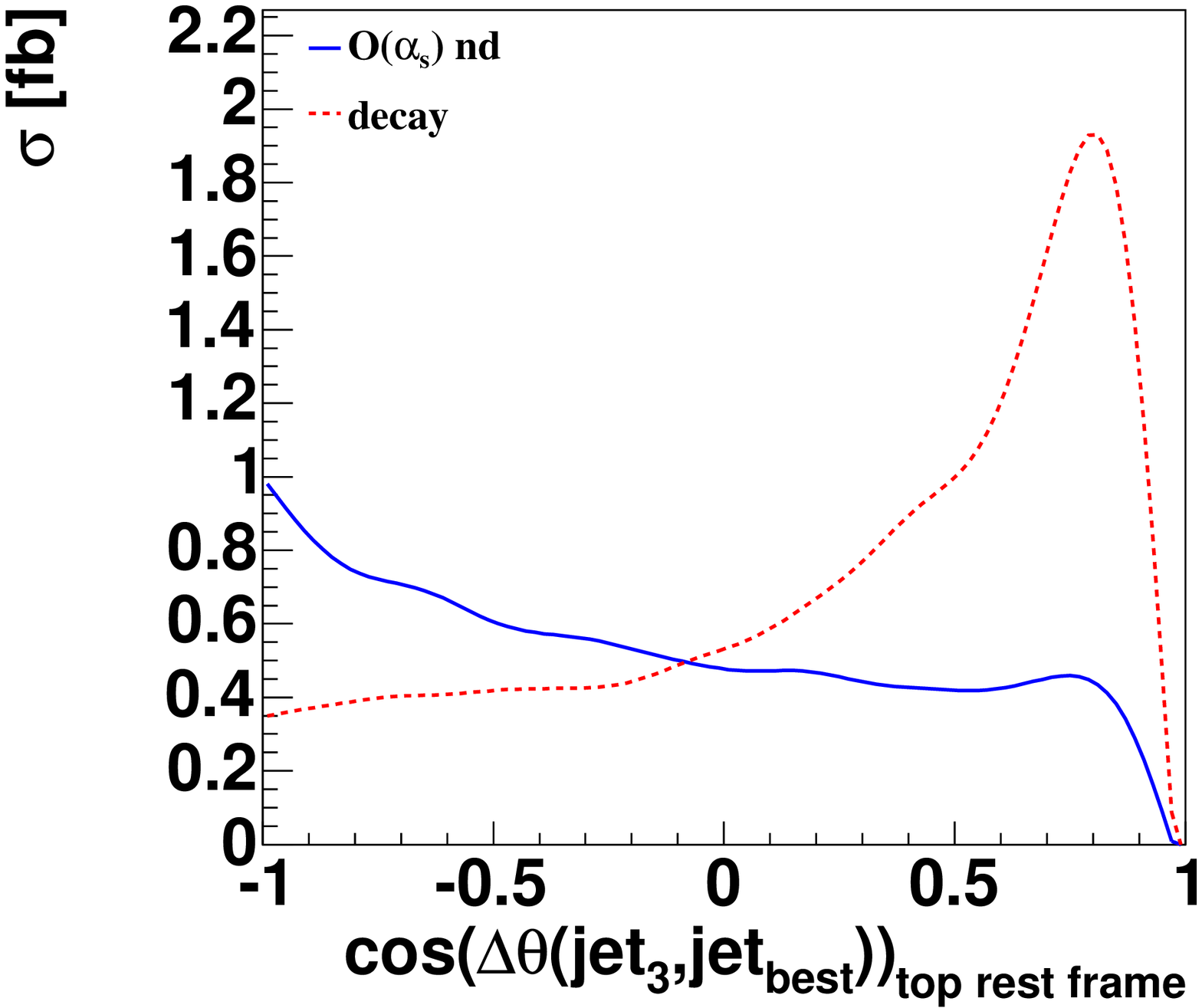}

\caption{Angular correlation $cos\theta$ between the gluon and the b quark
before any selection cuts using the full parton information (left)
and between the third jet and the best-jet after selection cuts (right).
The solid line shows all $\oalphas$ contributions except for the
decay part, while the dashed line shows only the $\oalphas$ decay
contribution.\label{fig:CosThetaJet3BgJet}}
\end{figure}

The left-hand side of Fig.~\ref{fig:CosThetaJet3BgJet} shows the
angular correlation $cos\theta$ between the gluon and the $b$~quark
at parton level before cuts.  The right-hand side of the same figure
shows the same correlation after event reconstruction between the
third jet and the best-jet. In this case there is a clear separation
between production-stage and decay-stage emission and the best-jet
algorithm can be used to separate the two because only those events
are included in the plot for which the best-jet algorithm consists
of exactly two jets.

\section{Single Top Production as Background to SM Higgs Searches\label{sec:Higgs}}

The s-channel single top quark process also contributes as one of
the major backgrounds to the SM Higgs searching channel $q\bar{q}\rightarrow WH$
with $H\rightarrow b\bar{b}$. In this case it is particularly important
to understand how $\oalphas$ corrections will change distributions
around the Higgs mass region. 

\begin{figure}
\includegraphics[scale=0.3]{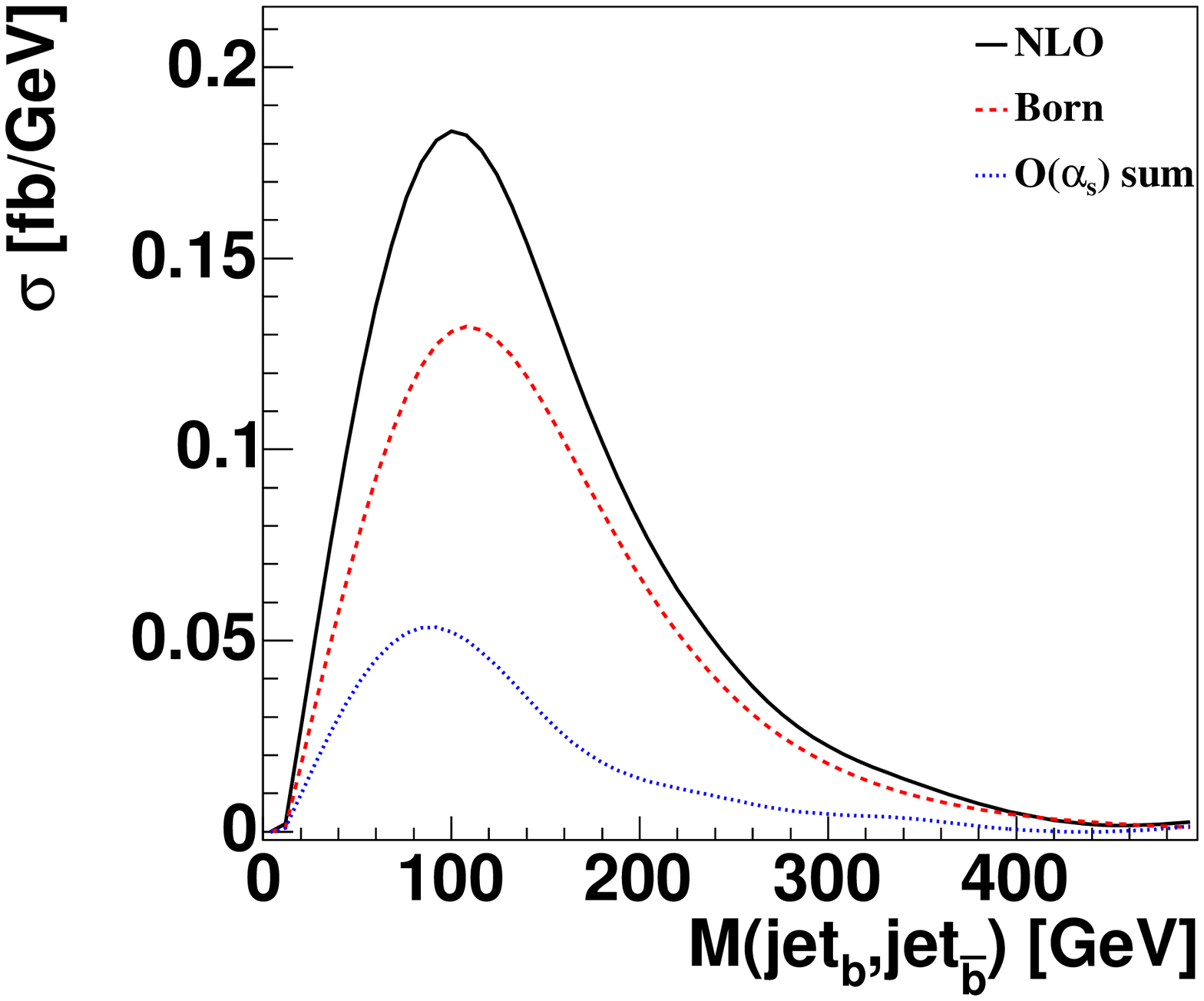}
\includegraphics[scale=0.3]{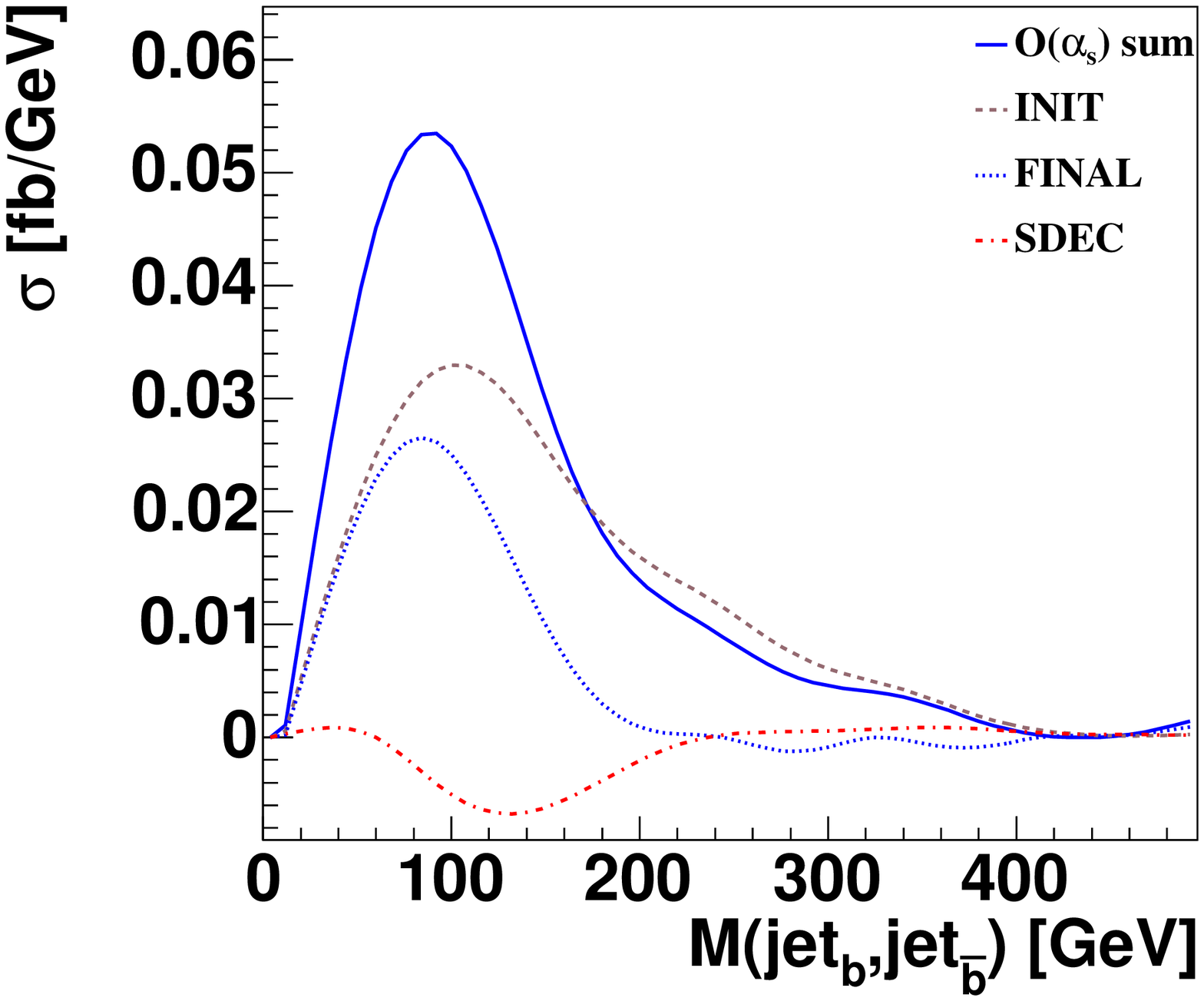}

\caption{Invariant mass of the ($b$-jet, $\bar{b}$-jet) system after selection
cuts, comparing Born level to $\oalphas$ corrections.\label{fig:bJetbbarJetMass}}
\end{figure}

Because of the scalar property of the Higgs boson, its decay products
$b$ and $\bar{b}$ have symmetric distributions. Fig.~\ref{fig:bJetbbarJetMass}
shows the invariant mass distribution of the $b$-jet, $\bar{b}$-jet
pair. For a Higgs signal, this invariant mass of the two reconstructed
$b$-tagged jets would correspond to a plot of the reconstructed Higgs
mass. Thus, understanding this invariant mass distribution will be
important to reach the highest sensitivity for Higgs boson searches
at the Tevatron. The figure shows that at $\oalphas$, the invariant
mass distribution not only peaks at lower values than at Born level,
it also drops off faster. This change in shape is particularly relevant
in the region focused on by SM Higgs boson searches of
$115\leq m_{b\bar{b}}\leq 130~{\rm GeV}$
where the Higgs associated production cross section is also at the $fb$ level. 
In particular the SDEC contribution,
while small overall, has a sizable effect in this region of the invariant
mass and will thus have to be considered in order to make reliable
background predictions for the Higgs boson searches. 

\begin{figure}
\includegraphics[scale=0.3]{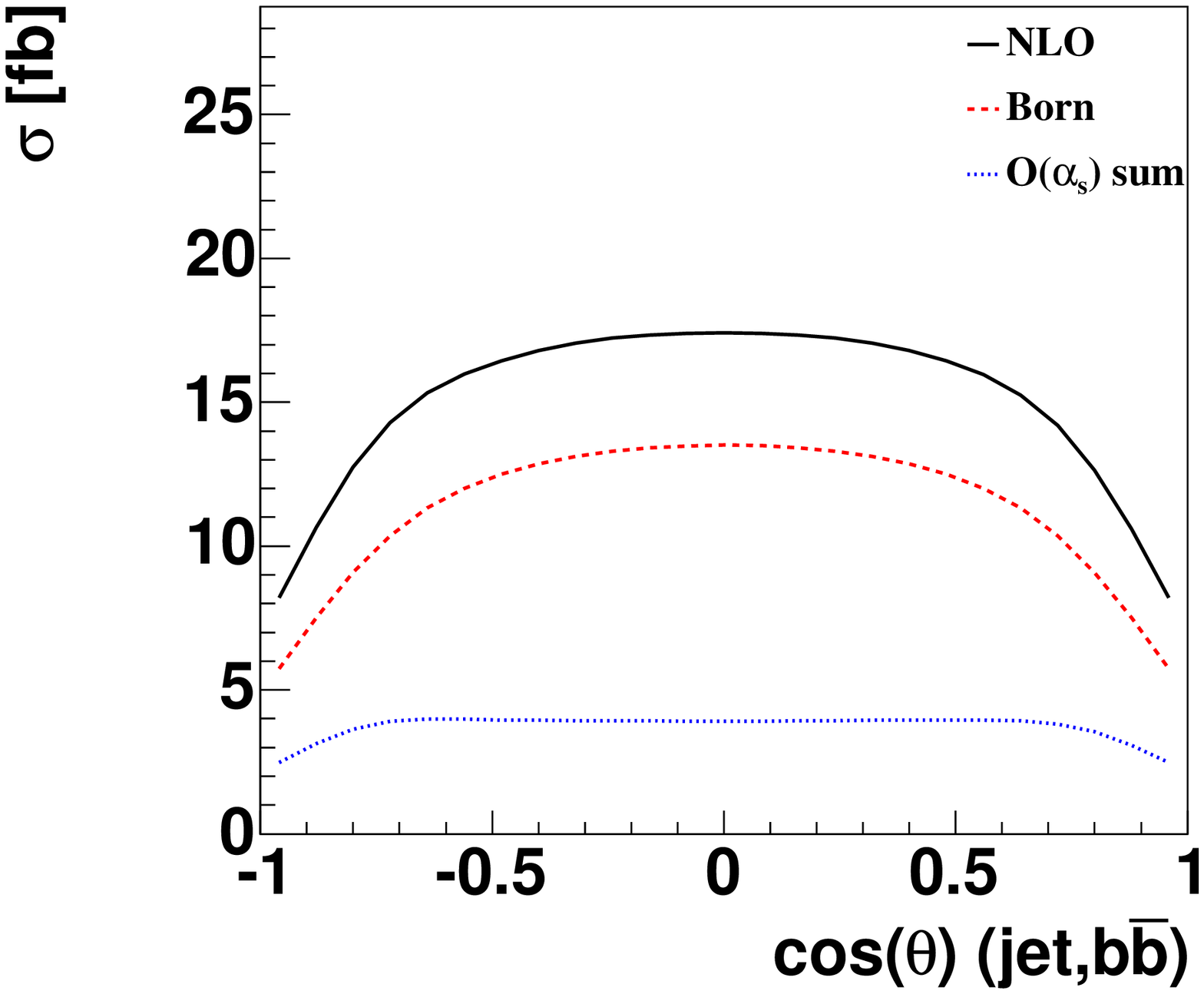}
\includegraphics[scale=0.3]{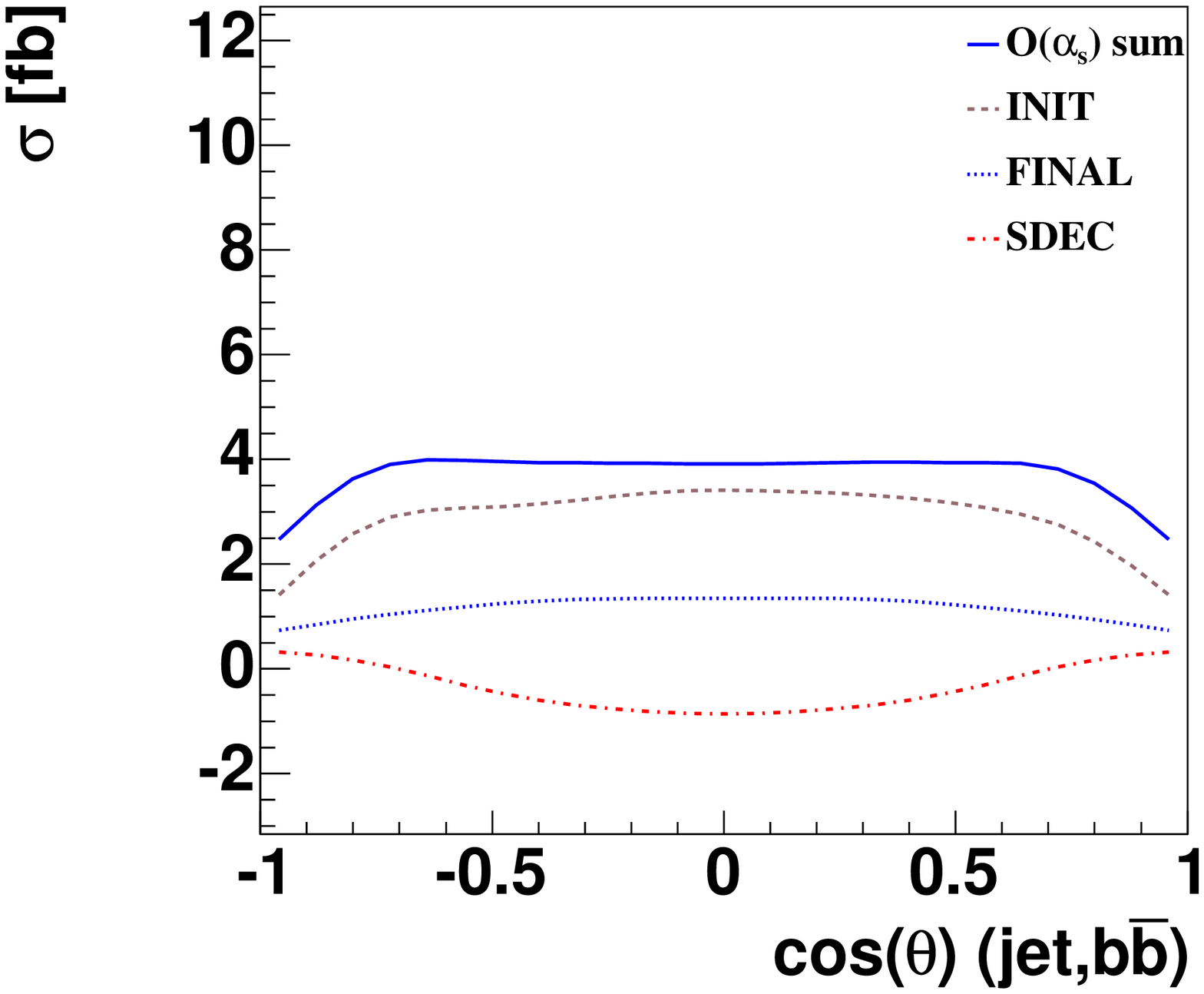}

\caption{Angular distance $cos\theta$ between a $b$-tagged jet and the ($b$
jet, $\bar{b}$ jet) system after selection cuts, comparing Born level
to $\oalphas$ corrections. \label{fig:bJetbbarJetCosTheta}}
\end{figure}

The higher order QCD corrections are evident in the 
distribution of $cos\theta$ for the two $b$-tagged jets, 
cf. Fig.~\ref{fig:bJetbbarJetCosTheta}. 
Here $\theta$ is the angle between the direction of a $b$-tagged
jet and the direction of the ($b$-jet, $\bar{b}$-jet) system, in
the rest frame of the ($b$-jet, $\bar{b}$-jet) system. Experiments
cannot distinguish between the $b$- and the $\bar{b}$-jets, we therefore
include both the $b$-jet and the $\bar{b}$-jet in the graph. This
distribution is generally flat at Born level, with a drop-off at high
$cos\theta$ due to jet clustering effects, and a drop-off at negative
$cos\theta$ due to kinematics. The $\oalphas$ corrections change
this distribution significantly and result in a more forward peak
of the distribution. 
In particular the SDEC contribution flattens out the $O(\alpha_s)$ 
corrections and therefore the s-channel distribution resembles more
that of the Higgs boson signal. This becomes important when the shape of
this distribution is used to extract the Higgs boson signal from 
various backgrounds.

\section{Conclusions\label{sec:Conclusions}}

We have presented a next-to-leading order study of s-channel single
top quark events at the Tevatron, including $\oalphas$ QCD corrections
to both the production and decay of the top quark. 
To obtain an accurate prediction of the inclusive rate of 
s-channel single top production, a ``modified''
narrow width approximation was adopted to link the production of the
top quark with its decay (and thus preserve top quark spin information)
instead of the usual narrow width approximation.
In the former, a Breit-Wigner distribution of top quark mass
is generated in the phase space generator of the Monte Carlo program, and
the generated mass is also used to calculate the scattering matrix elements.
Hence, the effect of the top quark width is included. In the latter, a fixed value
of top quark mass is used.
The impact of kinematical cuts on the acceptances has been
studied for several different sets of cuts. We found that the difference between 
Born level and NLO acceptances is about 5\% for two different sets of 
$p_T$ and $\eta$ cuts.
This difference becomes significantly larger when changing the jet clustering cone size and 
jet-lepton separation cut $R_{cut}$ from 0.5 to 1.0. Consequently, a 
constant K-factor cannot
be used to normalize Born-level distributions to NLO, especially for large 
values of $R_{cut}$.

We categorize the $\oalphas$ contributions to the s-channel single 
top process into three 
gauge invariant sets: the initial state corrections, the final state
corrections and the top quark decay corrections. Keeping track of the 
different categories facilitates the comparison between event generators and 
exact NLO predictions.
 
The $\oalphas$ corrections are significant in size and contribute
over 40\% of the inclusive cross section at NLO, a large fraction
of which in events with three reconstructed jets in the final state.
They also affect the shape of some of the important kinematical distributions
that might be used experimentally to separate the s-channel single
top signal from the various backgrounds. Among the higher order QCD corrections, 
the initial state correction dominates over the corrections from the final state and top decay 
processes in all inclusive single particle kinematical distributions. This implies
that in particular the initial state soft gluon resummation effects need to be modeled properly in
single top event generators. We have also found that the $\oalphas$ decay contribution, 
while small in size, has a significant impact on several distributions.

In order to study top quark properties such as the top quark polarization, we need to
reconstruct the top quark
by combining the reconstructed $W$ boson with one of the jets in the event. There is an ambiguity
in the choice of jet for this reconstruction due to the presence of two $b$ jets in the event
(the $b$ quark from the top quark decay and the $\bar{b}$ quark produced with the top quark). 
We use two different methods to resolve this ambiguity: the leading jet (highest $p_T$ jet)
and the best-jet (giving a $Wj$ invariant mass closest to 178GeV). 
Our study shows that the best-jet algorithm identifies the correct jet in about 80\% of the
events while the leading jet is the correct jet in only 55\% of the events. We thus choose
the best-jet algorithm to explore kinematical correlations in the event.

The top spin correlation is examined in both the helicity basis and optimal basis. 
At parton level, without any kinematical cuts, both bases show strong correlations,
with a polarization fraction of 98\% in the optimal basis and 82\% in the helicity basis.
The measured polarization as well as the difference in fraction of polarization
between the optimal basis and the helicity basis
is reduced after event reconstruction due to the event selection cuts and the
top quark reconstruction procedure (imperfect $W$ reconstruction). It is further reduced
by the $\oalphas$ corrections, resulting in a fraction of polarization of about 70\% in
both basis at NLO.

As a major irreducible background to $W^{\pm}H$ searches, the s-channel
single top quark process needs to be well understood in order to discover
the Higgs boson or set a bound on the Higgs boson mass. We showed
that the $\oalphas$ corrections have an effect on distributions that
are used in $W^{\pm}H$ searches, in particular in the Higgs mass region 
$115~{\rm GeV}\leq m_H \leq 130~{\rm GeV}$. The contribution from the top quark decay process, 
has a sizable effect in this region of the invariant mass 
distribution of the $b$ and $\bar{b}$ jets. The decay contribution also affects
the shape of other kinematical distributions, and reliable background predictions
in Higgs boson searches will therefor need to include higher order QCD corrections.

In this work we have studied the s-channel single top process at NLO and 
shown that there are 
significant differences compared to Born level in the cross section and several 
kinematical distributions that are relevant to select single top events at 
the Tevatron. 
A more realistic study including the analysis of background 
processes and the comparison of signal rates and distributions
predicted by various theory calculations 
will be presented elsewhere.

\begin{acknowledgments}
We thank Hong-Yi Zhou for collaboration in the early stage of this
project, and R. Brock for a critical reading of the manuscript and
discussions. We also thank Z. Sullivan for 
pointing out a numerical error in our previous result of the 
narrow width approximation calculation.
CPY thanks the hospitality of National Center for Theoretical
Sciences in Taiwan, ROC, where part of this work was completed. This
work was supported in part by NSF grant PHY-0244919. 
\bibliographystyle{plunsrt}
\clearpage\addcontentsline{toc}{chapter}{\bibname}\bibliography{./reference}
\end{acknowledgments}

\end{document}